%% file: TheoryUncert.tex
\documentclass[preprintnumbers,groupedaddress,nofootinbib,aps]{revtex4}
\pdfoutput=1

\usepackage{amsmath,amssymb,mathrsfs,graphicx}
\usepackage{multirow}
\usepackage{subfigure}
\usepackage{hyperref}
\usepackage{url}

\begin{document}
\input{declare}

\let\oldtabular\tabular
\renewcommand{\tabular}{\scriptsize\oldtabular}

\title{Decoupling Theoretical Uncertainties from Measurements of the Higgs Boson}

\author{Kyle Cranmer$^1$}
\author{Sven Kreiss$^1$}
\author{David L\'opez-Val$^2$}
\author{Tilman Plehn$^3$}

\affiliation{$^1$Center for Cosmology \& Particle Physics, New York University, USA}
\affiliation{$^2$Centre for Cosmology, Particle Physics \& Phenomenology, Universit\'e Catholique de Louvain, Belgium}
\affiliation{$^3$Institut f\"ur Theoretische Physik, Universit\"at Heidelberg, Germany}

\begin{abstract}
We develop a technique to present Higgs coupling measurements, which
decouple the poorly defined theoretical uncertainties associated to
inclusive and exclusive cross section predictions.  The technique
simplifies the combination of multiple measurements and can be used in
a more general setting.  We illustrate the approach with toy LHC Higgs
coupling measurements and a collection of new physics models.
\end{abstract}

\maketitle

\bigskip \bigskip \bigskip
\tableofcontents
\newpage

\section{Introduction}
\label{sec:intro}

The discovery of the Higgs
boson~\cite{higgs,atlas_discovery,cms_discovery,lecture} has initiated
a vigorous program of precision Higgs measurements at the LHC. 
The primary focus is on the couplings of this new state in terms of an effective 
 Lagrangian based on the Standard Model gauge structure. 
The key question is if this Lagrangian
is fully renormalizable, as predicted  in the Standard Model. 
Deviations in the properties of the newly found particle from
the Standard Model predictions would indicate 
the first hint for physics beyond the Standard Model.

These precision studies provide one of the key motivation for new
accelerators~\cite{extrapolations,ilc}.
Complicating the interpretation of both the current results and future
prospects of the LHC is the presence of  uncertainties on
the theoretical link between Lagrangian parameters and LHC observables.  In this
paper we outline a strategy to decouple the theoretical uncertainties
from the experimental results while retaining the ability to incorporate those uncertainties in a subsequent stage we refer to as \textit{recoupling}.
\bigskip

In reporting results, experimentalists strive to find a representation of the results
that is as free of theoretical assumptions as possible while still being
convenient for addressing specific theories of interest.  
 For example, the measurement of a cross-section in a well defined fiducial region
allows experiments to report results that are not tied to a specific
theory at the cost of requiring the reader to estimate a
model-dependent acceptance.  This approach has advantages, but is
difficult to generalize.  For example, extracting the Higgs couplings requires simultaneous inference of multiple production and decay modes from the combination of multiple searches with correlated experimental systematic uncertainties~\cite{duehrssen,atlas_coup,cms_coup,sfitter}.  
A convenient approach to combining multiple measurements is the 
Best Linear Unbiased Estimator (BLUE) technique~\cite{BLUE}; however, 
that approach is limited to  measurements that are well within the Gaussian regime.
The methodology presented here  addresses both of these challenges.

The first wave of Higgs coupling measurements~\cite{atlas_discovery,atlas_bosons,les_houches} involves \textit{signal strength}
modifiers $\mu_{pd}$ that scale the total rate of events for a given
combination of production $(p)$ and decay $(d)$ modes,
\begin{alignat}{5}
\mu_{pd} = \cfrac{\sigma_p \times \text{BR}_d}
                {(\sigma_p \times \text{BR}_d)^\text{SM}}
= \left(\cfrac{\sigma_p}{\sigma^\text{SM}_p}\right) \, 
  \left(\cfrac{\Gamma_d}{\Gamma^\text{SM}_d}\right) \,
  \left(\cfrac{\Gamma^\text{SM}_H}{\Gamma_H} \right) \; . 
\label{eq:strength}
\end{alignat}
The ATLAS
experiment has made such likelihood functions
available~\cite{atlas_bosons2}.
If we focus on measuring the Higgs couplings in the Standard Model
Lagrangian we can link the event rates to the set of shifted Higgs
couplings $g_x/g_x^\text{SM}$ to any Standard Model particle
$x$. Kinematic distributions (and thus selection efficiencies and detector acceptances)
 will not change as long as the Higgs
couplings are roughly in their Standard Model ranges.   While
$\sigma_p$ and $\Gamma_d$ can usually be directly linked to a specific
coupling $g_x$, the appearance of the total width forces us to make
non-trivial assumptions and induces strong correlations. The LHC Cross Section Working Group has defined a number of benchmark scenarios that specify how the production cross section and branching ratios are modified with respect to the standard model values~\cite{HiggsXS}, and the ATLAS and CMS collaborations are reporting their results in terms of these benchmark scenarios.\bigskip

Thus far, results for signal strengths and Higgs couplings include the
theoretical uncertainty as part of the Standard Model prediction.
Large uncertainties on the production %
appear because of unknown higher orders in the
perturbative QCD expansion. The problem with these theoretical
uncertainties is two-fold: first, the size of the associated uncertainty
in not known. Second, the uncertainties are not statistical in nature -- there is no random variable associated to missing higher orders.  While there has been effort to assign a Bayesian degree of belief to these uncertainties~\cite{cacciari} and an effort to complete the perturbative series~\cite{david_passarino}, we lack an objective probabilistic interpretation of these uncertainties.
While the field may be able to settle on a consensus for both the size and shape of these uncertainties, it is an area of debate and will certainly change with time as theoretical progress is made.  Therefore, our view is that we should develop a technique in which the theoretical uncertainties are factorized from the experimental result.\bigskip

In Section~\ref{sec:approach} we describe the approach allowing us to decouple and recouple  the theoretical uncertainties from
the experimental measurement. This approach is independent of our
specific application to the Higgs coupling measurements at the LHC. In
Section~\ref{sec:qcd} we briefly discuss the issues with theoretical
uncertainties at hadron colliders and how they affect the Higgs
couplings measurement.  Section~\ref{sec:example} demonstrates the procedure with a toy example based on the ATLAS results presented in Ref.~\cite{atlas_bosons}.  Next, we consider some specific new physics scenarios that can be tested through Higgs coupling measurements in
Section~\ref{sec:new_physics}. A simple example is worked in detail in Appendix~\ref{Sec:SimpleExample} and more details on the new physics models in Appendix~\ref{app:new_physics}.

\section{The Approach}
\label{sec:approach}

\subsection{The Statistical Model}
\label{sec:model}

In this Section we outline briefly the statistical modeling approach used by the LHC experiments following Ref.~\cite{ATLAS:2011tau}.  Once the statistical model has been constructed, the LHC experiments employ the profile likelihood ratio to define confidence intervals~\cite{Cowan:2010js} on the parameters of interest.

The coupling measurements require defining several disjoint categories of events, indexed by $c$, which satisfy specific selection criteria designed, in part, to be particularly sensitive to a particular production or decay mode.  Each category has associated to it an expected (observed) number of events $\nu_c$ ($n_c$). Each category may also have some discriminating variable(s) $x$, such as an invariant mass, and a corresponding probability density function $f(x)$. The data associated to the $c^\text{th}$ category is denoted $\mathcal{D}_c = \{x_1, \dots, x_{n_c}\}$.  In general the expected number of events and their distribution will depend on both the signal strength parameters $\vec\mu$ and nuisance parameters $\vec\alpha$. The nuisance parameters $\vec\alpha$ parametrize both theoretical and experimental uncertainties. Typically, the expected number of events is written
\begin{equation}\label{eq:def_nu}
\nu_c(\vec\mu,\vec\alpha) = \sum_{p,d} \mu_{pd} s_{cpd}(\vec\alpha)  + b_c(\vec\alpha) \; ,
\end{equation}
where $b_c(\vec\alpha)$ is the background in this category, and
\begin{equation}\label{eq:def_s}
s_{cpd}(\vec\alpha) = L(\vec\alpha) \; \sigma_{p}^\text{SM}(\vec\alpha) \; 
 \text{BR}_d^\text{SM}(\vec\alpha) \; \epsilon_{cpd}(\vec\alpha) 
\end{equation}
is the expected Standard Model signal for production mode $p$ and decay mode $d$ predicted by the product of the integrated luminosity, cross section, branching ratio, and selection efficiency, each of which may depend on theoretical uncertainties parametrized by $\vec\alpha$.
 
Constraint terms associated with systematic uncertainties are described as $f_i(a_i | \alpha_i)$, where $\alpha_i$ are nuisance parameters and $a_i$ are auxiliary or control measurements designed to estimate those nuisance parameters.  
In the case of  experimental uncertainties, there are often real auxiliary measurements that are summarized by  $f_i(a_i | \alpha_i)$.  However, in the case of most theoretical uncertainties, the auxiliary measurement does not truly exist and an \textit{ad hoc} $f_i(a_i | \alpha_i)$ is introduced for convenience. The full likelihood function used by the experiments~\cite{Aad:2012an} can be written as a product of the main experimental measurement and the constraint terms
\begin{equation}\label{Eq:statmodel}
L_\text{full}( \vec\mu, \vec\alpha) = \underbrace{\prod_{c\in \text{category}} \left[ \Pois(n_c | \nu_c(\vec\mu, \vec\alpha) )\prod_{e=1}^{n_c} f_c(x_e | \vec\mu, \vec\alpha) \right] }_{\equiv L_\text{main}(\vec\mu,\vec\alpha)} \underbrace{ \prod_{i\in \text{syst}} f_i(a_i | \alpha_i)}_{\equiv L_\text{constr}(\vec\alpha)} \; .
\end{equation}

Typically, confidence intervals are then defined by contours of the profile likelihood ratio
\begin{equation}\label{eq:profile}
\lambda(\vec\mu) = \frac{L(\vec\mu, \hat{\hat{\alpha}}(\vec\mu))}{L(\hat{\vec\mu}, \hat{\vec\alpha})}
\end{equation}
where $\vec{\hat{\hat{\alpha}}}(\vec\mu)$ is the conditional maximum likelihood estimate and 
${\hat{\vec\mu}}, \hat{\vec\alpha}$ are the unconditional maximum likelihood estimates~\cite{Cowan:2010js}.

\subsection{The Effective Signal Strength}
\label{sec:signal_strength}

We are interested in inferring the values of the signal strength parameters $\vec\mu$, which scale the signal expectation $s_{cpd}$; however,  the presence of experimental and theoretical uncertainties mean that the signal and background expectations are functions of the nuisance parameters as in Eq.\eqref{eq:def_nu}.  Alternatively, we can introduce effective scale factor with respect to some fixed reference scenario $\vec\alpha_0$, so the expected number of events can be re-written
\begin{alignat}{5}
\nu_c(\vec\mu,\vec\alpha) &= \;
\sum_{p,d} \mu_{pd} \; s_{cpd}(\vec\alpha) + b_c(\vec\alpha) 
\notag \\
&\to 
\sum_{p,d} \muefff{cpd}(\vec\mu,\vec\alpha) \; s_{cpd}(\vec\alpha_0) + b_c(\vec\alpha_0)  \;.
\label{eq:def_mueff}
\end{alignat}
The key conceptual jump is to realize that we can think of $\muefff{cpd}$ not as a function, but as a well-defined parameter free of theoretical uncertainty that we can infer directly.

While the signal strength parameters $\mu_{pd}$ we ultimately want to infer are independent of the details of the individual analysis categories, the effective signal strength $\muefff{cpd}$ is specific to the $c^\text{th}$ category due to the selection efficiency (and, more generally, the distributions $f_c(x|\vec\mu,\vec\alpha)$).  In particular, selection requirements that leverage exclusive or differential properties of a specific production and decay will introduce category-specific theoretical uncertainties.  \bigskip

State-of-the art Higgs property measurements can include hundreds of categories of events. The full likelihood $L_\text{full}$ defined in Eq.\eqref{Eq:statmodel} encodes a detailed description of the correlated effect of common experimental systematic uncertainties.  In practice, we want to find some coarse graining of the many categories into a few groups so that we can ignore the category index within each group. By taking a common category-weighted signal strength factor we can suppress the category index and make the replacement $\muefff{cpd} \to \muefff{pd}$. By doing so, we can define
\begin{equation}\label{eq:LeffDef}
L_\text{eff}(\vecmueff) \equiv L_\text{main}(\vec\mu=\vecmueff,\vec\alpha=\vec\alpha_0) \; .
\end{equation}
 The goal is to show that by providing $L_\text{eff}(\vecmueff)$, the reparametrization $\vecmueff(\vec\mu,\vec\alpha)$, and the constraint terms $f_i(a_i| \alpha_i)$ that we can recouple these ingredients and approximate the full likelihood 
\begin{equation}\label{Eq:Lrecouple}
 L_\text{full}(\vec\mu,\vec\alpha)  \approx  L_\text{recouple}(\vec\mu,\vec\alpha) \equiv L_\text{eff}(\vecmueff(\vec\mu,\vec\alpha)) \cdot  L_\text{constr}(\vec\alpha) \; . 
 \end{equation}

In the case of an inclusive cross section uncertainty $\muefff{cpd}$ is the same for all $c$ ---  in which case we say the effect of that uncertainty is \textit{category-universal}.  If all uncertainties are category-universal, then it is possible for this approach to be exact.  More generally the grouping of categories will lead to $\vecmueff(\vec\mu,\vec\alpha)$ encoding some weighted effect from the individual categories.  
We will discuss some examples in the next section.

\subsection{Reparametrization Templates}
\label{sec:templates}

The art of this approach lies in choosing a template for the reparametrization in which the coefficients of the template can be effectively deduced from the likelihood.  We treat the likelihood as a ``black box'' since the diversity and complexity of statistical models created by experimentalists and encompassed by Eq.\eqref{Eq:statmodel} is so diverse.  In the case that the reparametrization $\vecmueff(\vec\mu,\vec\alpha)$ is category-universal, which can be trivially achieved if the ingredients are explicitly provided for each category, this reformulation of the likelihood can be exact.\bigskip

For example, a natural way to parametrize the dependence of the expected signal due to uncertainties that modify inclusive production cross sections is
\begin{equation}\label{Eq:inclusiveProdTemplate}
s_{cpd}(\vec\alpha) =  s_{cpd}(\vec\alpha_0) \left [ 1+\sum_i \eta_{pi} (\alpha_i-\alpha_{0,i}) \right ] \; (\forall c,d)  \; ,
\end{equation}
which is equivalent to
\begin{equation}\label{Eq:inclusiveProdTemplateMu}
\muefff{pd}(\vec\mu,\vec\alpha) =  \mu_{pd}    \left [ 1+\sum_i \eta_{pi} (\alpha_i-\alpha_{0,i}) \right ] \;  (\forall c,d) \;.
\end{equation}
In this situation, $\muefff{pd}(\vec\mu,\vec\alpha)$ is not linear in the full set of parameters, but is bi-linear in $(\mu_{pd},\vec\alpha)$. This $\mu$ scaling is important for capturing the behavior of the likelihood away from the maximum likelihood estimate and distinguishes this approach from techniques such as BLUE~\cite{BLUE}. 

In the case of uncertainties that only affect the background through 
\begin{equation}
b_c(\vec\alpha) = b_c(\vec\alpha_0) \left [ 1+\sum_i \phi_{ci} (\alpha_i-\alpha_{0,i}) \right ] \; (\forall p,d)
\label{Eq:backgroundTemplate}
\end{equation}
the equivalent form of the effective signal strength is 
\begin{equation}
\muefff{pd}(\vec\mu,\vec\alpha) =  \mu_{pd}  + \frac{b_c(\vec\alpha_0)}{s_{cpd}(\vec\alpha_0)} \left [ \sum_{i} \phi_{ci} (\alpha_i-\alpha_{0,i}) \right ] \; (\forall p,d)\;,
\end{equation}
which is linear in $\mu_{pd}$ and $\vec\alpha$.  Because  $b_c(\vec\alpha_0)/s_{cpd}(\vec\alpha_0)$ is a constant,  this pre-factor can be absorbed into $\phi_{ci}$ and the category-weighted effect would simply be written $\phi_i$.  

Another example is motivated by the 
large uncertainty associated to gluon-fusion (ggF) Higgs production   with two additional jets, which populates the categories meant to isolate weak boson fusion (VBF). We would expect the uncertainty to modify the $\muefff{c,p=\text{VBF},d}$ signal strength for weak boson fusion, but be proportional to $\mu_{\text{ggF}}$.  Thus we should anticipate templates of the form
\begin{equation}
\muefff{pd}(\vec\mu,\vec\alpha) =  \mu_{pd}  + \sum_{i,p'} \mu_{p'd} \, \eta^{p'}_{pi}\, (\alpha_i-\alpha_{0,i})  \;  .
\label{Eq:genericTemplate}
\end{equation}
Combining these three situations, a fairly general template would be
\begin{equation}
\muefff{pd}(\vec\mu,\vec\alpha) =  \mu_{pd}    + \sum_{i,p' } \mu_{p'd}  \, \eta^{p'}_{pi}\, (\alpha_i-\alpha_{0,i}) + \sum_i \phi_{i} (\alpha_i-\alpha_{0,i})  \;  ,
\label{Eq:generalTemplate}
\end{equation}
where we can identify $\eta_{pi} = \eta_{pi}^p$ from Eq.\eqref{Eq:inclusiveProdTemplate} and $b_c(\vec\alpha_0)\phi_{ci}/s_{cpd}(\vec\alpha_0)=\phi_{i}$.  This general template involves $n_\alpha (n_p^2 + 1)$ coefficients (per grouping of categories).  \bigskip

The LHC experiments must cope with additional complications.  First is the fact that uncertainties on $\alpha_i$ can be large, and linear extrapolation of the effect of this uncertainty on a signal or background rate can lead to unphysical expectations like $s_{cpd}<0$ or $b_c<0$.  To cope with this, the LHC experiments typically implement log-normal priors (constraint terms), which are implemented via reparametrization so that $f_i(a_i|\alpha_i)$ is a Gaussian distribution and $s_{cpd}(\alpha_i)$ is an exponential response function.  This approach has two advantages:  it ensures $s_{cpd}(\alpha_i) > 0$ and it allows for multiple signal expectations with different sensitivities to a common source of  uncertainty to have a similar log-normal behavior  parametrized by $\alpha_i$. 

Second, experiments often have a few numbers with which to parametrize the signal (and background) expectations.  Typically, this is based on a nominal $s_{cpd}(\alpha_{i}=0)$ and  ``$\pm 1\sigma$'' variations on $s_{cpd}(\alpha_i=\pm1)$ (using a conventional scaling of $\alpha_i$).  Often the variation of $s_{cpd}(\alpha_i=\pm1)$ with respect to $s_{cpd}(\alpha_{i}=0)$ is asymmetric, which requires some assumptions about the intermediate behavior and the use of \textit{ad hoc} interpolation algorithms, such as second degree polynomial or higher degree polynomials that match the exponential extrapolation up to the second derivative~\cite{Conway:2011in,Cranmer:2012sba}.  

Finally, to ensure the positivity of $s_{cpd}(\vec\alpha)$ under the joint effect of several sources of uncertainty, it is common that the interpolation/extrapolation template is multiplicative over the nuisance parameters.  In generic terms, this often leads to signal expectations parametrized as
\begin{equation}\label{Eq:multiplicativeTemplate}
s_{cpd}(\vec\alpha) =  s_{cpd}(\vec\alpha_0) \prod_i I(\alpha_i) \; ,
\end{equation}
where $I(\alpha_i)$ is some interpolation/extrapolation function based on $s_{cpd}$ evaluated at several points in $\alpha_i$.  These \textit{ad hoc} choices influence the resulting inference and further strengthen the motivation to decouple theoretical uncertainties from the presentation of experimental results.

\subsection{Determining the Coefficients of the Reparametrization Template}
\label{sec:coeffs}

Following Eq.\eqref{Eq:Lrecouple}, our goal is to show that by providing $L_\text{eff}(\vecmueff)$, the reparametrization $\vecmueff(\vec\mu,\vec\alpha)$, and the constraint terms $f_i(a_i, \alpha_i)$ we can ``recouple'' the ingredients to approximate the original likelihood $L_\text{full}(\vec\mu,\vec\alpha)$. Below we develop two approaches to determine the coefficients of these reparametrization templates.

\paragraph{Via the local covariance matrix}

A direct path towards this goal is to use a re\-para\-metrization template like the ones introduced in Sec.~\ref{sec:templates} and determine the coefficients that reproduce the local covariance structure around the maximum likelihood estimate $(\hat{\vec\mu}, \hat{\vec\alpha})$.
The maximum likelihood estimate is defined by $-\nabla \ln L = 0$, while the Hessian of $-\ln L$ captures the local covariance structure and is referred to as the observed Fisher information matrix or precision matrix.  In particular, the observed Fisher information is an estimate of the inverse of the covariance matrix
\begin{equation}
V^{-1}_{ij} = - \partial_i \partial_j  \ln L \; .
\end{equation}
It is helpful to factorize the full likelihood in Eq.\eqref{Eq:statmodel} as a product of the main measurement and the constraint terms for the nuisance parameters $\vec\alpha$.  This is equivalent to decomposing the full information matrix into two parts
\begin{equation}\label{eq:InfoMatrixDecomp}
V^{-1}_\text{full} = V^{-1}_\text{main}  + V^{-1}_\text{constr} \;.
\end{equation}
By fixing $\vec\alpha=\vec\alpha_0$ the experiments can provide the (profile) likelihood for the effective signal strength $L_\text{eff}(\vecmueff)$ via Eq.\eqref{eq:LeffDef}.  We denote the Fisher information for this effective signal strength as  $V^{-1}_\text{eff}$.
Reparametrizing $L_\text{eff}(\vecmueff)$ via $\vecmueff \to \vecmueff(\vec\mu,\vec\alpha)$ implies the following transformation to the Fisher information matrix
\begin{equation}\label{eq:InfoMatrixReparam}
V^{-1}_\text{eff}(\vecmueff) \to V^{-1}_\text{eff}(\vec\mu,\vec\alpha) = J^T V^{-1}_\text{eff} J \;,
\end{equation}
where the Jacobian of the transformation is defined by
\begin{equation}
J = \frac{\partial( \vecmueff)}{\partial( \vec\mu, \vec\alpha)} \; .
\end{equation}
Even though $V^{-1}_\text{eff}$ may be a constant matrix, the Jacobian in Eq.\eqref{eq:InfoMatrixReparam} depends on $(\vec\mu,\vec\alpha)$ so that the resulting information matrix need not be constant.  This is an important point as the analysis below is only sensitive to the linear approximation of  $\vecmueff(\vec\mu,\vec\alpha)$ at $(\hat{\vec\mu}, \hat{\vec\alpha})$ although the template may encode important non-linear behavior.
Combining Eq.\eqref{eq:InfoMatrixDecomp} and
Eq.\eqref{eq:InfoMatrixReparam} gives us
\begin{equation}\label{eq:InfoMatrixEquality}
V^{-1}_\text{full} = J^T V^{-1}_\text{eff} J  + V^{-1}_\text{constr} \;.
\end{equation}
The final stage of this procedure is either to check if Eq.\eqref{eq:InfoMatrixEquality} holds for a particular reparametrization $\vecmueff \to \vecmueff(\vec\mu,\vec\alpha)$ or to determine the coefficients for a template by imposing this equality.

\bigskip

Perhaps the most intuitive way to think of the effect of a given uncertainty is the shift it induces in the best fit value of the signal strengths $\vec\mu$.   This way of thinking requires keeping the $\vec\alpha$  fixed and considering the likelihood is only a function of $\vec\mu$.  For example, Fig.~\ref{fig:counting_signalStrength}(b) shows the shift in the $L(\vec\mu)$ contour due to a shift in the gluon fusion inclusive cross section.  We denote the best fit $\vec\mu$ with fixed $\vec\alpha$ as 
\begin{equation}\label{eq:mufix}
\hat{\vec\mu}^\text{fix}(\vec\alpha) \equiv \text{argmax}_{\vec\mu} \; L_\text{full}(\vec\mu,\vec\alpha) 
\end{equation}
with $\hat{\vec\mu}^\text{fix}(\hat{\vec\alpha}) = \hat{\vec\mu}$.  Because of the definition $L_\text{full}(\vec\mu,\vec\alpha_0) \equiv L_\text{eff}(\vec\mu)$ in  Eq.\eqref{eq:LeffDef}, this implies that $\hat{\vec\mu}^\text{fix}(\vec\alpha_0) = \hat{\vec\mu}^\text{eff}$.  Finally, if the main measurement is not able to measure the nuisance parameters, \textit{i.e.} there is flat direction in the likelihood or a degeneracy between $\vec\mu$ and $\vec\alpha$, then $\vec\alpha_0=\hat{\vec\alpha}$.  By using a specific template --- in this case Eq.\eqref{Eq:inclusiveProdTemplateMu} --- we can equate  $\mu^\text{eff}_p(\hat{\mu}_p^\text{fix}, \alpha_i) = \hat{\mu}_p^\text{fix}(1+\eta_{ip} \alpha_i) = \hat{\mu}_p$ and
 then explicitly evaluate the partial derivative that quantifies the shift to the best fit value of $\vec\mu$
\begin{equation}\label{Eq:SolveViaPartial}
\left . \frac{\partial \hat{\mu}^\text{fix}_p}{\partial \alpha_i}  \right |_{\hat{\vec\mu},\hat{\vec\alpha}} =  - \hat\mu_p \eta_{ip}\; .
\end{equation}
These partial derivatives are visualized as vectors in the signal strength plane in Fig.~\ref{fig:visEtas}.\bigskip

Another way to arrive at Eq.\eqref{Eq:SolveViaPartial}  is to approximate the likelihood in the neighborhood of the maximum likelihood estimate as a multivariate Gaussian  $G(\vec\mu,\vec\alpha | \hat{\vec\mu},\hat{\vec\alpha}, \Sigma)$. The conditional distribution of $\vec\mu$ given $\vec\alpha$ is also a multivariate Gaussian with mean given by
\begin{equation}
\hat{\vec\mu}^\text{fix}(\vec\alpha) = \hat{\vec\mu} + \Sigma_c \Sigma_\alpha^{-1}(\vec\alpha - \hat{\vec\alpha}) \; ,
\end{equation}
where $\Sigma_c$ is the upper-right sub-block of the full covariance matrix $\Sigma$
 \begin{equation}
(V^{-1}_\text{full})^{-1}= \Sigma = \left[\begin{array}{cc}
\Sigma_\mu & \Sigma_c \\
\Sigma_c^T & \Sigma_\alpha \\\end{array}\right] \; .
\end{equation}
In situations that the main measurement does not constrain or pull on the nuisance parameters, $\Sigma_\alpha$ is just the covariance matrix associated to the constraint term defined in Eq.\eqref{Eq:statmodel}. In general $\Sigma_c$ will depend both on the constraint terms and the main measurement.  Clearly, the conditional likelihood with $\vec\alpha$ fixed as in Eq.\eqref{eq:mufix} is independent of the constraint term, thus the product $(\Sigma_c \Sigma_\alpha^{-1})$ can only depend on the details of the main measurement. Through the techniques developed for regression in the general linear model~\cite{KendallCh29} one can show that
\begin{equation}\label{Eq:SolveViaCov}
\left . \frac{\partial \hat{\mu}^\text{fix}_p}{\partial \alpha_i}  \right |_{\vec\alpha=\hat{\vec\alpha}} =(\Sigma_c \Sigma_\alpha^{-1})_{ip}  = - \hat\mu_p \eta_{ip}\; ,
\end{equation}
where the right-most equivalence is specific to the template of Eq.\eqref{Eq:inclusiveProdTemplateMu} and explicit computation requires an assumption about the constraint terms even though the result is independent of those assumptions.%
  This is shown explicitly for a simple example in Appendix~\ref{Sec:SimpleExample}.  Note, one can eliminate the constraint terms and only consider the main measurement, in which case $\Sigma$ in Eq.\eqref{Eq:SolveViaCov} is a singular matrix and one must use the pseudo-inverse.  While this approach is  mathematically cumbersome, it is equivalent to Eq.\eqref{Eq:SolveViaPartial} and Eq.\eqref{Eq:SolveViaFisher}.\bigskip

A third approach is to impose Eq.\eqref{eq:InfoMatrixReparam} with $V^{-1}_\text{constr}$ subtracted from both sides
\begin{eqnarray}\label{Eq:SolveViaFisher}
V^{-1}_\text{main} = J^T V^{-1}_\text{eff} J \; .
\end{eqnarray}
This allows us to proceed without assumptions on the  constraint terms, which we are trying to decouple from the procedure.  
The upper-right  sub-bloc of this matrix leads to a system of equations that can be used to determine the coefficients of the template.  Again in the case of template of Eq.\eqref{Eq:inclusiveProdTemplateMu}, these linear equations
 provide the same solutions for $\eta_{ip}$ as Eq.\eqref{Eq:SolveViaPartial} and Eq.\eqref{Eq:SolveViaCov}.   These equations are solved explicitly for a simple example in Appendix~\ref{Sec:SimpleExample}.

It is worth noting that in the case of an arbitrary template $\vecmueff(\vec\mu,\vec\alpha; \vec\eta)$, these equations fix the linear behavior at $\vec\alpha_0$ (\textit{i.e.} $\partial \vecmueff / \partial \alpha_i |_{\vec\alpha=\vec\alpha_0}$), which may be non-trivially related to the coefficients $\vec\eta$.  For example, this is the case for the most common parametrizations encompassed by Eq.\eqref{Eq:multiplicativeTemplate} and used by the LHC experiments and  to implement asymmetric uncertainties~\cite{Cranmer:2012sba,Conway:2011in}. \bigskip

The information matrix from the full likelihood function has $(n^2+n)/2$ independent components, where $n=n_p+n_\alpha$ is  the sum of the number of parameters of interest and nuisance parameters. An $n_p \times n_p$ sub-block describes $\Sigma_\mu$ or $V_\text{eff}^{-1}$ and a $n_\alpha \times n_\alpha$ sub-block describes $\Sigma_\alpha$ or $V^{-1}_\text{constr.}$.  Thus, there are a remaining $n_p\times n_\alpha$ numbers in the local covariance encoded in $\Sigma_c$ or $V^{-1}_\text{main}$  that can be used to determine the coefficients of the template. This counting matches precisely for the templates of Eq.\eqref{Eq:inclusiveProdTemplate} and Eq.\eqref{Eq:backgroundTemplate}, but leaves ambiguity for the more general template of Eq.\eqref{Eq:genericTemplate}. 

\paragraph{The art of choosing the template}

As mentioned above, the local information provided by the observed information matrix evaluated at the maximum likelihood estimate is only sensitive to the linear behavior of  $\vecmueff(\vec\mu,\vec\alpha)$ at $(\hat{\vec\mu}, \hat{\vec\alpha})$.  For example Eq.\eqref{Eq:inclusiveProdTemplate} and Eq.\eqref{Eq:backgroundTemplate} can have the same linear behavior at $(\hat{\vec\mu}, \hat{\vec\alpha})$ though the effect of an uncertainty $\alpha_i$ scales with $\vec\mu$ in the former and not in the latter.  This difference is illustrated in Fig.~\ref{fig:TemplateComparison}.
Thus, while both templates will be equivalent locally, if one wants to capture the higher-order corrections encoded in the different templates some additional information will be required.  This information can either be injected by hand or by using information about the likelihood away from $(\hat{\vec\mu}, \hat{\vec\alpha})$, as described below.  For instance, one may know that a particular nuisance parameter $\alpha_i$ is associated to the uncertainty in the inclusive cross section for the $p^\text{th}$ production mode, thus the coefficients of the general template can be restricted by hand to $\eta_{pi}^{p'}=\delta_{pp'}\eta_{ip}$ and $\phi_{ip}=0$. In another extreme case,  the uncertainty on gluon fusion production with two jets primarily affects the inference of $\muefff{p=\text{VBF}}$ signal strength for vector boson fusion, but the size of the shift is proportional to $\mu_{\text{ggF}}$. Here the general template would be constrained so that $\eta_{pi}^{p'}=0$ unless $p'= \text{ggF}$ and $p= \text{VBF}$ or vice versa.  We explore a simple model for this situation in Scenario C of Appendix~\ref{Sec:SimpleExample}.

\paragraph{Via a global learning approach}

Ideally we would have a formalism that would work with the black box likelihood $L_\text{full}(\vec\mu,\vec\alpha)$ and a general template like Eq.\eqref{Eq:genericTemplate} without having to introduce by hand restrictions on that template as described above.  In order to do that we must introduce  information about the likelihood away from $(\hat{\vec\mu}, \hat{\vec\alpha})$.   A flexible approach to that problem is based on the ideas of machine learning and function approximation in which one aims to minimize a loss function with respect to some model parameters (in this case the template coefficients).  The loss function needs to be a scalar evaluated over the $(\vec\mu,\vec\alpha)$ space, which in frequentist terms has no measure. However, from the point of view of decision theory, one can introduce some weighting over the parameter space (without regarding it as a Bayesian prior) and evaluate 
\begin{equation}\label{eq:loss}
\text{Loss}( \vec\eta) = \int d\vec\mu d\vec\alpha \, \pi(\vec\mu,\vec\alpha) \, \left | L_\text{full}(\vec\mu,\vec\alpha) -L_\text{recouple}(\vec\mu,\vec\alpha; \vec\eta) \right |^2
\end{equation}
The choice of the weighting function $\pi(\vec\mu,\vec\alpha)$ is arbitrary, but a reasonable choice is the Bayesian posterior with respect to some baseline constraint terms interpreted as a  prior on $\vec\alpha$, which leads to
\begin{equation}
\pi(\vec\mu,\vec\alpha) \propto  L_\text{main}(\vec\mu,\vec\alpha) L_\text{constr}(\vec\alpha) \; .
\end{equation}
This approach will put the highest weight for $L_\text{recouple}$ to approximate $L_\text{full}$ near the best fit point ($\hat{\vec\mu},\hat{\vec\alpha})$ and lesser weight as one moves away from it.  Importantly, this minimum loss solution can be found numerically and is well defined even when the number of parameters in the template is larger than $n_p\times n_\alpha$.  Furthermore, this approach may be more robust in the case of very complicated likelihood functions where the numerical accuracy of the covariance matrix, information matrix, and partial derivatives needed in Eqs.~\eqref{Eq:SolveViaPartial}, \eqref{Eq:SolveViaCov}, and \eqref{Eq:SolveViaFisher} may be poor. 

 In situations where additional experimental uncertainties $\alpha_\text{exp}$ have been profiled in providing $L_\text{eff}(\vecmueff)$ --- a situation discussed in more detail in Sec.~\ref{sec:combinations} --- one must take care that the loss function makes the comparison for equivalent values of the profiled nuisance parameters. For example, when creating $L_\text{eff}(\vecmueff)$, one can keep track of the profiled values $\vec{\hat{\hat{\alpha}}}_\text{exp}(\vecmueff)$ and then in Eq.\eqref{eq:loss} make the replacement
\begin{equation}\label{eq:learningWithAlphaExp}
L_\text{full}(\vec\mu,\vec\alpha)  \to  L_\text{full}(\vec\mu,\vec\alpha, \,\vec{\hat{\hat{\alpha}}}_\text{exp}(\vecmueff(\vec\mu,\vec\alpha))\, )  .
\end{equation}

We demonstrate the effectiveness of this approach in Sec.~\ref{sec:example} and Scenarios B and C of Appendix~\ref{Sec:SimpleExample}.\bigskip

\paragraph{Software}
The software implementation of the reparametrization templates described in Sec.~\ref{sec:templates} as well as the three strategies  for determining the coefficients of those templates from the local covariance matrix and the learning approach described in described in Sec.~\ref{sec:coeffs} is available at Ref.~\cite{githubdecouple}. Experiments can use this software on their full \texttt{RooFit/RooStats}~\cite{RooFitRooStats} models and obtain the effective likelihood $L_\text{eff}(\vecmueff)$ as well as the reparametrization $\vecmueff(\vec\mu,\vec\alpha)$ for publication. These ingredients can be supplied in a technology independent format enabling others to perform the recoupling stage, modify constraint terms associated to theoretical uncertainties,  combine multiple results, and create likelihood scans in benchmark models.

\subsection{Grouping of Categories and Combinations} 
\label{sec:combinations}

In Section~\ref{sec:signal_strength} we discuss  the coarse graining of the many categories into a  few groups. By taking a common category-weighted signal strength factor we are able to suppress the category index and make the replacement $\muefff{cpd} \to \muefff{pd}$ for each group.  In practice, this grouping is often based on the Higgs decay mode indexed by $d$.  For example, ATLAS provided the profile likelihood $\lambda(\mu_\text{ggF}, \mu_\text{VBF})$ for the three decay modes $d=\gamma\gamma, WW, ZZ$~\cite{atlas_bosons2}.  These profile likelihoods are colloquially referred to as ``likelihoods''; however, they have eliminated several nuisance parameters via profiling as defined in Eq.\eqref{eq:profile}.  In addition to the common theoretical uncertainties, these three likelihoods share common experimental systematic uncertainties.  Na\"ively combining these likelihoods will thus double count the common constraint terms $f_i(a_i|\alpha_i)$ and lead to an artificial reduction in the uncertainty.  This effect can be seen by comparing a na\"ive  combination of the three individual profile likelihoods provided by ATLAS with the official ATLAS combined result (which avoids the erroneous double counting). The same effect is demonstrated with a toy example in Fig.~\ref{fig:counting_kVkF_combined_comparison}.\bigskip

The decouple/recouple procedure described here has focused on theoretical uncertainties for reasons that will be elaborated in Sec.~\ref{sec:qcd}; however, the same technique can be used for experimental uncertainties.  In particular,  sources of systematic uncertainties that are anticipated to be common to other analysis should not be profiled or it will not be possible to avoid double counting of constraint terms.  The luminosity uncertainty is an example of an experimental uncertainty that should not be profiled, but instead the effect of the uncertainty should be parametrized in the template $\vecmueff(\vec\mu,\vec\alpha)$.  As a rule of thumb, one can safely profile uncertainties that are specific to the categories in a given analysis, but should include common (correlated) sources of uncertainty in the reparametrization.\bigskip

\subsection{Anticipating Higher Dimension Operators and Other Effects}

At this point, the Higgs coupling analysis is primarily focused on small deviations in the coefficients of the SM operators.  This justifies the scaling of the SM Higgs expectations as $\mu_{pd} s_{cpd}$, since the efficiency and acceptance of the modified signal in the $c^\text{th}$ category is not affected.  The presence of non-SM operators will generically change the pattern of the signal across the categories in a way specific to the operators under consideration~\cite{d6}. In particular, operators that modify kinematic distributions will affect the cut efficiencies and acceptances $\epsilon_{cpd}(\vec\alpha)$ of Eq.\eqref{eq:def_s}  One approach to anticipate these studies is to characterize the effect of a deviation in the number of events in each individual category.  In particular, by considering a perturbation of the form
\begin{equation}
\nu_c(\vec\mu,\vec\alpha) = \sum_{p,d}\mu_{pd} s_{cpd}(\vec\alpha) +b_c(\vec\alpha) + \alpha_{c}  \;,
\end{equation}
where $\alpha_c$ is some addition or reduction to the number of events in that specific category.  This would add one nuisance parameter for each category, each of which  would have the effect of shifting $\hat{\vec\mu}^\text{eff}$.  These $\alpha_c$  can be seen as a basis for possible new physics effects and can be incorporated into $\vecmueff(\vec\mu,\vec\alpha)$ using the same formalism.  To use this information in the context of a specific new physics model, one would need to estimate the perturbation to the number of events in each category ($\alpha_c$) through knowledge of the details of the selection for that category. If that is possible, then one can parametrize each $\alpha_c$ in terms of the coefficients of these new operators. This may involve parametrizing several $\alpha_c$ in terms of just a few operator coefficients.

\section{Theoretical Uncertainties}
\label{sec:qcd}

In this section we focus on the dominant theoretical uncertainties plaguing Higgs coupling measurements at the LHC.
There are several sources of
uncertainties, which can be associated with the theoretical description
of LHC processes~\cite{lecture}: 
\begin{enumerate}
\item the inclusive rate for a given production and decay,  for
  example the total cross section $\sigma_\text{tot} (gg \to H+X)$.
\item efficiency and acceptance of kinematic cuts for the signal process, for
  example cuts on the transverse momentum of the Higgs boson or a central jet veto.
\item the parton densities describing the initial state, which have both experimental and theoretical uncertainties.
\item the parton shower, hadronization, underlying event, and other key
  parts of the Monte Carlo description of signal and background  events.
\end{enumerate}
Linking  event rates in a given kinematic region
to perturbative theoretical predictions
might be considered the critical problem for the LHC's precision Higgs program.
According to the above list it can be separated into two parts: first,
the total Higgs production cross section parametrically depends on the
Higgs coupling in a given renormalization scheme. This relation
is based, for example, on collinear factorization, includes a small number
of scales, and can be computed in perturbative QCD. 
 However, this connection cannot be replaced or checked experimentally, which
means that Higgs coupling measurements will eventually be
limited by theory predictions~\cite{extrapolations,ilc}. 

Kinematic cuts linked to Higgs-plus-jet production~\cite{higgs_pt},
tagging jets in weak boson fusion~\cite{tagging,cjv},
or advanced analysis methods lead to additional complications.
They do not automatically obey collinear factorization and induce a large
number of energy scales and possibly large logarithms~\cite{manchester,neubert,gavin,scet,scaling}.
Already for the relatively safe inclusive Higgs production cross section,
the next-to-next-to-leading order corrections more than double the
predicted number of produced Higgs
bosons~\cite{higgs_nlo,higgs_nnlo,higgs_nnnlo,higgs_resum}.
The source of these problems is the
choice of protons as colliding particles, combined with the large
value of the strong coupling constant, which leads to a poor 
convergence of the perturbative QCD description. \bigskip

Aside from the poor convergence in perturbative QCD the corresponding
theoretical uncertainties face a more fundamental problem: there does
not exist any well--defined estimate for an uncertainty on a
production cross section computed in perturbative QCD. Traditionally, we
derive a range of allowed cross section values using a variation of
the factorization and renormalization scales. This is based on the
fact that these scales are artifacts of the perturbative expansion, so
the scale dependence has to vanish once we include all orders in
perturbation theory. This recipe captures some of the effects of the
theoretical uncertainties, but we know from the Drell--Yan process or
Higgs production in gluon fusion that it does not give a conservative
estimate. For Higgs production at the LHC we know that equal 
variation of the factorization and renormalization scales leads 
to a cancellation of the two scale dependencies as one possible 
reason for the underestimate of the theoretical uncertainties~\cite{higgs_nnlo}.
On the other hand, a separate variation of the factorization and
renormalization scales ruins the theoretical description of the
scale--dependent parton densities in terms of resummed large logarithms
of transverse jet momenta~\cite{lecture}. Independently varying the two scales makes sense phenomenologically, but not in terms of a formal QCD description.
A recent, promising approach might be to
combine the scale variation with an extrapolation of the perturbative
series~\cite{david_passarino}. However, no matter what recipe we
choose, it is clear that the size of any theoretical uncertainty is
poorly defined.\bigskip

In addition to the overall size of the uncertainty, most statistical analysis
techniques require the uncertainty to be quantified in the likelihood, \textit{i.e.}  
the constraint terms of Eq.\eqref{Eq:statmodel}. 
 Frequentist interpretation of the 
likelihood requires one to be able to identify some  
random observable who's probability is defined with a corresponding ensemble.
No traditional ensemble defined by repeated observation exists for theoretical uncertainties.
Missing higher order terms in a perturbative prediction are not random in nature, there is no ensemble, thus $L_\text{constr}(\vec\alpha)$ is ill-defined. 
Attempts to provide a meaningful degree of belief for the perturbative uncertainties~\cite{cacciari}
requires a Bayesian interpretation and attributes a fundamental meaning to the 
perturbative QCD expansion.

In a practical sense we must make some choice for the likelihood $L_\text{constr}$ even if it fundamentally ill-defined.  Unfortunately, the various choices have their own unique pathologies.
The often--used Gaussian and log-normal distributions inject information by
ascribing a preferred (peak) value and allowing for
long tails into what are untenable values for the theoretical prediction.
Alternatively, we can try to avoid introducing a peak by adopting a constant likelihood function, as proposed in the \textsc{RFit} scheme~\cite{rfit,sfitter}. The choice of 
distributions might not appear numerically relevant, but in combination with 
a profile likelihood, it can lead to significant differences. For instance, the combined effect of  uncertainties described in the \textsc{RFit} scheme add linearly rather than in quadrature~\cite{lecture}.\bigskip

Given that the size as well as the shape of the theoretical
uncertainties entering the Higgs couplings measurements are subject to
variations in time, geography, and personal taste, the best solution
is  to decouple them from the experimental result as proposed in
Section~\ref{sec:approach}. This allows for efficient tests of
different assumptions on the theoretical uncertainties as well as an
efficient implementation of a perceived or actual improvement.

\section{A Toy Example}
\label{sec:example}

We now consider a toy example that is representative of the current ATLAS results for $H\to \gamma\gamma, WW, ZZ$~\cite{atlas_bosons}.  The statistical model here is based purely on number of events in various categories without including discriminating variable distributions for  $m_{\gamma\gamma}$, $m_{4\ell}$, or $m_{T}$ (the terms $f_c(x|\vec\mu,\vec\alpha)$ in Eq.\eqref{Eq:statmodel}).     Each decay mode groups together several categories of events that together provide sensitivity to the underlying production modes. We model the uncertainty on the signal expectation from luminosity, parton distribution functions, the inclusive gluon fusion cross section, and the uncertainty on the cross section for gluon fusion in association with two or more jets.

The $H\to\gamma\gamma$ likelihood includes a simplified version of the 14 categories  considered by ATLAS including the low- and high-$p_T$ categories, the low- and high-mass 2-jet categories, the high-${E}\!\!\!\!/_T$ significance category, and the lepton-tagged category.  The $H\to ZZ\to4\ell$ likelihood includes three ggF-like categories (for $4\mu$, $2e2\mu$, and $4e$) as well as a VBF-like and a $VH$-like category.  The $H\to WW\to \ell\nu\ell\nu$ likelihood includes 0-, 1-, and 2-jet categories.  The \texttt{HistFactory} script and \texttt{RooFit/RooStats} workspace for this toy model can be found at Ref.~\cite{DOIforToyModel}. \bigskip

\begin{figure}[b]
	\centering
	\subfigure{ \includegraphics[width=0.45\textwidth]{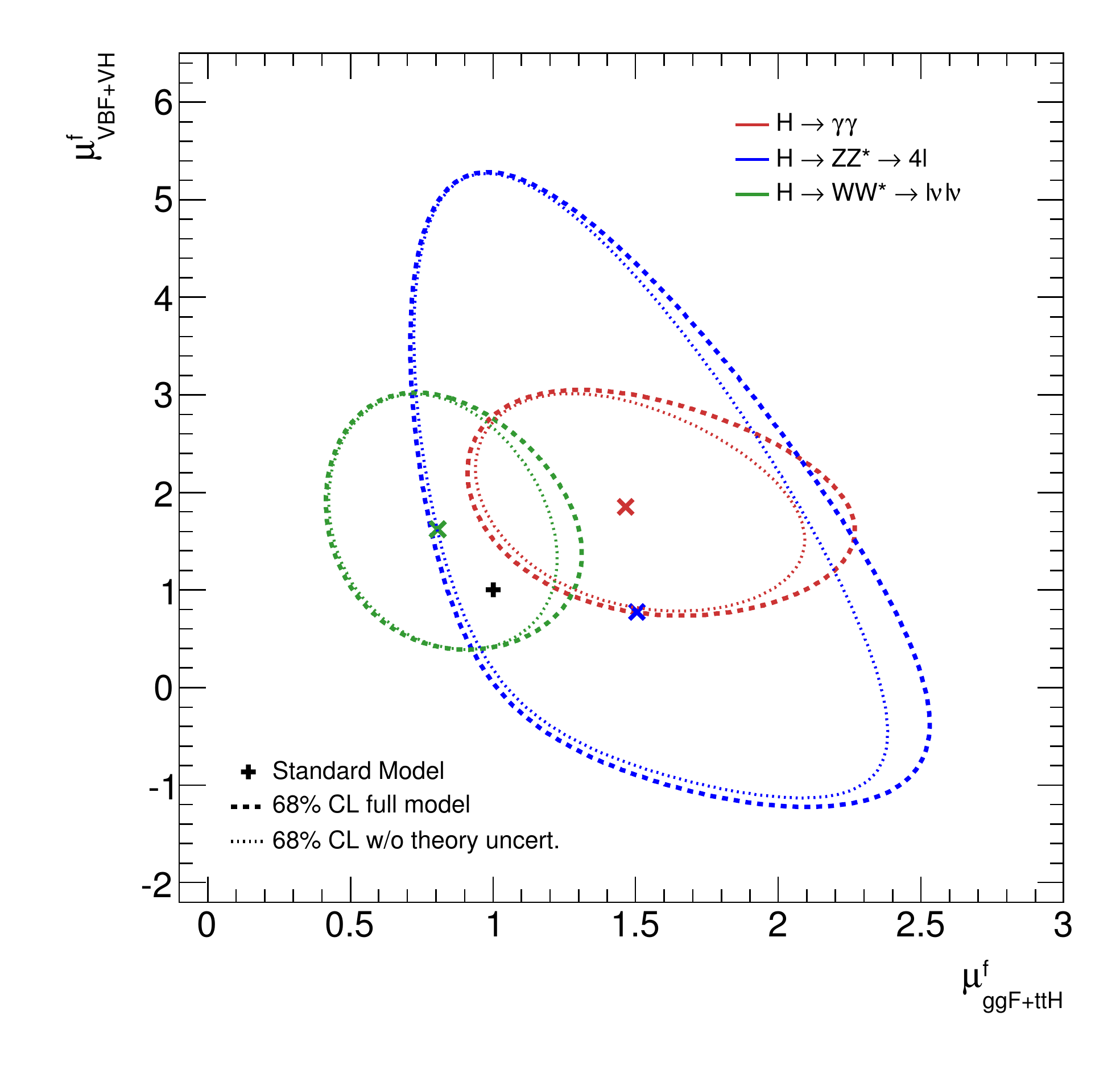} }
	\subfigure{ \includegraphics[width=0.45\textwidth]{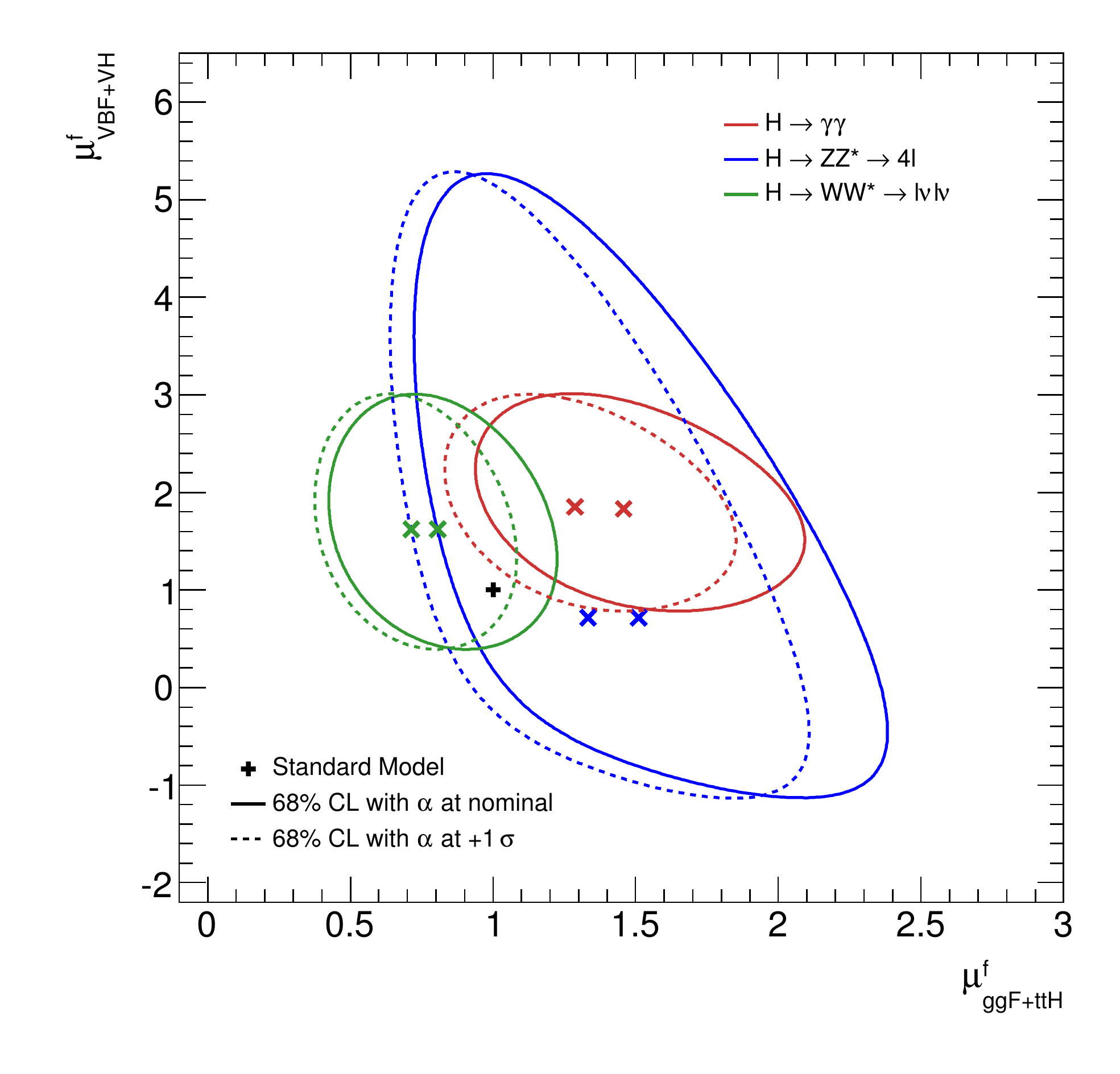} }
\caption{Likelihood contours for three different Higgs decays: (a) with (solid) and without (dashed) theoretical uncertainties; (b) without theoretical uncertainties for the nominal gluon fusion cross section (solid) and a shifted value (dashed) estimated from QCD scale variations.}
	\label{fig:counting_signalStrength}
\end{figure}

\begin{figure}[htb]
	\centering
	\subfigure{ \includegraphics[width=0.3\textwidth]{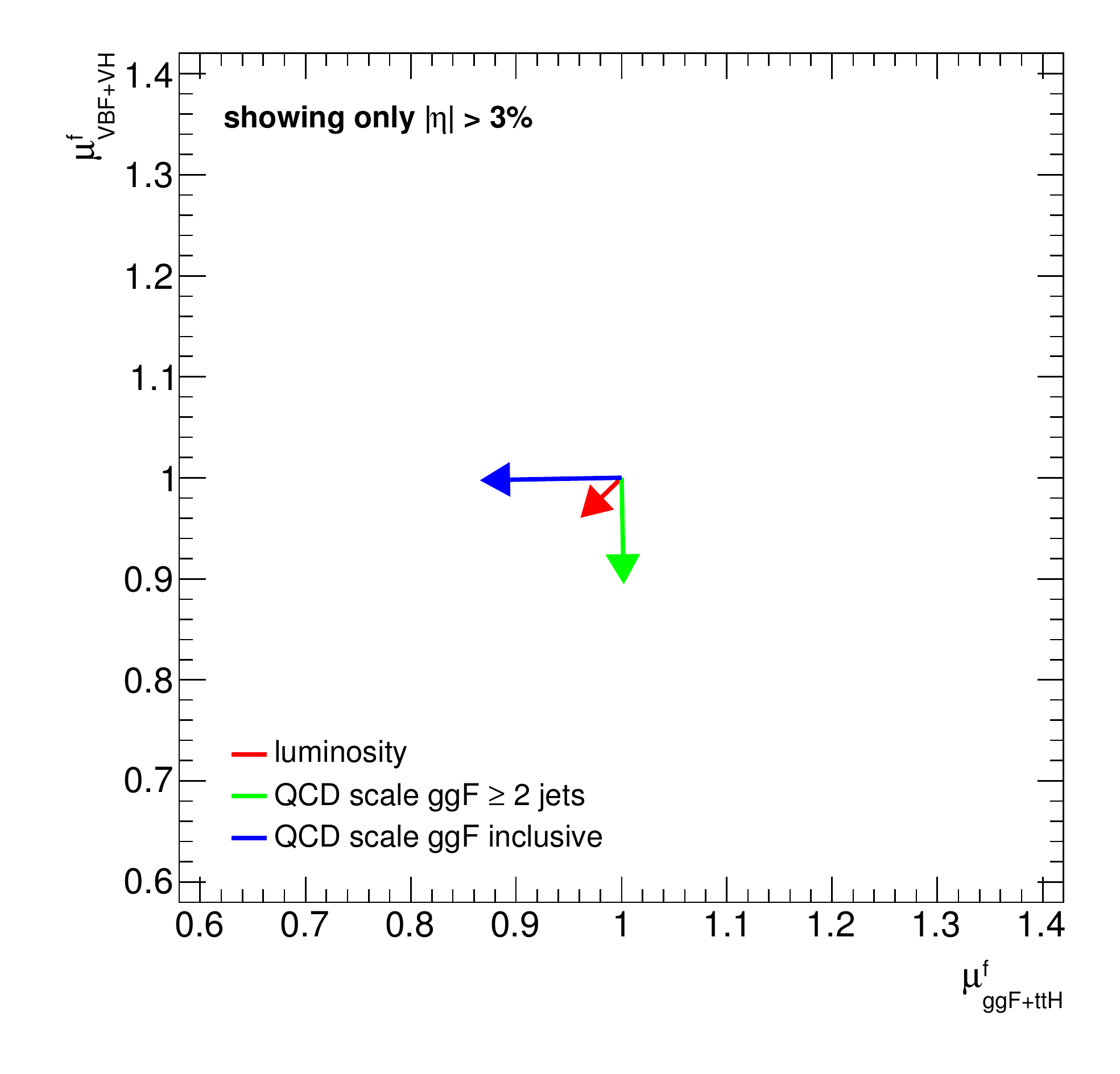} }
	\subfigure{ \includegraphics[width=0.3\textwidth]{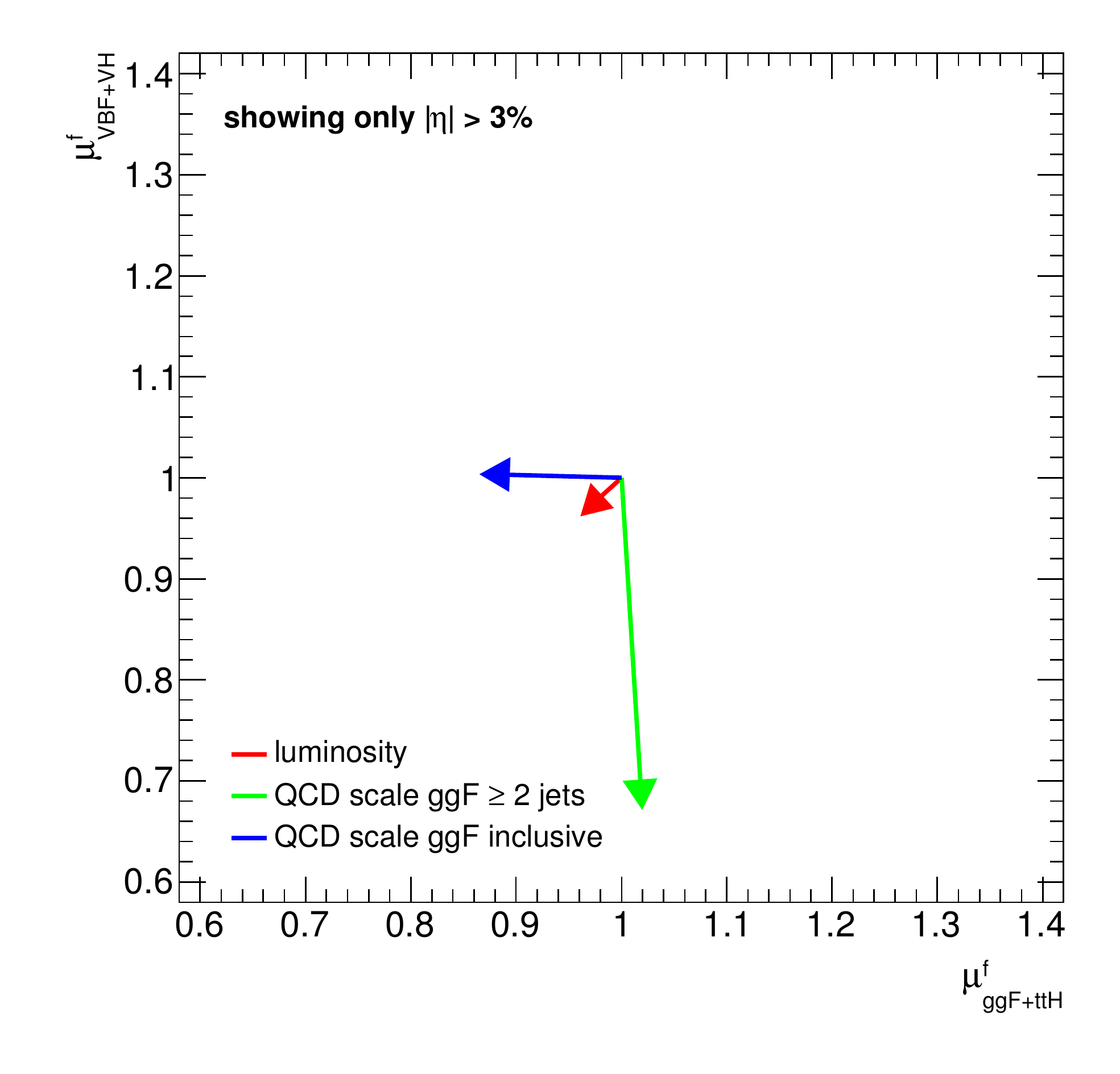} }
	\subfigure{ \includegraphics[width=0.3\textwidth]{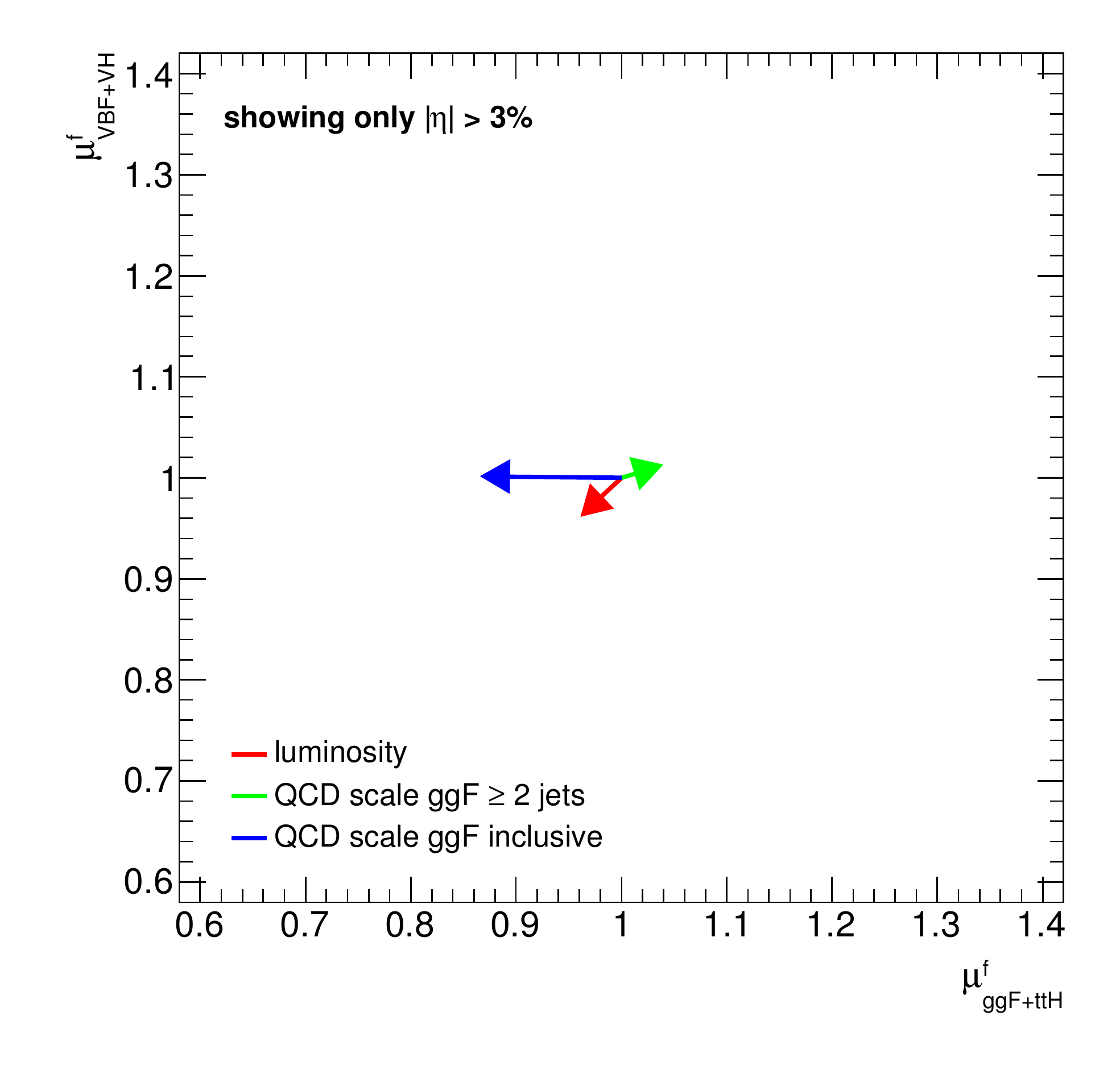} }
	\caption{Visualization of $\partial \vec\mu^\text{fix}/\partial \alpha_i$ for the (a) $H\to\gamma\gamma$, (b) $H\to ZZ\to 4\ell$, and (c) $H\to WW\to \ell\nu\ell\nu$ likelihoods. }
	\label{fig:visEtas}
\end{figure}

Figure~\ref{fig:counting_signalStrength}(a) shows the likelihood contours for the three different decays with and without theoretical uncertainties, which are modeled using Gaussian constraint terms and a linear response as in Eq.~\eqref{Eq:inclusiveProdTemplate}. Fig.~\ref{fig:counting_signalStrength}(b) shows the shift to the contours without theory uncertainty due to fixing the inclusive gluon fusion cross section to its ``$+1\sigma$'' value estimated from QCD scale variation.  The larger gluon fusion cross section leads to a smaller inferred value for $\mu_\text{ggF}$.  This can be repeated for each of the nuisance parameters $\alpha_i$ as in Eq.\eqref{Eq:SolveViaPartial}.  The corresponding partial derivatives are visualized in Fig.~\ref{fig:visEtas}.

\begin{figure}[b!]
	\centering
	\subfigure{ \includegraphics[width=0.45\textwidth]{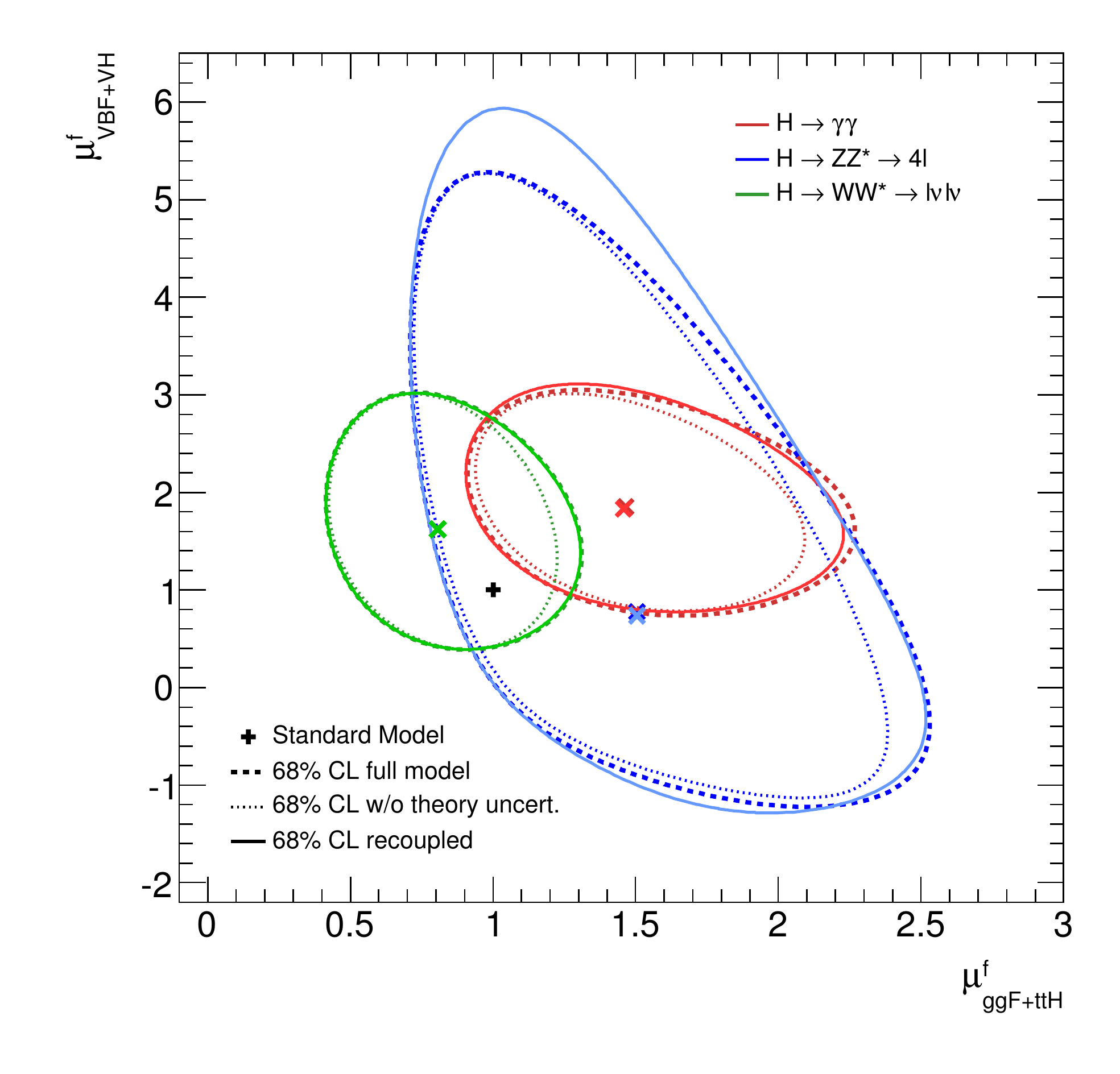}}
	\subfigure{ \includegraphics[width=0.45\textwidth]{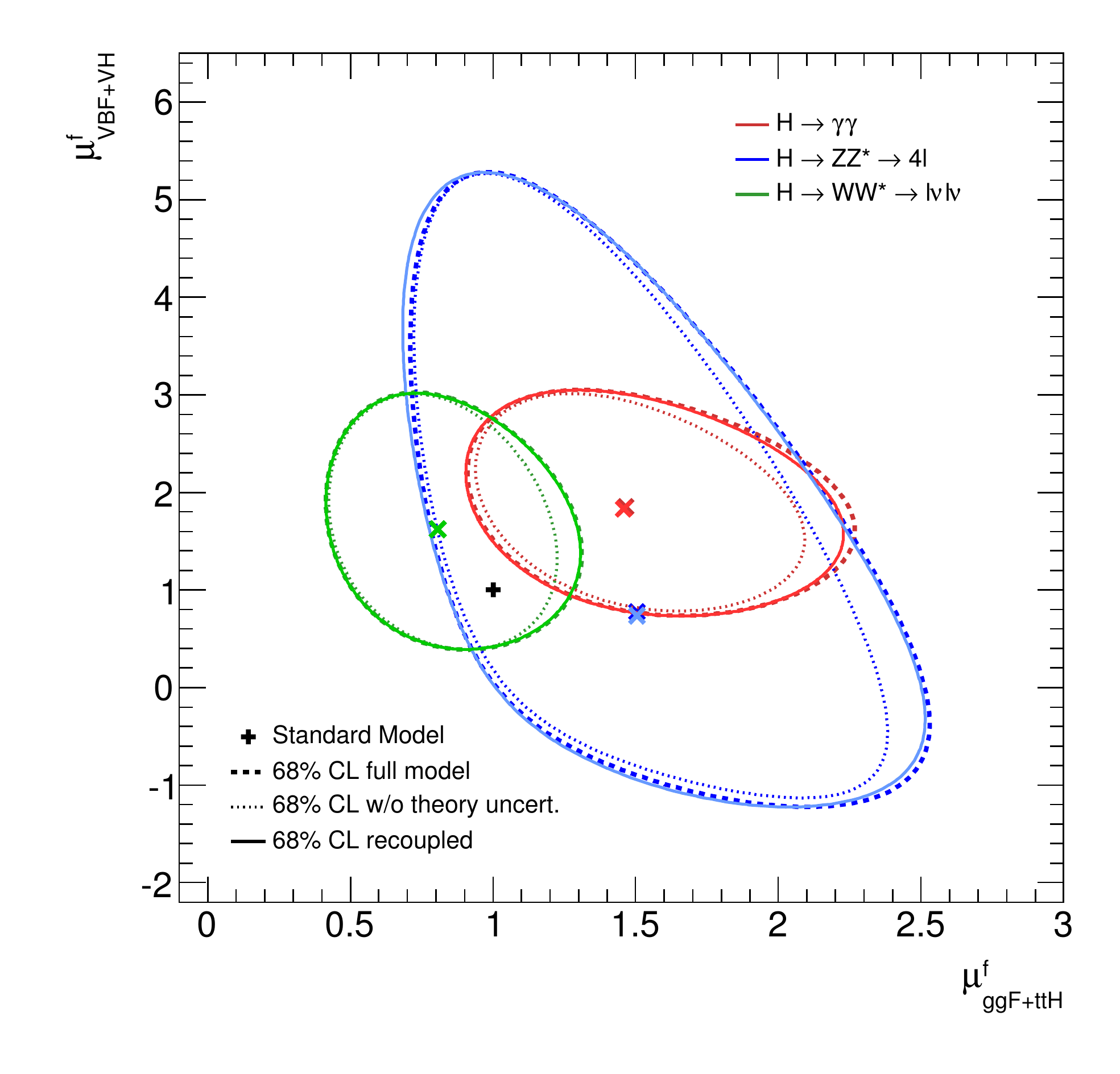} }
	\caption{Comparison of effective model using different templates with same local covariance }
		\label{fig:TemplateComparison}
\end{figure}

Next we demonstrate the \textit{recoupling} stage based on the local covariance structure at $(\hat{\vec\mu},\hat{\vec\alpha})$.   If we choose the reparametrization template in Eq.\eqref{Eq:inclusiveProdTemplateMu} we can solve for the $\eta_{pi}$ coefficients that reproduce the  local covariance structure using either Eq.\eqref{Eq:SolveViaPartial}, Eq.\eqref{Eq:SolveViaCov}, or Eq.\eqref{Eq:SolveViaFisher}.  Having determined the $\eta_{pi}$ coefficients, the reparametrization $\vecmueff(\vec\mu,\vec\alpha)$ is specified and we can create the decoupled likelihood via Eq.\eqref{Eq:Lrecouple}.  Fig.~\ref{fig:TemplateComparison}(a) compares the full likelihood to the recoupled likelihood using this template. The result is not bad, the effect of the theoretical uncertainties have been recovered; however, there is a significant discrepancy for $H\to ZZ\to 4\ell$ at the top of the contour.  \bigskip

The source of this discrepancy between $L_\text{full}$ and $L_\text{recouple}$ contours is the non-category universal nature of the theoretical uncertainty associated to gluon-fusion+$\ge$2 jets, which primarily affects the VBF-like categories. Correspondingly, the effect of this uncertainty is a shift in the $\mu_\text{VBF}$ direction, as illustraited in Fig.~\ref{fig:visEtas}. The reparametrization template in Eq.\eqref{Eq:inclusiveProdTemplateMu} implies that this uncertainty would scale with $\mu_\text{VBF}$,  though in the full model this source of uncertainty scales with gluon fusion.  At the top of the contour, $\mu_\text{VBF}\approx 5$, while $\mu_\text{ggF}\approx 1$, which leads to a large inflation on the effect of this uncertainty.  

As discussed in Sec.~\ref{sec:coeffs} the local covariance structure of $L_\text{recouple}$ is only sensitive to the linear behavior of the template, so this physical insight must either be put in ``by hand'' or one should instead use more global information about the likelihood as in the learning approach.   To illustrate the ``by hand'' approach, we can  consider the more general template in Eq.\eqref{Eq:generalTemplate}.  The general template can be constrained so that $\eta_{pi}^{p'}=0$ unless $p'= \text{ggF}$ and $p= \text{VBF}$ (or vice versa), where $i$ is the index for the nuisance parameter associated to the gluon fusion $+\ge 2$ jet cross section.  In addition, we set  $\phi_i=0$ for all $i$. The same procedures are followed to determine the coefficients that reproduce the local covariance structure for this template.  The resulting contour is shown in Fig.~\ref{fig:TemplateComparison}(b), where the agreement  is improved, particularly near the top of the $H\to ZZ\to 4\ell$ contour.  The alternative learning approach works with the general template in Eq.\eqref{Eq:generalTemplate} without restricting the terms by hand.  The optimized values of the coefficients  leads to even better agreement of the contours compared to those shown in  Fig.~\ref{fig:TemplateComparison}(b). \bigskip

\begin{figure}[t]
	\centering
	\subfigure{ \includegraphics[width=0.45\textwidth]{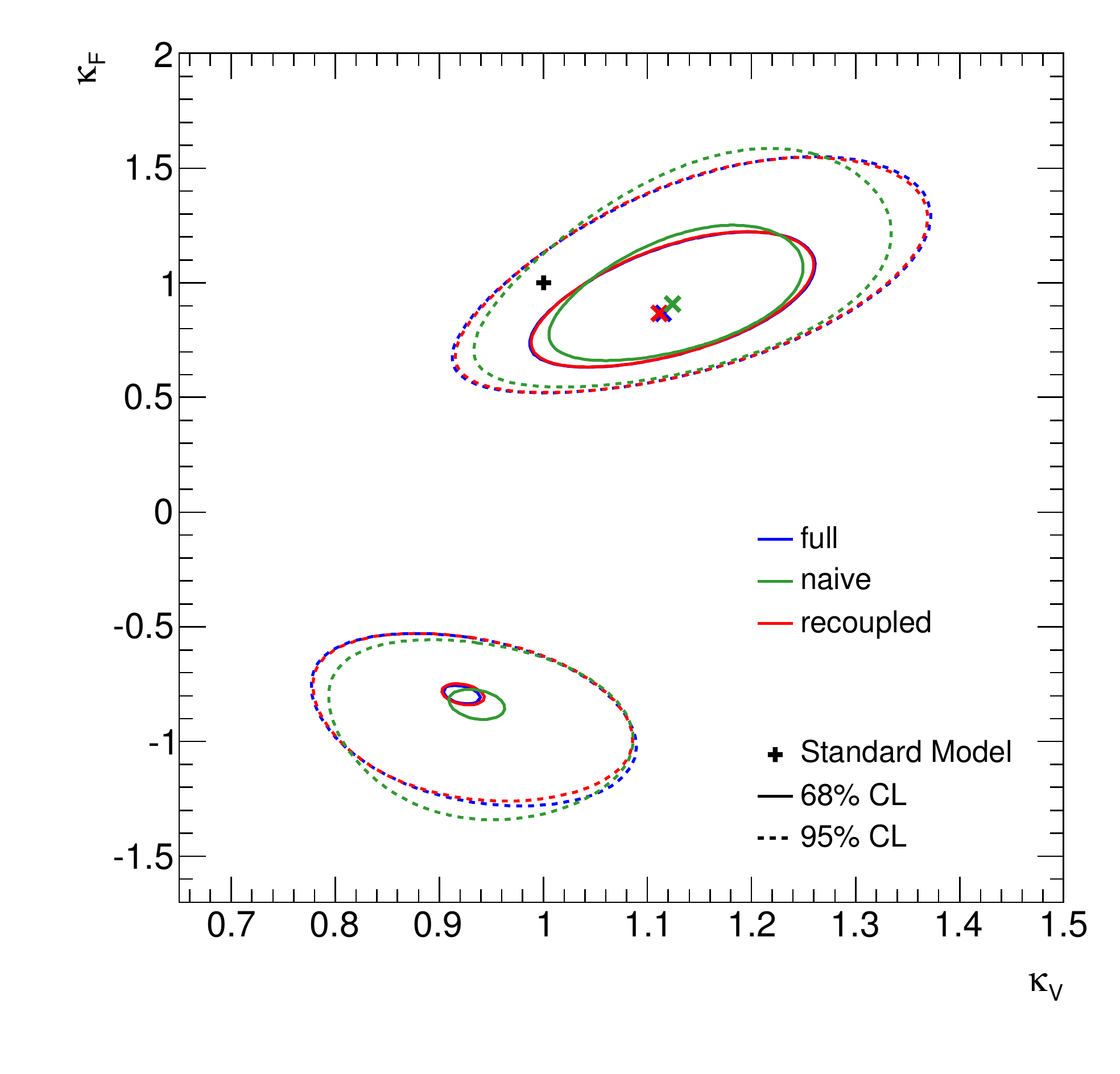}}
	\subfigure{ \includegraphics[width=0.45\textwidth]{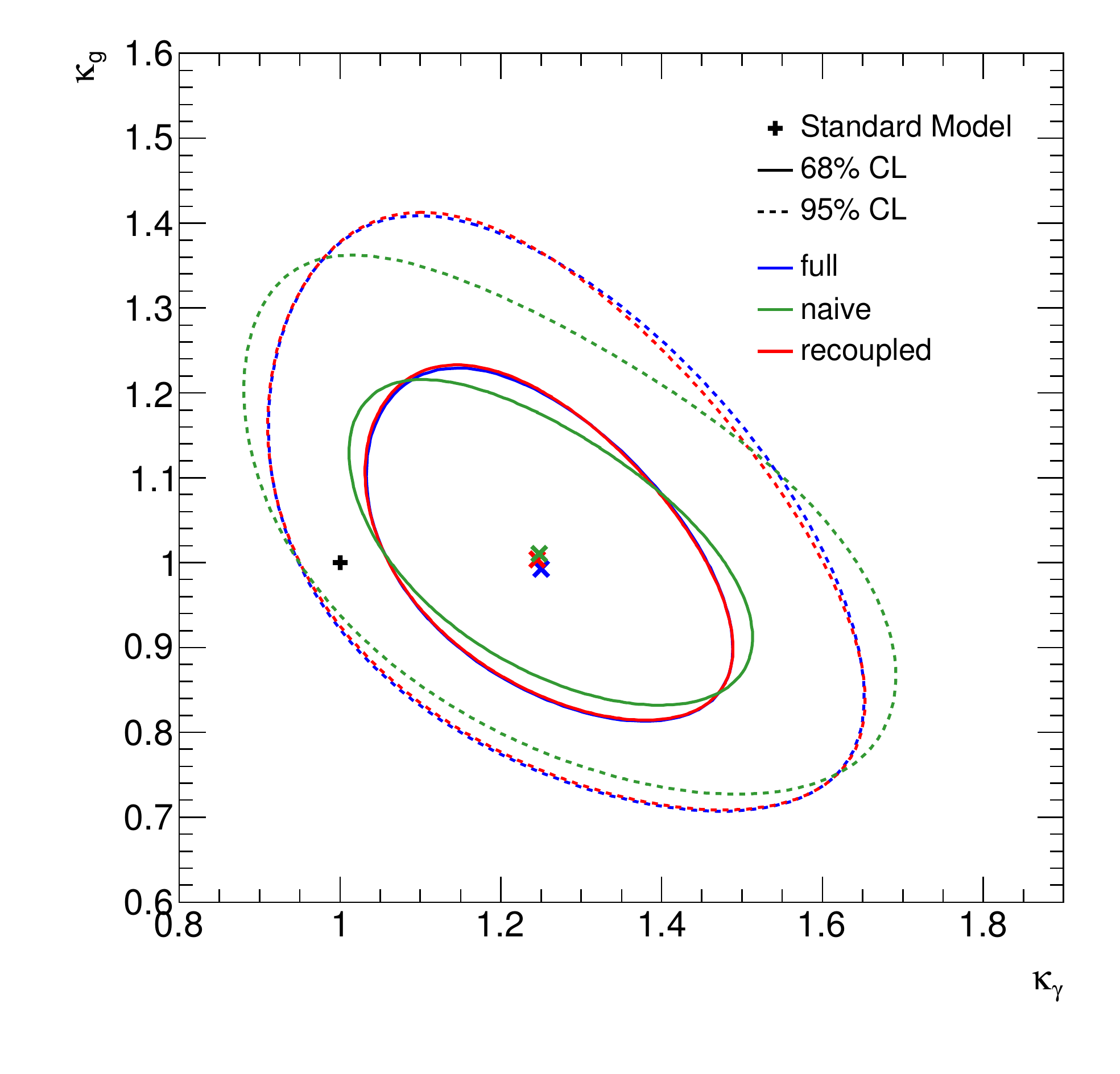} } \\
		\subfigure{ \includegraphics[width=0.45\textwidth]{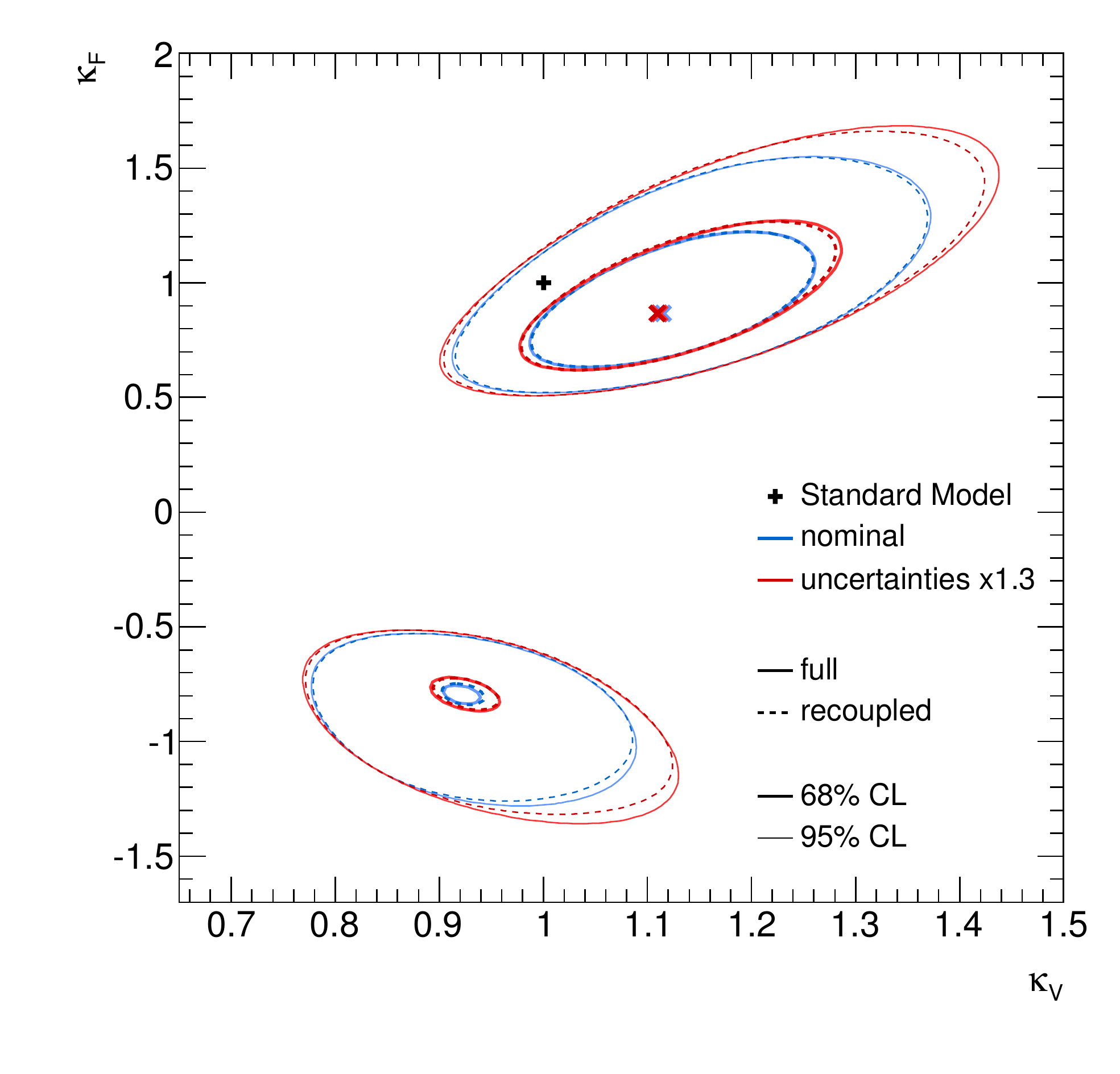}}
	\subfigure{ \includegraphics[width=0.45\textwidth]{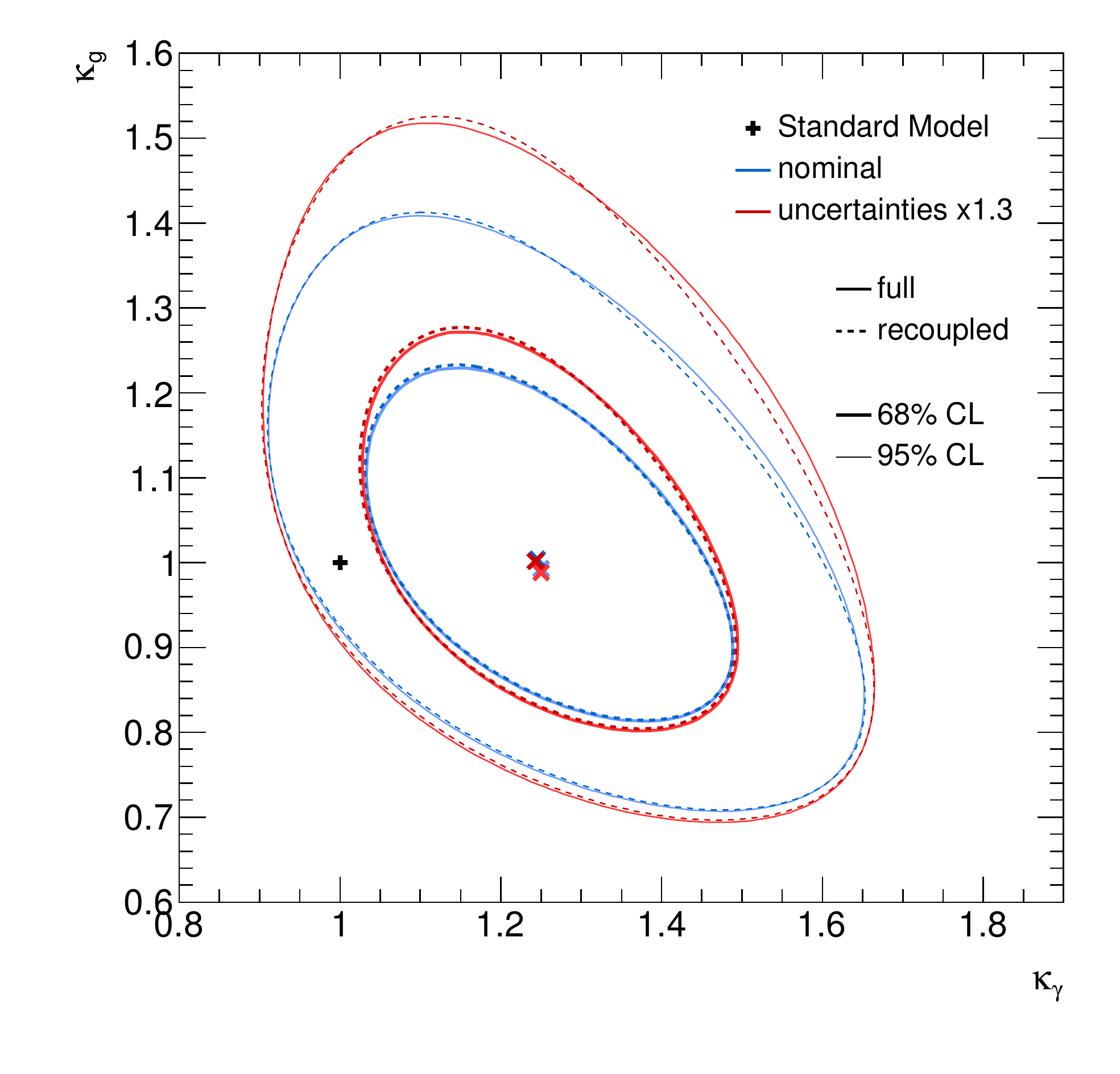} }
	\caption{Various comparisons of the combined $\gamma\gamma, ZZ, WW$ likelihoods in the $(\kappa_V, \kappa_F)$ (left) and $(\kappa_\gamma , \kappa_g)$ (right) planes. Top: comparison of the full combined likelihood, a na\"ive combination with inconsistent profiling and double counted constraint terms, and the combination of recoupled likelihoods with consistent profiling and without double counted constraint terms. Bottom: comparison of the full combined likelihood and the combination of recoupled likelihoods using the nominal uncertainties and a modified  constraint terms with uncertainties inflated by 30\%.}
		\label{fig:counting_kVkF_combined_comparison}
\end{figure}

Next,  we combine these three likelihoods. As discussed in Sec.~\ref{sec:combinations}, the experiments typically present results grouped by decay mode.  In this example, the three likelihoods share common sources of systematic uncertainty.  Na\"ively combining these likelihoods double counts the common constraint terms $f_i(a_i|\alpha_i)$ and leads to an artificial reduction in the uncertainty.  This effect can be seen in the top plots of Fig.~\ref{fig:counting_kVkF_combined_comparison}, which compare a na\"ive  combination to the the full combined result that avoids the erroneous double counting. There is no meaning to a combined contour in the signal strength plane $\mu_{\text{ggF},d}-\mu_{\text{VBF},d}$ due to the decay index $d$.  Thus,  two simple 2-parameter benchmark models are used to present the result. Fig.~\ref{fig:counting_kVkF_combined_comparison}(a) shows a benchmark model that dictates all $\mu_{p,d}$ based on the scaling of the fermionic couplings $g_f = \kappa_f g_f^\text{SM}$ and weak boson couplings $g_V = \kappa_V g_V^\text{SM}$.  Similarly, Fig.~\ref{fig:counting_kVkF_combined_comparison}(b) considers a simple 2-parameter benchmark in which we scale the effective $Hgg$ by a factor $\kappa_g$ and the effective $H\gamma\gamma$ coupling by a factor $\kappa_\gamma$~\cite{HiggsXS}.  The contours from the naive combination are considerably smaller in the $\kappa_V$ and $\kappa_g$ directions, leading to poor agreement with the full combination (based on all 22 categories of events without double counting constraint terms).  

In contrast, the recoupling approach allows one to avoid double counting constraint terms and for a consistent profiling over the common sources of uncertainty (both theoretical and experimental), which leads to an improved agreement.   In addition, the learning approach of Eq.\eqref{eq:loss} has been used for $L_\text{recouple}$ in Fig.~\ref{fig:counting_kVkF_combined_comparison}.  Note, in this example the sources of experimental uncertainty that are unique to one of the decays were profiled in $L_\text{eff}(\vecmueff)$, as would be done by the experiments.
 Thus, there is a small effect that is neglected due to the fact that the profiling of these analysis-specific uncertainties is slightly affected by identifying the common $\vec\alpha$ in the combination.
\bigskip

Perhaps the greatest benefit of this approach is that the theoretical uncertainties have been decoupled from the experimental result encoded in $L_\text{eff}(\vecmueff)$.  Moreover, the recoupling procedure in Eq.\eqref{Eq:Lrecouple} allows one to replace the constraint terms in $L_\text{constr}(\vec\alpha)$ with some other constraint terms $L'_\text{constr}(\vec\alpha)$. The bottom plots of Fig.~\ref{fig:counting_kVkF_combined_comparison} demonstrate the change by using the same $L_\text{eff}$ and reparametrization template, but with uncertainties inflated by 30\%. This recouped approach is compared to the full model where the same modification is made to the constraint term. In addition,  an example of replacing Gaussian constraint terms with the \textsc{Rfit} scheme is given in Appendix~\ref{Sec:SimpleExample}. 

\section{New Physics Effects vs Theoretical Uncertainty}
\label{sec:new_physics}

Any coupling measurement based on a Lagrangian description of
underlying field theory assumes a set of operators, specifically those
describing the Higgs interactions. In the Standard Model this includes
the renormalizable Higgs Lagrangian with the tree-level interactions
to massive gauge bosons and fermions.  The effective Higgs couplings 
to photons and gluons are loop--induced, but they avoid the
decoupling theorem and are only suppressed by $1/v$. This means that
at LHC energies they have to be included in the description of the
Higgs signatures. Additional higher-dimensional operators can be
included in the analysis once the corresponding measurements separate
them from deviations in the renormalizable Higgs operators~\cite{d6}.
The prefactors of each operator, combined with a renormalization
scheme, constitute the measurable couplings.  When extracting the
couplings of a light Higgs boson with mass 126~GeV we follow an
effective theory approach. This allows us to only consider
modifications of the light Higgs boson while keeping the ultraviolet
properties of a functioning, partly decoupled Higgs
sector~\cite{sfitter_2hdm}.  We analyze extensions of the SM Higgs
sector based on Higgs coupling modifications, written
as~\cite{sfitter,sfitter_2hdm,couplings,higgs_signals}
\begin{alignat}{6}
 g_x &\equiv g_x^\text{SM} \; (1 + \Delta_x) \notag \\
 g_\gamma &\equiv g_\gamma^\text{SM} \; (1 + \Delta_\gamma^\text{SM} + \Delta_\gamma) \; .
\end{alignat} 
For loop-induced Higgs couplings it is important that we first define
the deviation due to a shift in the SM-like loops and then separately
treat additional particles in the loop.  The Higgs signal strength is
given in Eq.\eqref{eq:strength}.  The three ratios of the production
rate, the decay width, and the total width depend on one or more Higgs
couplings.  Because the total Higgs width cannot be measured at the
LHC we assume that it is given by the sum of all observable partial
widths, where we align the relevant second generation modes with the
corresponding third generation decay measurements.\bigskip

In different new physics models we describe the leading deviations
from the Standard Model in terms of one free parameter $\xi$.  For
more details and a complete set of references we refer to
Appendix~\ref{app:new_physics} and to Ref.~\cite{sfitter_2hdm}.
Because new physics effects which violate custodial symmetry are
unlikely to be discovered in the Higgs sector, we simplify our example
analysis by assuming $\Delta_W = \Delta_Z \equiv \Delta_V$. In all
models considered below we also find $\Delta_V < 0$.  For example
two--Higgs--doublet models can violate both of these features at the
loop level, so eventually we should release this simplification. We
choose $\xi$ such that in the custodial limit the (effective) Higgs
coupling to gauge bosons gets modified as
\begin{alignat}{6}
\frac{g_V}{g_V^\text{SM}} =
1 - \frac{\xi^2}{2} + \mathcal{O}(\xi^3)
\qqquad \text{or} \qqquad
\Delta_V = - \frac{\xi^2}{2} + \mathcal{O}(\xi^3) \; .
\label{eq:def_xi}
\end{alignat}
%

\paragraph{Dark singlet}

Dark singlet models include an additional scalar $S$ which couples to
the Higgs through dimension-4 portal
interactions~\cite{dark_singlet_orig}.  The extra scalar does not form
a VEV, and its decays are precluded by a discrete $Z_2$
symmetry. Because it couples to the Higgs it leads to an invisible
decay width and hence impacts the Higgs measurements through an
invisible width,
\begin{alignat}{6}
\Gamma_\text{inv} = \xi^2 \Gamma_\text{SM}
\qqquad \text{with} \qquad
\mu_{p,d} 
= \dfrac{\Gamma_\text{SM}}{\Gamma_\text{SM} + \Gamma_\text{inv}}
= 1 - \xi^2 + \mathcal{O}(\xi^3) < 1 \; .
\label{eq:scale_ds}
\end{alignat}
In this case the scaling pattern of Eq.\eqref{eq:def_xi} does not hold
for the actual coupling $g_V$, but for its apparent value from the
rate measurement.

\paragraph{Additional singlet}

In the presence of an additional $SU(2)_L$ singlet with non--zero VEV,
its mixing with the Higgs boson is described by an angle $\sin \theta
\equiv \xi$~\cite{portal_orig}.  All Higgs couplings to fermions and
gauge bosons are rescaled by the common factor
\begin{equation}
1 + \Delta_x = \cos\theta = \sqrt{1 - \xi^2} 
\qqquad \text{with} \qquad
\mu_{p,d} = 1 - \xi^2 + \mathcal{O}(\xi^3) < 1 \; .
\label{eq:scale_singlet}
\end{equation}
The phenomenological equivalence of the dark singlet, the singlet
mixing, and the simplest strongly interacting single form factor
models can only be broken by an observation of invisible Higgs decays.

\paragraph{Composite Higgs}

Minimal composite Higgs models (MCHM) describe the Higgs boson as a
pseudo--Nambu--Goldstone boson in a new strongly interacting sector
with a spontaneously broken global symmetry~\cite{composite,mchm}. In
the Randall--Sundrum picture this global symmetry has to include the
local gauge groups of the Standard Model and can in addition include a
global custodial symmetry of the Standard Model.  This way one of the
states in the Higgs--sector will be light, while the others reside at
$f \gg m_H$.  The light Higgs couplings are shifted proportional to
the ratio\footnote{Note that for consistency reasons this definition
  of $\xi$ differs from the original $\xi = (v/f)^2$.}
\begin{alignat}{6}
\xi = \frac{v}{f} \; .
\end{alignat}
Depending on the symmetry structure this Goldstone--protected strongly
interacting Higgs sector predicts different coupling patterns for
fermions and gauge bosons.  In the MCHM5 setup the ratio of production
rates scales like
\begin{alignat}{5}
\frac{\mu_{\text{VBF},d}}{\mu_{\text{GF},d}} 
&= \left( \frac{1 + \Delta_V}{1 + \Delta_f} \right)^2 
 = \frac{\phantom{xxxx} (1-\xi^2)^2 \phantom{xxxx}}{(1-2\xi^2)^2} 
 = 1 + 2 \xi^2 + \mathcal{O}(\xi^3) \; .
\label{eq:scale_mchm}
\end{alignat}
%

\paragraph{Additional doublet}

In the most general setup with an additional Higgs
doublet~\cite{2hdm}, the Yukawa--aligned 2HDM, the different Higgs
couplings to the heavy fermions vary independently.  However, the
light Higgs couplings to the massive gauge bosons are universally
modified by the mixing angle $\cos(\beta-\alpha) \equiv \xi$,
\begin{equation}
1+ \Delta_V = \sin (\beta - \alpha) 
= \sqrt{ 1 - \xi^2}
\label{eq:2hdm_scaling}.
\end{equation}
The decoupling parameter $\xi$ parameterizes the distance from the
Standard Model limit. In the fermion sector four setups accommodate
the flavor symmetry of the Standard Model~\cite{2hdm_flavor}
\begin{itemize}
\item[--] type-I, where all fermions couple to just one Higgs doublet
  $\Phi_2$
\item[--] type-II, where up-type (down-type) fermions couple
  exclusively to $\Phi_2$ ($\Phi_1$)
\item[--] lepton--specific, with a type-I quark sector and a type-II
  lepton sector
\item[--] flipped, with a type-II quark sector and a type-I lepton
  sector
\end{itemize}
These natural flavor conserving models 
correspond to particular cases of the aligned 2HDM
with specific alignment angles.
The modification factors $\Delta_{t,b,\tau}$ depend on the mixing
angles $\alpha$ and $\beta$. We vary $\sin(\beta-\alpha)$ while fixing
$\tan\beta =1.5$. We also fix the charged Higgs contribution to
$g_\gamma$ via $m_{H^\pm} = 500$~GeV and $\tilde{\lambda} = 1$ for 
the relevant self coupling defined in Ref.~\cite{sfitter_2hdm}. This
choice leads to a small charged Higgs contribution with the same sign
as the top loop. It sharpens the destructive interference with the $W$
loop and yields $\mu_{p,\gamma\gamma} < 1$ in the limit $\xi \to
0$. 

\paragraph{MSSM}
\label{sec:mssm}

The Higgs sector of the MSSM is a particular case of a type-II 2HDM
with a supersymmetric Higgs potential~\cite{mssm}. This identifies the
Higgs self couplings with gauge couplings, so neither the additional 
heavy Higgs masses 
nor the mixing angle $\alpha$ are free quantities anymore.
The decoupling limit can be described in terms of the general 2HDM, as
defined in Eq.\eqref{eq:2hdm_scaling}. The only difference is that for
the MSSM the gauge couplings are related to the heavy Higgs masses as
\begin{alignat}{5}
 \xi^2 = \cos^2(\beta-\alpha) \simeq
 \cfrac{m^2_{\hzero}\,(m_Z^2-m^2_{\hzero})}{m^2_{\Azero}(m^2_{\Hzero}-m^2_{\hzero})}
 \simeq \cfrac{m_Z^4\,\sin^2(2\beta)}{m^4_{\Azero}} \; ,
 \label{eq:mssm-decoup}
\end{alignat}
entering the gauge boson couplings just as in
Eq.\eqref{eq:2hdm_scaling}.  These tree--level Higgs mass and coupling
patterns may be strongly modified by the inclusion of quantum
effects~\cite{m_h}. In the computation of the decoupling parameter
$\xi$ we therefore use $\alpha_\text{eff}$, defined as the mixing
angle of the two scalar Higgs states with the loop--corrected mass
matrix. The ratio of the two vacuum expectation values is fixed to
$\tan \beta = 10$ . The resulting size of $\xi$ is limited by the
range $m_{\Azero} = 200-1000$~GeV, which will mean that we never reach
$\xi = 0.2$.

We use the MSSM Higgs boson cross sections and branching ratios given
by {\sc FeynHiggs}~\cite{feynhiggs}.  The Higgs signal strengths are
defined after identifying the (lightest) Higgs masses in both models
$m^\text{SM}_{\PHiggs} = m_{\hzero}^\text{MSSM} = 126$~GeV. We show a
benchmark scenario with a maximum light Higgs mass generated through
large stop mixing $(m_h^\text{max})$.  Compared to the general 2HDM,
the possible departures from the linear correlations are milder. 
This can be
understood from the more constrained 2HDM potential in the MSSM case, 
which implies a
fast transition to the decoupling regime~\cite{sfitter_2hdm}.  The largest deviations
arise in the low-$m_{\Azero}$ regime or for light stop and stau
masses.

\paragraph{Signal strengths}

\begin{figure}[t]
\begin{center}
\includegraphics[width=0.32\textwidth]{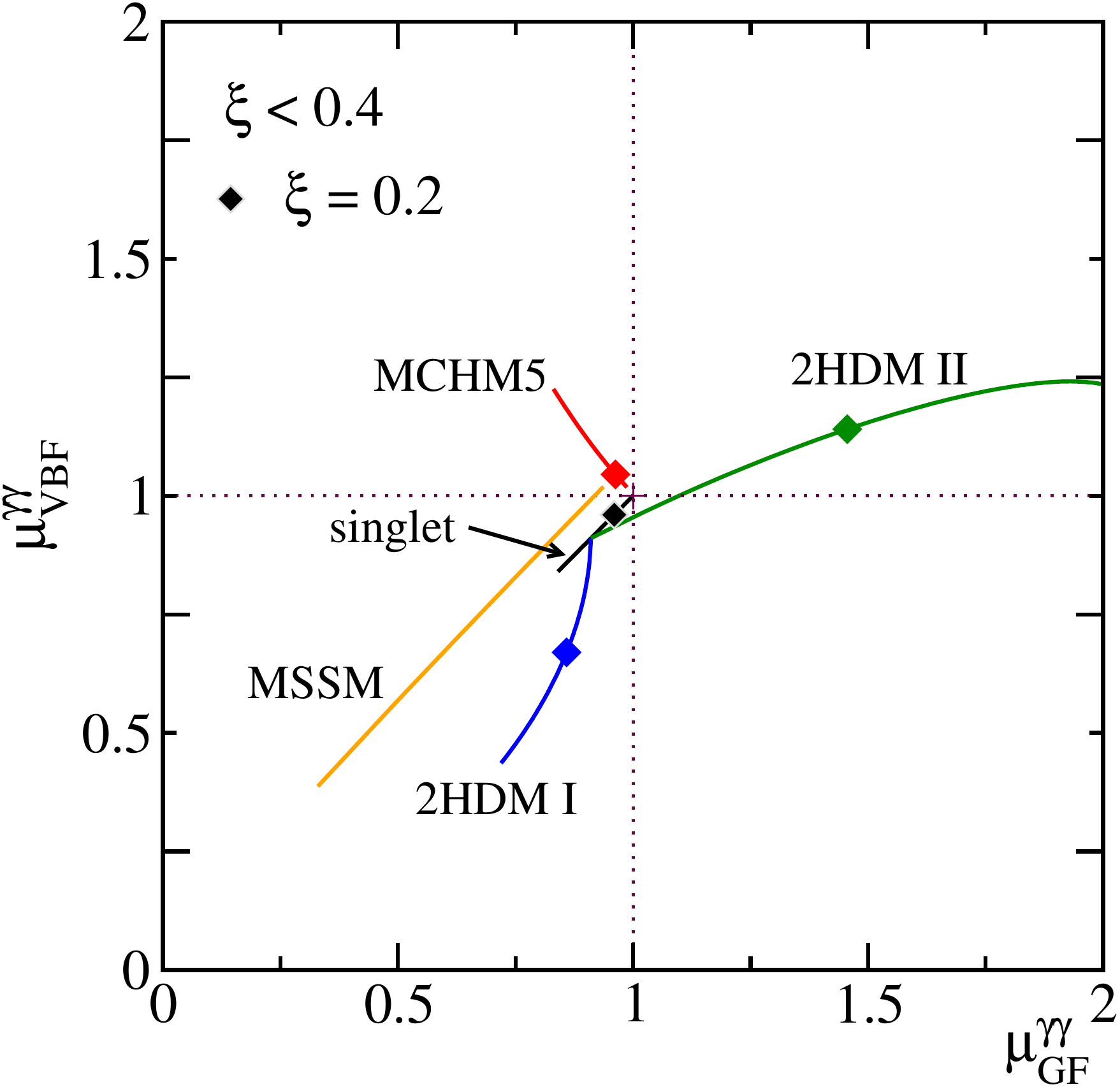} 
\includegraphics[width=0.32\textwidth]{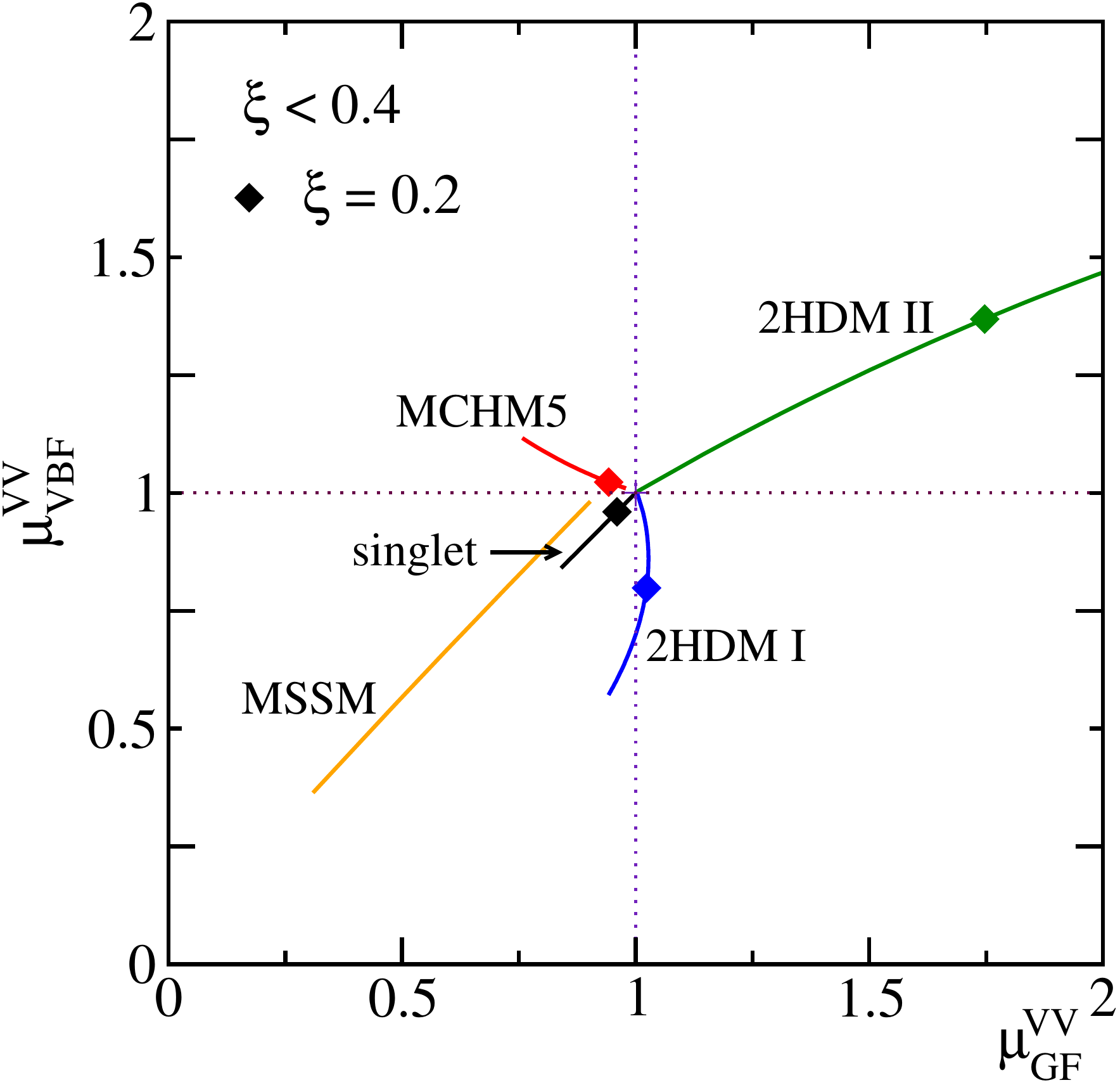} 
\includegraphics[width=0.32\textwidth]{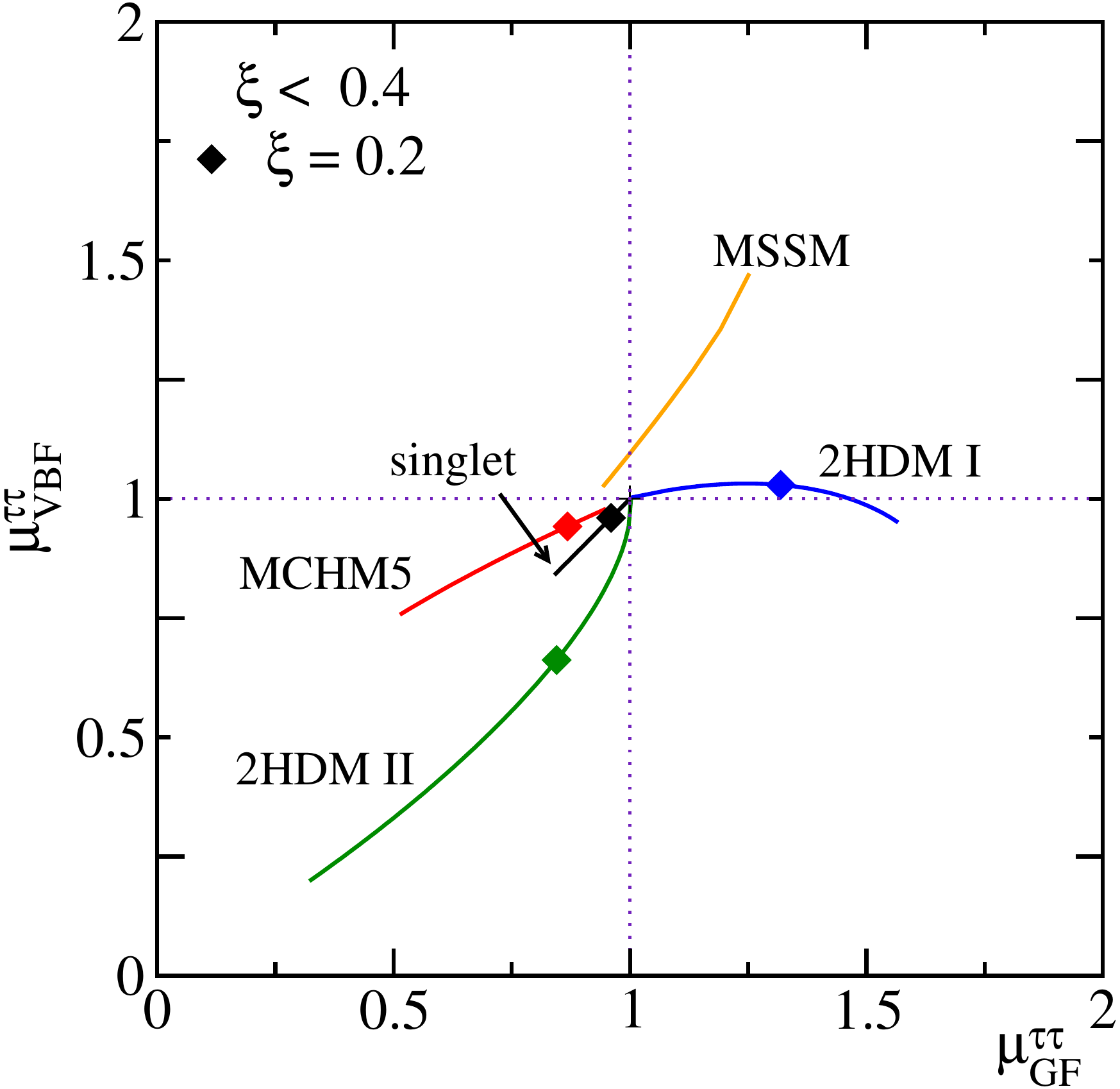}
\end{center}
\caption{Decay--diagonal correlations of signal strengths
  $\mu_{\text{GF},d}$ vs $\mu_{\text{VBF},d}$ for $d = \gamma\gamma, VV, 
   \tau\tau$ in different models.  The coupling variation is
  limited to $\xi < 0.4$ and the value $\xi = 0.2$ is singled
  out. The slight deviations from a complete decoupling are 
  discussed in the text.}
\label{fig:comparison}
\end{figure}

For the above described modifications to a Standard--Model--like Higgs
sector we find simple patterns in the two-dimensional plane of
coupling strengths.  In Figure~\ref{fig:comparison} we compare the
correlated modifications as functions of the decoupling parameter
$\xi$. In general, the behavior in the $H \to VV$ and $H \to \gamma
\gamma$ decay planes should be similar, as long as the loop--induced
Higgs--photon coupling is dominated by the $W$ loop.  We focus on $\xi
< 0.4$, corresponding to a modification of $g_V$ by 8\%, which the LHC is sensitive to with sufficient luminosity.  In addition, we
mark a deviation by $\xi=0.2$, equivalent to a 2\% coupling
deviation. The latter could be considered the target of a linear
collider analysis.\bigskip

For the simplest models, the dark singlet and the singlet mixing, all
correlations follow a straight diagonal line towards reduced coupling
strengths $\mu_{p,d} < 1$. This is due to the simple mixing pattern
and the net scaling of the LHC event rates as $\sigma \times
\text{BR} \propto g^2$. The same pattern appears for the simplest
strongly interacting models with a single Higgs form factor. For the
more complex strongly interacting model MCHM5, we find both correlated
($\mu_{\text{GF},\tau\tau} - \mu_{\text{VBF},\tau\tau}$) and
anti--correlated ($\mu_{\text{GF},VV} - \mu_{\text{VBF},VV}$)
patterns. It predicts an increased number of weak boson fusion events
whenever the couplings $\Delta_V$ in the production process decrease
more slowly than the coupling $\Delta_f$ in the total width, as shown
in Eq.\eqref{eq:scale_mchm}.

Larger departures from the Standard Model are possible in the 2HDM,
thanks to its more flexible coupling structure.  In the type-I setup
there is essentially no mechanism to increase the number of weak boson
fusion events as compared to the Standard Model, because of the
increase in the Higgs width combined with the reduced gauge boson
coupling. This is described in more detail in Appendix~\ref{app:new_physics}.
 For gluon fusion production combined with a fermionic decay the
suppression by the total width can be compensated by the production
and decay couplings. In the type-II setup both signal strengths can,
unexpectedly, be enhanced for bosonic Higgs decays. The reason is a
strongly decreased partial Higgs width to bottoms and taus which
cannot be generated in the type-I model.  For fermionic decays the
direct link between the bottom and tau down-type Yukawas leads to a
systematically decreased event rate. The fact that for $\xi \to 0$ the
2HDM rates do not match the Standard Model is linked to the finite
contribution of the charged Higgs to the effective photon--Higgs
coupling.

Finally, the MSSM as a constrained type-II 2HDM shows limited signal
strength variations because of the supersymmetric constraints.  Unlike
 the general 2HDM even in the type-II setup the MSSM does not allow
for a free variation of the two parameters $\alpha$ and $\beta$, which
affect the Yukawa couplings in a complicated manner.  Departures from
the decoupling limit in this case lie below $\xi \lesssim 0.2$ for the
considered parameter space configurations.  In the MSSM deviations
from the Standard Model in the limit $\xi \to 0$ arise through
contributions of the sfermions and the charged Higgs to both the
effective gluon--Higgs and photon--Higgs couplings. In that sense the
parameter $\xi$ does not fully track down the decoupling limit
for the loop--induced Higgs couplings, similarly to the 2HDM case. 
The effect of a shifted bottom
Yukawa is not sufficient to overcome the reduction in $g_V$, which
means that unlike for the 2HDM both signal strength deviations for the $VV$ and
$\gamma \gamma$ decays are negative, leading to an (almost linear) correlated suppression. 
Moreover, we see that the typical
deviations in the signal strengths can be achieved for small values
$\xi<0.2$ in the MSSM, because quantum effects dominate over
the mere tree--level rescaling $\Delta_V$.

\paragraph{A Heuristic for Robustness to Theory Uncertainty}

One of the great challenges of the Higgs coupling program is to understand what type of deviation from the Standard Model prediction would be compelling enough to make a claim for new physics.  It is clear that the pattern of deviations in the various production and decay modes carries much more information than considering them individually.  Furthermore, the inability to measure the total width of the Higgs necessitates either assumptions on the total width or consideration of various ratios in which the total width cancels. Both approaches lead to strong correlations in the inferred couplings.  If the theoretical uncertainties were well defined and statistical in nature, the significance of any given deviation could be readily assessed by standard statistical methods.  However, the ill-defined nature of theoretical uncertainties is beyond the scope of rigorous statistical procedures.  

For example, if one were to see a $4\sigma$ deviation from the Standard Model that could be reduced to a $2\sigma$ deviation by inflating a theoretical uncertainty by a factor of two or by changing from a Gaussian constraint to an \textsc{Rfit} constraint, then the deviation would most likely be met with healthy skepticism by the community. However, a deviation that is orthogonal to the effect of a theoretical uncertainty is much more robust.\bigskip

This motivates a heuristic to evaluate the robustness of an observed deviation $\hat{\vec\mu}$ to a theoretical uncertainty parametrized by $\alpha_i$, which we denote $R_i(\vec\mu)$.  
We want a larger deviation from the Standard Model to be reflected as a larger value for this robustness heuristic, so we begin with the length of the vector $|\vec\mu -\vec{1}|$. We also want the robustness to be larger as the deviation from the Standard Model $(\vec\mu -\vec{1})$  becomes more orthogonal to the shift in the signal strength induced from varying $\alpha_i$ as defined in Eq.\eqref{Eq:SolveViaPartial} and denoted  as 
$ \partial_{\alpha_i} \,\vec\mu^\text{fix}$. Since the magnitude of the theoretical uncertainty is poorly defined, it is natural that we only consider the angle between $(\vec\mu -\vec{1})$ and  $\partial_{\alpha_i} \,\vec\mu^\text{fix}$. This leads to the robustness heuristic
\begin{equation}\label{eq:robustness_obs}
R_i(\vec\mu) = \frac{|\vec\mu-\vec{1}|^2 \, | \partial_{\alpha_i} \,\vec\mu^\text{fix}|}{(\vec\mu-\vec{1}) \cdot (\partial_{\alpha_i} \,\vec\mu^\text{fix})} \; .
\end{equation}

\begin{figure}[t]
\includegraphics[width=0.32\textwidth]{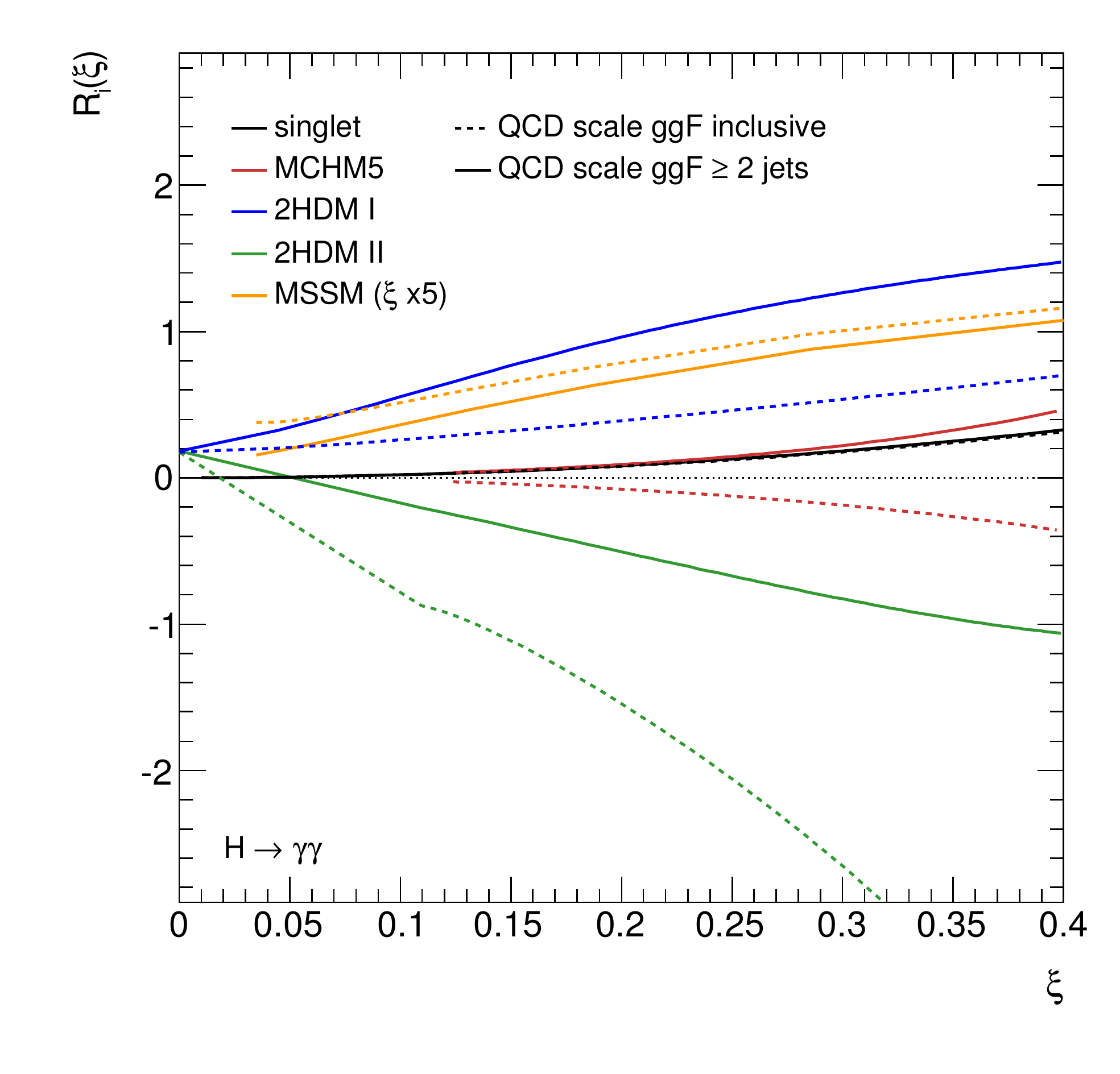} 
\includegraphics[width=0.32\textwidth]{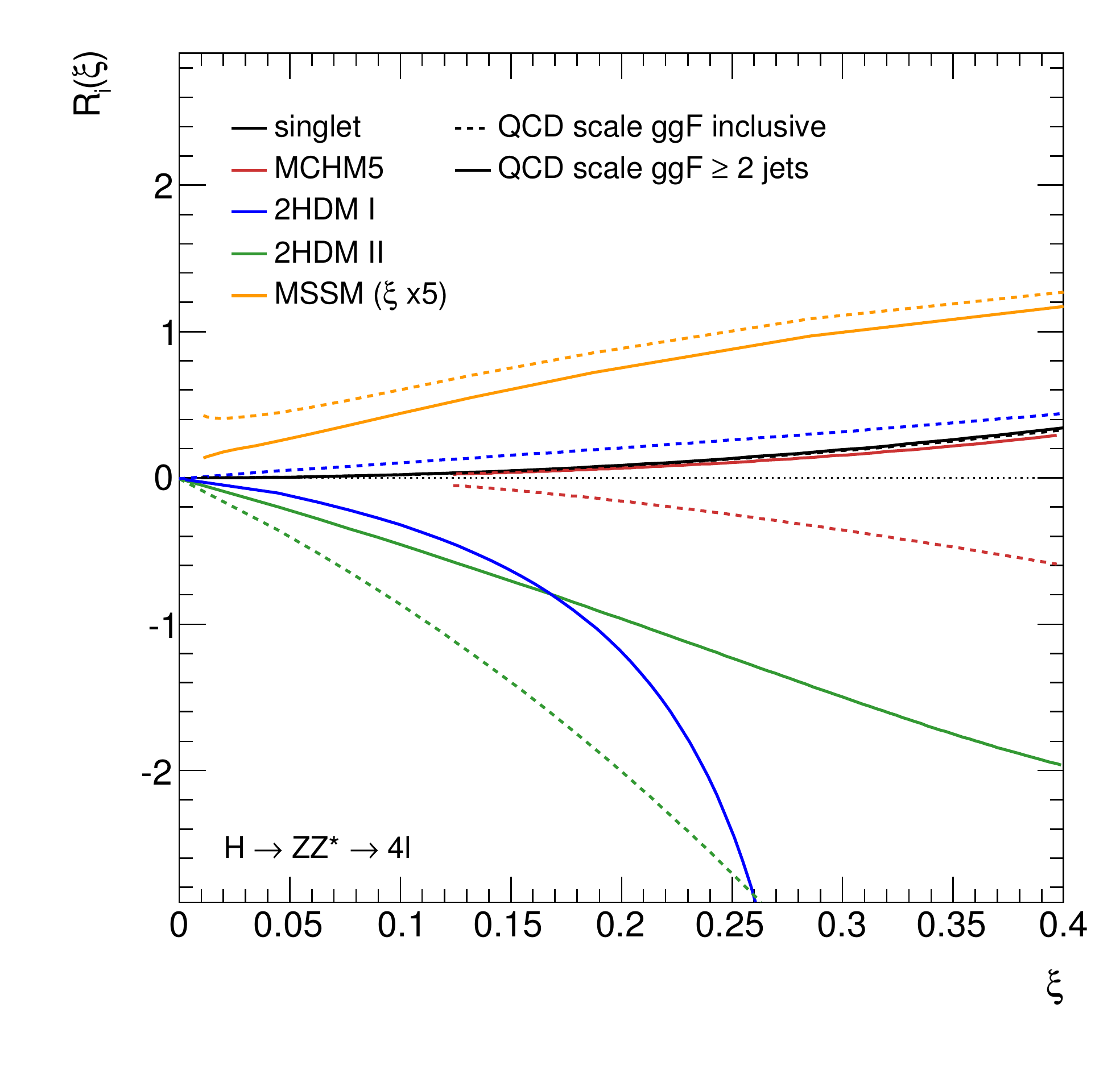} 
\includegraphics[width=0.32\textwidth]{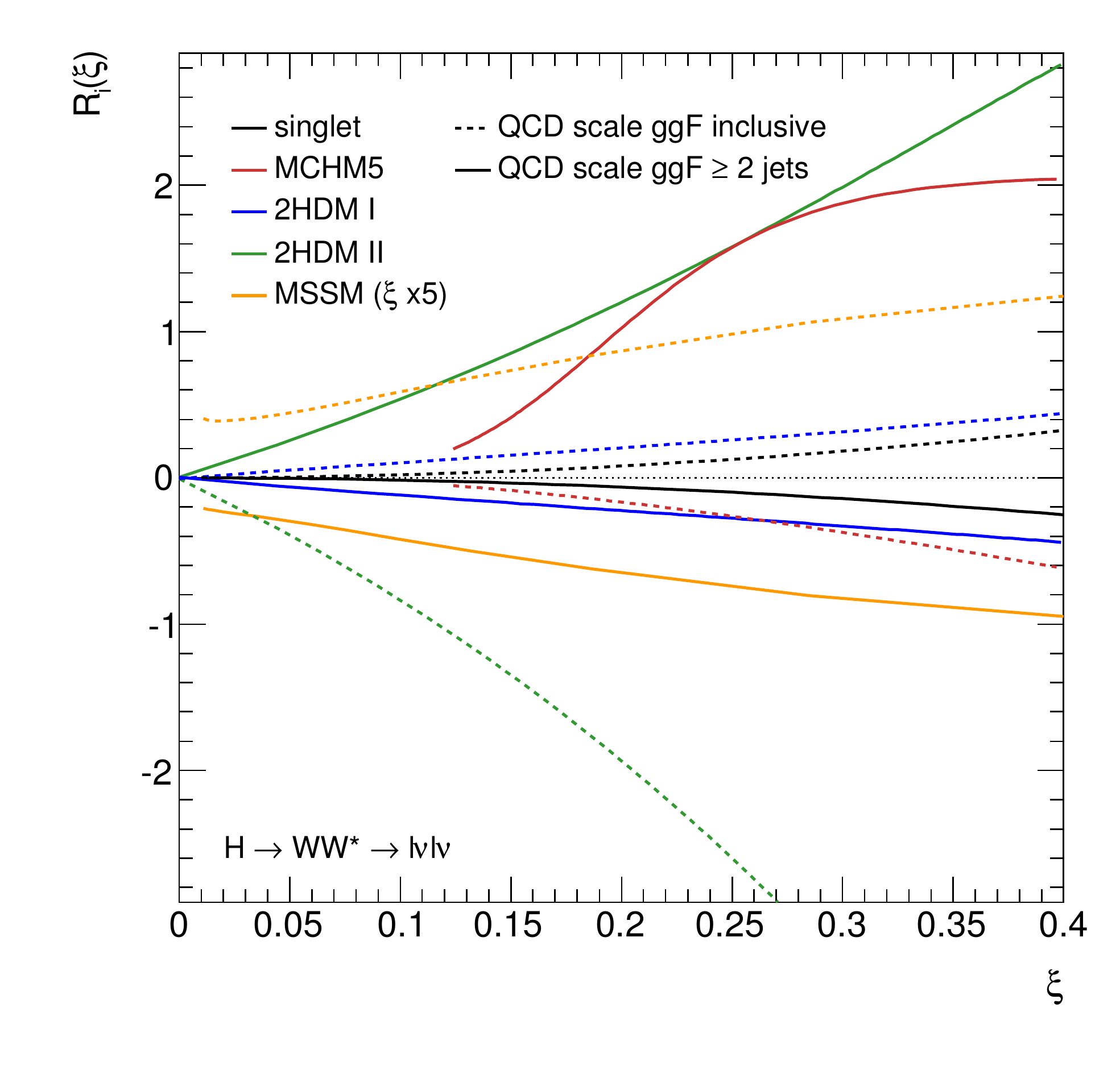}
\caption{The sensitivity heuristic $R_i(\xi)$ evaluated for various new physics models and the theoretical uncertainties $i$ associated to the gluon fusion cross section for $\ge 0$-jets and $\ge 2$-jets.}
\label{fig:robustness}
\end{figure}

This same  heuristic can be used to assess an expected departure from the Standard Model based on a new physics model parametrized by $\xi$ via composition $R_i(\vec\mu(\xi))$.  Figure~\ref{fig:robustness} shows the heuristic for the new physics effects in Fig.~\ref{fig:comparison} combined with the theoretical uncertainties in the toy model illustrated in Fig.~\ref{fig:visEtas}. As 
expected from the discussion of the different models, the different 
two--Higgs--doublet models can be distinguished from the Standard
Model even for $\xi\sim 0$. While for the type-I model the VBF topology is more robust with 
respect to potential QCD effects, deviations due to a type-II model are
more robust to the inclusive Higgs production rate. The most reliable 
signature for the strongly interacting MCHM5 model would be observed in 
weak--boson--fusion Higgs production with a decay $H \to WW$. For the 
MSSM the $\xi$ axis is rescaled, because deviations $\xi \gtrsim 0.1$ are 
hardly generated in our scan over MSSM spectra. The robustness of a
supersymmetric Higgs sector is roughly equal for the inclusive and VBF 
Higgs topologies.

\section{Conclusion}

Motivated by the fact that ill-defined theoretical uncertainties will eventually be the limiting factor in Higgs
coupling measurements at the LHC, we have developed a technique to decouple the theoretical
uncertainties from the experimental results while retaining the ability to incorporate those
uncertainties in a subsequent stage we refer to as \textit{recoupling}.  

This approach is amenable to simultaneously measuring multiple quantities, such as a vector of signal
strength parameters $\vec\mu$ for different Higgs production and decay signatures.  Moreover, the technique lends itself well to combinations with
several common sources of uncertainty that induce correlations among the contributing
measurements. In that respect it is similar to the Best Linear Unbiased Estimator (BLUE)~\cite{BLUE} technique, but not restricted to
Gaussian measurements or linear response to the source of uncertainty.  We considered a toy example modeled after the current ATLAS Higgs coupling measurements where  the measurements are not in the Gaussian regime and these non-linear effects are important for approximating the full likelihood function.

One of the most powerful features of this approach is that it allows one to change the assumptions
on both the magnitude and the shape of the uncertainty in the recoupling stage, which may occur
long after the experimental groups have released their results. This includes the ability to introduce \textit{a priori} correlations in the source of the systematics, which might have been neglected originally.  

These same capabilities would be possible if the experiments published the full statistical model using \texttt{RooFit/RooStats}~\cite{RooFitRooStats} as suggested in Refs.~\cite{PresentationOfResults}; however, the approach outlined here is less technology dependent.  The conceptual picture of these uncertainties leading to shifts in the inferred values of the parameters is intuitive and provides convenient visualizations as in Fig.~\ref{fig:visEtas}. \bigskip

In order for the experiments to present their results in this way, they would need to
\begin{itemize}
 \item publish the effective likelihood $L_\text{eff}(\vec\mueff)$ profiling only parameters that are not anticipated to be common to other measurements,
 \item publish the reparametrization template $\vec\mueff(\vec\mu,\vec\alpha)$, and
 \item document the conventions that establish meaning to the nuisance parameters $\vec\alpha$.
 \end{itemize}
In return, our approach allows for a flexible treatment of systematic uncertainties and removes the burden of choosing a description of theoretical uncertainties from the experimental groups. It also allows for future improvements in the theoretical description of Higgs processes at the LHC to be easily incorporated.
We discussed multiple strategies to determine the reparametrization template in Sec.~\ref{sec:approach}.

In the absence of a reliable measure of theoretical uncertainties
the key question becomes how easily an apparent deviation of
experimental measurements from the Standard Model description can be
explained by a change in the assumed theoretical uncertainties.  If we
describe new physics effects in the Higgs sector as one-parameter
deviations from the Standard Model decoupling limit in the signal
strength planes, the directions of these deviations can be compared to
the effects of a change in the theoretical uncertainties in the same
planes. The direct comparison of these two possible explanations for
an experimental observation leads us to the robustness heuristic
presented in Eq.\eqref{eq:robustness_obs}.\bigskip

While this paper focuses on the application of our approach to Higgs coupling measurements and the theoretical uncertainties associated with QCD, the technique is quite general and may find broad applications. 

\bigskip
\begin{center}
{\bf Acknowledgements}
\end{center}

The authors thank M. D\"uhrssen for discussion and encouragement early in the process of this project.  We thank M. Rauch for presenting preliminary results on our behalf.  We also thank Jamison Galloway for feedback on an early draft. DLV is supported by the F.R.S.-FNRS  ``Fonds de la Recherche Scientifique" (Belgium). TP would like to thank the CCPP at New York University for their hospitality and acknowledges BMBF support
under project number 05H12VHE. KC and SK are supported by the US National Science Foundation grants PHY-0854724 and PHY-0955626. Finally, we would like to thank a set of stunningly incompetent as well as unprofessional referees\footnote{For additional information please consult the original layout of the paper.} for turning us to a journal where this paper was received with great appreciation.

\newpage
\appendix

\section{A Worked Example}
\label{Sec:SimpleExample}

In this Appendix we work through a simple example that illustrates explicitly the procedure described in Sec.~\ref{sec:approach}. A \texttt{Mathematica} notebook carrying out these calculations can be found in Ref.~\cite{DOIforToyModel}. We consider two Gaussian measurements (indexed by $c=1,2$), which can be thought of as approximating the Poisson distribution for number counting analyses or approximating the maximum likelihood estimator of some more complicated analysis.  We consider two signal processes (indexed by $p=1,2$) with nominal (\textit{e.g.} Standard Model) expectations $s_{cp}$ and signal strength modifiers $\mu_p$.  Finally, we consider one systematic effect parametrized by $\alpha$ that shifts the signal expectation from the nominal $\alpha_0=0$ so that the expectation is
\begin{equation}\label{eq:exNu}
\nu_c(\mu_1,\mu_2,\alpha) = \sum_{p=1,2} \mu_p s_{cp} (1+\eta_{cp} \alpha)  + b_c \; .
\end{equation}
Including a Gaussian constraint term for the parameter $\alpha$ leads to the full likelihood
\begin{equation}\label{eq:exmodel}
L_\text{full}( \nu_1, \nu_2, \alpha) = \Gaus(x_1 | \nu_1, \sigma_1)\,\Gaus(x_2 | \nu_2, \sigma_2)\,\Gaus(a=0 | \alpha, \sigma_\alpha) \; .
\end{equation}
The maximum likelihood estimators are
\begin{eqnarray}\label{eq:exMLE}
\hat{\mu}_1 &=& 
\frac{s_{12} (x_2-b_2) - s_{22} (x_1-b_1)}{
s_{12} s_{21} - s_{11} s_{22}} \\ \nonumber
\hat{\mu}_2 &=&\frac{s_{21} (x_1-b_1) - s_{11} (x_2-b_2)}{
s_{12} s_{21} - s_{11} s_{22}} \\ \nonumber
\hat{\alpha}&=&0
\end{eqnarray}
The full Fisher information matrix is straightforward to calculate, but cumbersome to write explicitly.  Results below only require the off-diagonal  block  elements 
\begin{eqnarray}\tiny
V^{-1}_{\text{full}\, 13} &=&
   \frac{1}{
  \sigma_1^2 \sigma_2^2} \, \Big [ b_1 \eta_1 s_{11} \sigma_2^2 + 
    \eta_2 \hat\mu_2 (s_{11} s_{12} \sigma_2^2 + \sigma_1^2 s_{21} s_{22})  \\
       && + 
    \eta_1 (b_2 \sigma_1^2 s_{21} + 2 \hat\mu_1 (s_{11}^2 \sigma_2^2 + \sigma_1^2 s_{21}^2)   + \hat\mu_2 (s_{11} s_{12} \sigma_2^2 + \sigma_1^2 s_{21} s_{22}) - s_{11} \sigma_2^2 x_1 - 
       \sigma_1^2 s_{21} x_2)  \Big ] \notag \\ 
V^{-1}_{\text{full}\, 23}  & =& \frac{1}{
  \sigma_1^2 \sigma_2^2} \, \Big [ b_1 \eta_2 s_{12} \sigma_2^2 + 
    \eta_1 \hat\mu_1 (s_{11} s_{12} \sigma_2^2 + \sigma_1^2 s_{21} s_{22})  \notag \\
       && +    \eta_2 (b_2 \sigma_1^2 s_{22} + \hat\mu_1 (s_{11} s_{12} \sigma_2^2 + \sigma_1^2 s_{21} s_{22})  + 
       2 \hat\mu_2 (s_{12}^2 \sigma_2^2 + \sigma_1^2 s_{22}^2) - s_{12} \sigma_2^2 x_1 - 
       \sigma_1^2 s_{22} x_2) \Big ] \notag
\end{eqnarray}

\bigskip

In order to decouple the uncertainty, the experiment would provide the likelihood for an effective signal strength with respect to the nominal prediction $\alpha_0=0$.  The effective likelihood would be based on
\begin{equation}
\nu_c(\muefff{1},\muefff{2}) = \sum_{p=1,2} \mu_p^\text{eff} s_{cp}   + b_c \;,
\end{equation}
which has the same maximum likelihood estimates above and the following information matrix
\begin{equation}
V^{-1}_\text{eff} = \left[\begin{array}{cc}
\frac{s_{11}^2}{\sigma_1^2} + \frac{s_{21}^2}{\sigma_2^2}& \frac{s_{11} s_{12}}{\sigma_1^2} + \frac{s_{21} s_{22}}{\sigma_2^2} \\
\frac{s_{11} s_{12}}{\sigma_1^2} + \frac{s_{21} s_{22}}{\sigma_2^2} & \frac{s_{22}^2}{\sigma_2^2 + \frac{s_{12}^2}{\sigma_1^2}}\end{array}\right]
\end{equation}
The effective likelihood has the form:
\begin{equation}
L_\text{eff}(\muefff{1},\muefff{2}) \propto G ( \hat{\mu}_1,\hat{\mu}_2\, | \, \muefff{1},\muefff{2} , V^{-1}_\text{eff}) \;.
\end{equation}

\bigskip

Now we must choose a reparametrization template to be used to recouple the the uncertainty due to $\alpha$.  First let us choose the template of Eq.\eqref{Eq:inclusiveProdTemplateMu}, so that $\mu^\text{eff}_p = \mu_p (1+\eta_p \alpha)$.  Note, that the original model in Eq.\eqref{eq:exmodel} had four $\eta_{cp}$ while the template only has two $\eta_{p}$. If the effect of $\alpha$ is category universal so that $\eta_{c=1,p} = \eta_{c=2,p}$, then recoupling based on the this template will reproduce the full model exactly; however, in the more general situation $\eta_{c=1,p} \ne \eta_{c=2,p}$ it will not.

Now we solve for the coefficients of the template based on the local covariance matrix as described in Sec.~\ref{sec:coeffs}.  We outlined three equivalent approaches based on Eqs.\eqref{Eq:SolveViaPartial}, \eqref{Eq:SolveViaCov}, and \eqref{Eq:SolveViaFisher}. Let us demonstrate the last of these three approaches.  Based on the template of Eq.\eqref{Eq:inclusiveProdTemplateMu}, the reparametrization will lead to the Jacobian    
\begin{equation}
J = \frac{\partial( \muefff{1}, \muefff{2})}{\partial( \mu_1, \mu_2,\alpha)} = \left[\begin{array}{ccc}
(1+\eta_1 \alpha) & 0 & \mu_1 \eta_1 \\
0 & (1+\eta_2 \alpha) & \mu_2 \eta_2
\end{array}\right]
\end{equation}
As in Eq.\eqref{Eq:SolveViaFisher}, we use this Jacobian to relate the information matrix of the effective likelihood and the main measurement
\begin{equation}\label{eq:InfoMatrixReparam3}
V^{-1}_\text{main}(\mu_1,\mu_2,\alpha) = J^T V^{-1}_\text{eff} J \;.
\end{equation}
 The $(i=\mu_p,j=\alpha)$ sub-block of this matrix leads to the following system of linear equations
\begin{eqnarray}
\hat\mu_1 \eta_1 V^{-1}_\text{eff\ 11}  + \hat\mu_2 \eta_2 V^{-1}_\text{eff\ 12}  &=& V^{-1}_\text{main\ 13}\\ \nonumber
\hat\mu_1 \eta_1 V^{-1}_\text{eff\ 12}  + \hat\mu_2 \eta_2 V^{-1}_\text{eff\ 22} &=& V^{-1}_\text{main\ 23} \; ,
\end{eqnarray}
which can easily be inverted to provide solutions for $\eta_p$. In practice one would work with numerical representations of the maximum likelihood estimators and Fisher information matrices, but here we present the result symbolically. 
{\small
\begin{eqnarray}\label{eq:exEtaSolutions}
\eta_1 &=&  \Big [ b_1 s_{22} (-\eta_{12} s_{12} s_{21} - \eta_{21} s_{12} s_{21} + \eta_{22} s_{12} s_{21} + 
         \eta_{11} s_{11} s_{22}) + 
      b_2 s_{12} (\eta_{21} s_{12} s_{21} - (\eta_{11} - \eta_{12} + \eta_{22}) s_{11} s_{22}) \notag \\
      &+ &
      \eta_{12} s_{12} s_{21} s_{22} x_1 + \eta_{21} s_{12} s_{21} s_{22} x_1 - 
      \eta_{22} s_{12} s_{21} s_{22} x_1 - \eta_{11} s_{11} s_{22}^2 x_1 - \eta_{21} s_{12}^2 s_{21} x_2     \notag \\
      &+ &
      \eta_{11} s_{11} s_{12} s_{22} x_2 - \eta_{12} s_{11} s_{12} s_{22} x_2 + 
      \eta_{22} s_{11} s_{12} s_{22} x_2 \Big ] / 
      \Big [ (s_{12} s_{21} - s_{11} s_{22}) (b_2 s_{12} - b_1 s_{22} + 
        s_{22} x_1 - s_{12} x_2) \Big ] \notag \\
       && \\
\eta_2 &= &  \Big [ b_2 s_{11} (\eta_{11} s_{12} s_{21} - \eta_{12} s_{12} s_{21} - \eta_{21} s_{12} s_{21} + 
         \eta_{22} s_{11} s_{22}) + 
      b_1 s_{21} (\eta_{12} s_{12} s_{21} - (\eta_{11} - \eta_{21} + \eta_{22}) s_{11} s_{22}) \notag \\
      &- & 
      \eta_{12} s_{12} s_{21}^2 x_1 + \eta_{11} s_{11} s_{21} s_{22} x_1 - 
      \eta_{21} s_{11} s_{21} s_{22} x_1 + \eta_{22} s_{11} s_{21} s_{22} x_1 - 
      \eta_{11} s_{11} s_{12} s_{21} x_2  \notag \\
      &+&  \eta_{12} s_{11} s_{12} s_{21} x_2 +
      \eta_{21} s_{11} s_{12} s_{21} x_2 - 
      \eta_{22} s_{11}^2 s_{22} x_2  \Big ]  / 
       \Big [ (s_{12} s_{21} - s_{11} s_{22}) (-b_2 s_{11} + b_1 s_{21} - 
        s_{21} x_1 + s_{11} x_2)  \Big ] \notag 
\end{eqnarray}
}%
Note, in the category-universal situation, the solution simplifies to $\eta_p = \eta_{1p}=\eta_{2p}$ as expected.  In the non-category-universal situation it is possible for $\eta_p$ to lie outside of the range $[\eta_{1p},\eta_{2p}]$.

\paragraph{Three example scenarios}

To make these examples more explicit, Table~\ref{tab:scenarios} specifies three scenarios for the coefficients in Eq.\eqref{eq:exNu}.  Scenarios A and B are meant to be representative of LHC Higgs measurements in which the first category is gluon-fusion-like with significant background and signal  dominated by the first production mode ($p=1=\text{ggF}$), while the second category is VBF-like with negligible background and signal dominated by the second production mode ($p=2=\text{VBF}$).  The only difference between  Scenarios A and B are the $\eta_{cp}$ that quantify the response to the source of uncertainty. Since Scenario A is category-universal we achieve an exact reproduction of the full likelihood, while for Scenario B we don't expect exact results. Scenario C is meant to probe the extreme case in which the uncertainty for the first production mode only affects the expectation in the second category (which is representative of the theory uncertainty of gluon-fusion $+\ge 2$jets.

\begin{table}[htdp]
\caption{The values for the coefficients in Eq.\eqref{eq:exNu} used to define three example scenarios.}
\begin{center}
\begin{tabular}{c||c||c|c|c|c|c||c|c|c|c|c||c|c|c|c} \hline
Scenario & $\sigma_\alpha$ & $s_{11}$ & $s_{12}$ & $b_{1}$ & $x_{1}$ & $\sigma_{1}$ & $s_{21}$ & $s_{22}$ & $b_{2}$ & $x_{2}$ & $\sigma_{2}$ & $\eta_{11}$ & $\eta_{12}$& $\eta_{21}$ & $\eta_{22}$  \\ \hline \hline
A &  1&
    45& 5&  50& 100&  10&
   10& 90&  0&  100&  10&
   0.2& 0.2&   0.2&  0.2 \\ \hline
B &  1&
   45&  5&  50&  100&  10&
   10&  90& 0&  100&  10&
    0.1& 0.2&  0.3&  0.2 \\ \hline
C & 1&
   45&  5& 50&  100&  10&
   40&  60& 0&  100&  10&
   0 &  0&  0.2&  0\\ \hline \hline
\end{tabular}
\end{center}
\label{tab:scenarios}
\end{table}%

\begin{figure}[htb]
	\centering
	\subfigure{ \includegraphics[width=0.31\textwidth]{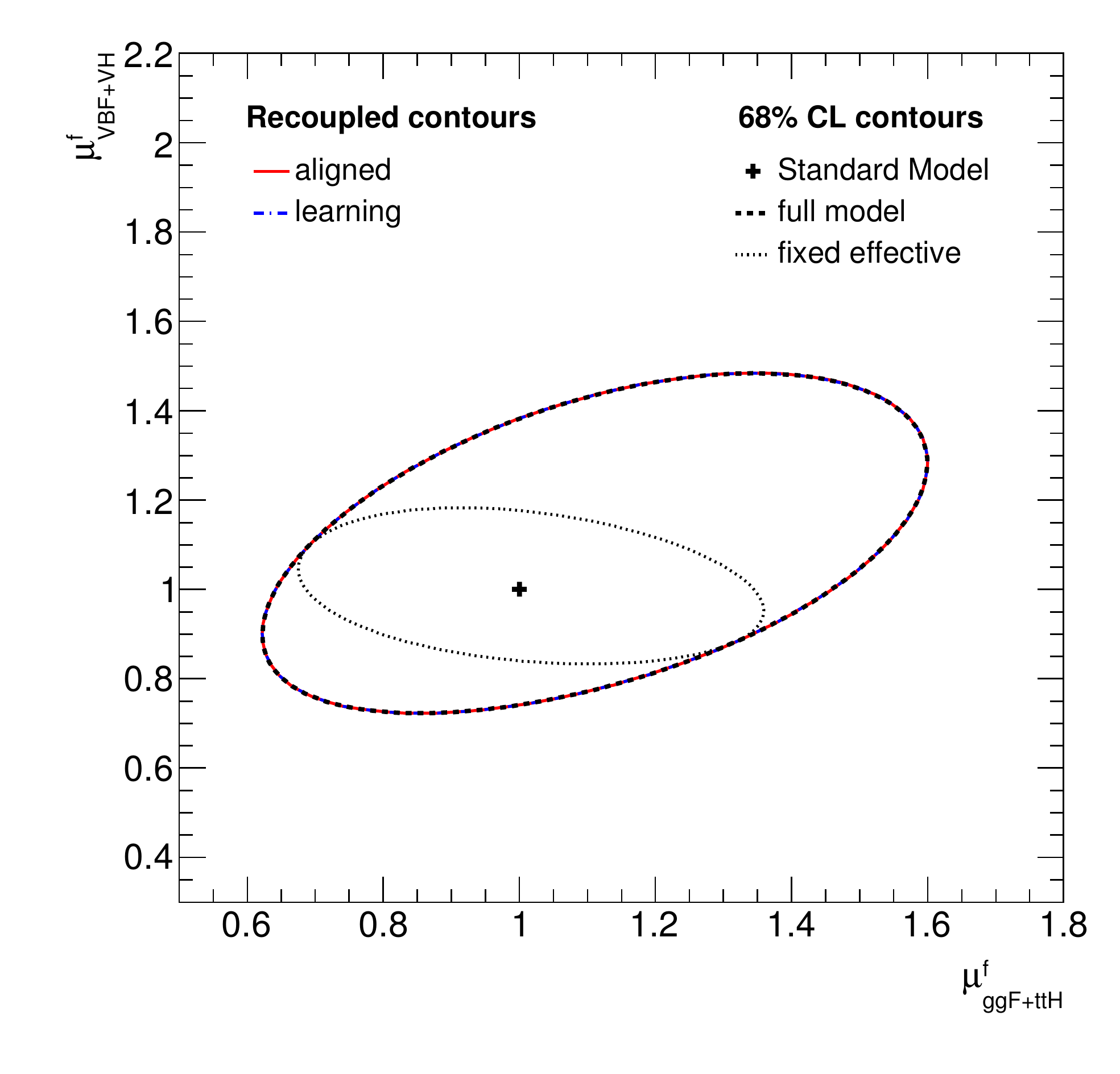} }
	\subfigure{ \includegraphics[width=0.31\textwidth]{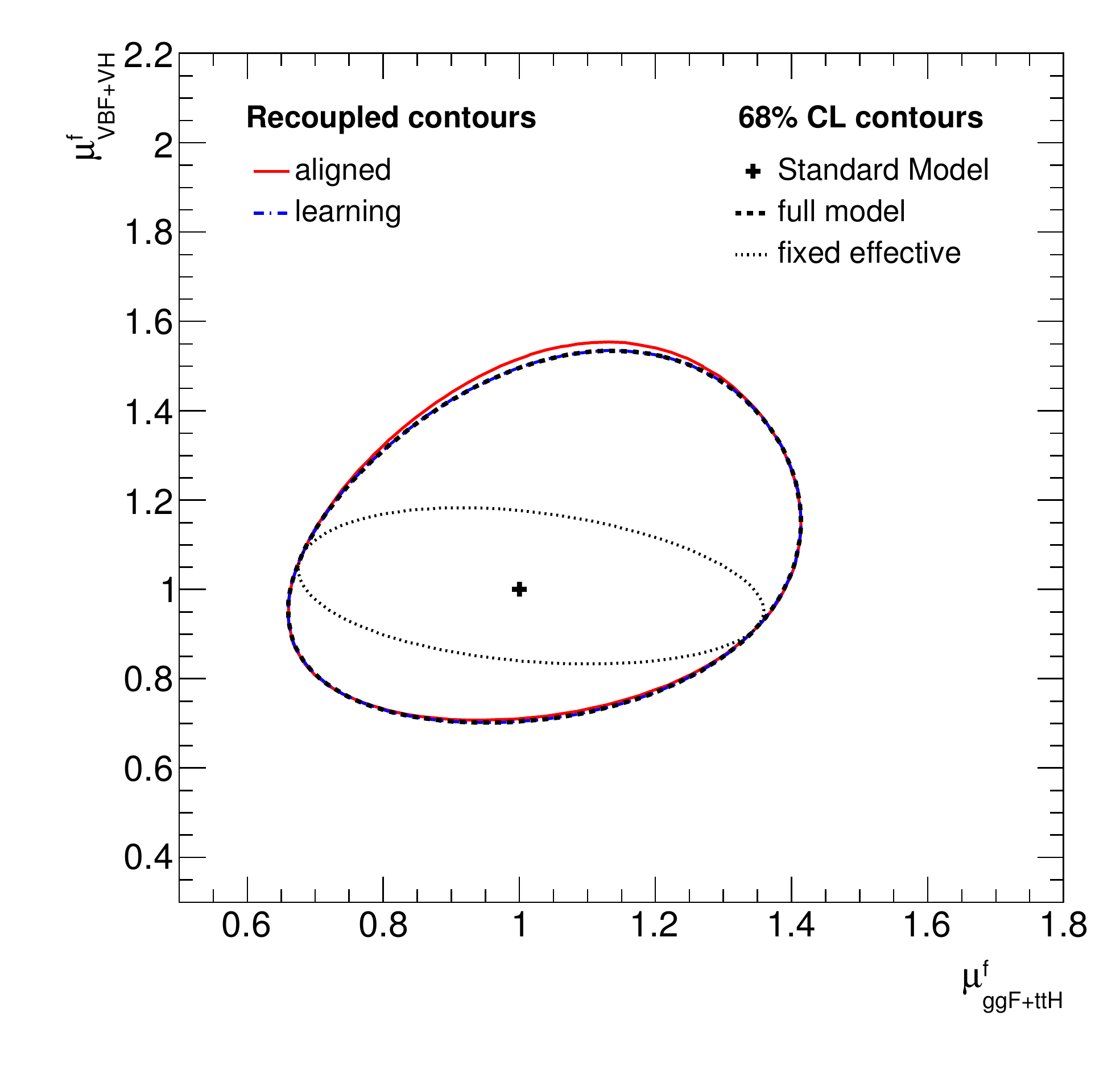} } 
	\subfigure{ \includegraphics[width=0.31\textwidth]{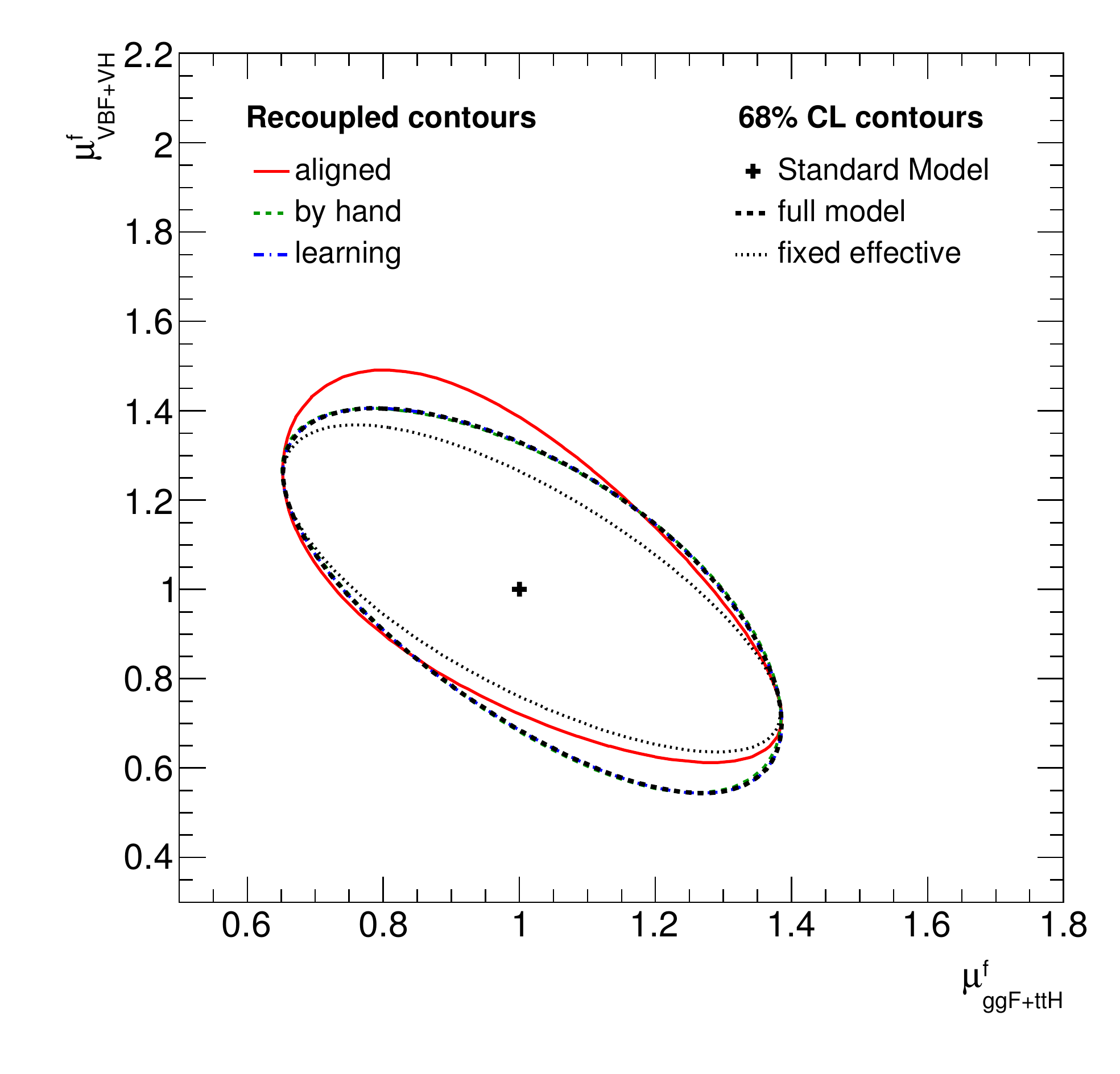} }\\
		\subfigure{ \includegraphics[width=0.31\textwidth]{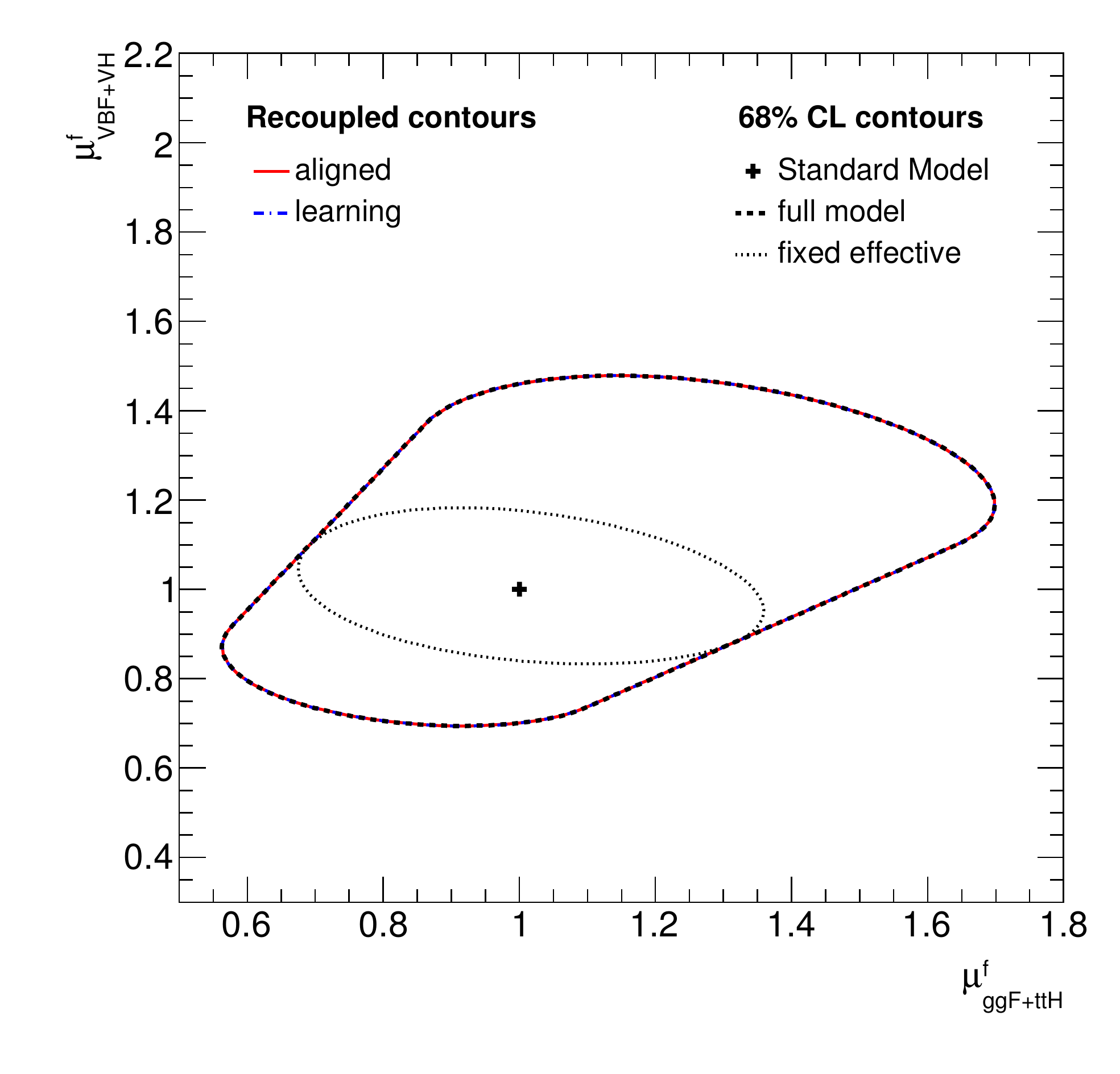} }
	\subfigure{ \includegraphics[width=0.31\textwidth]{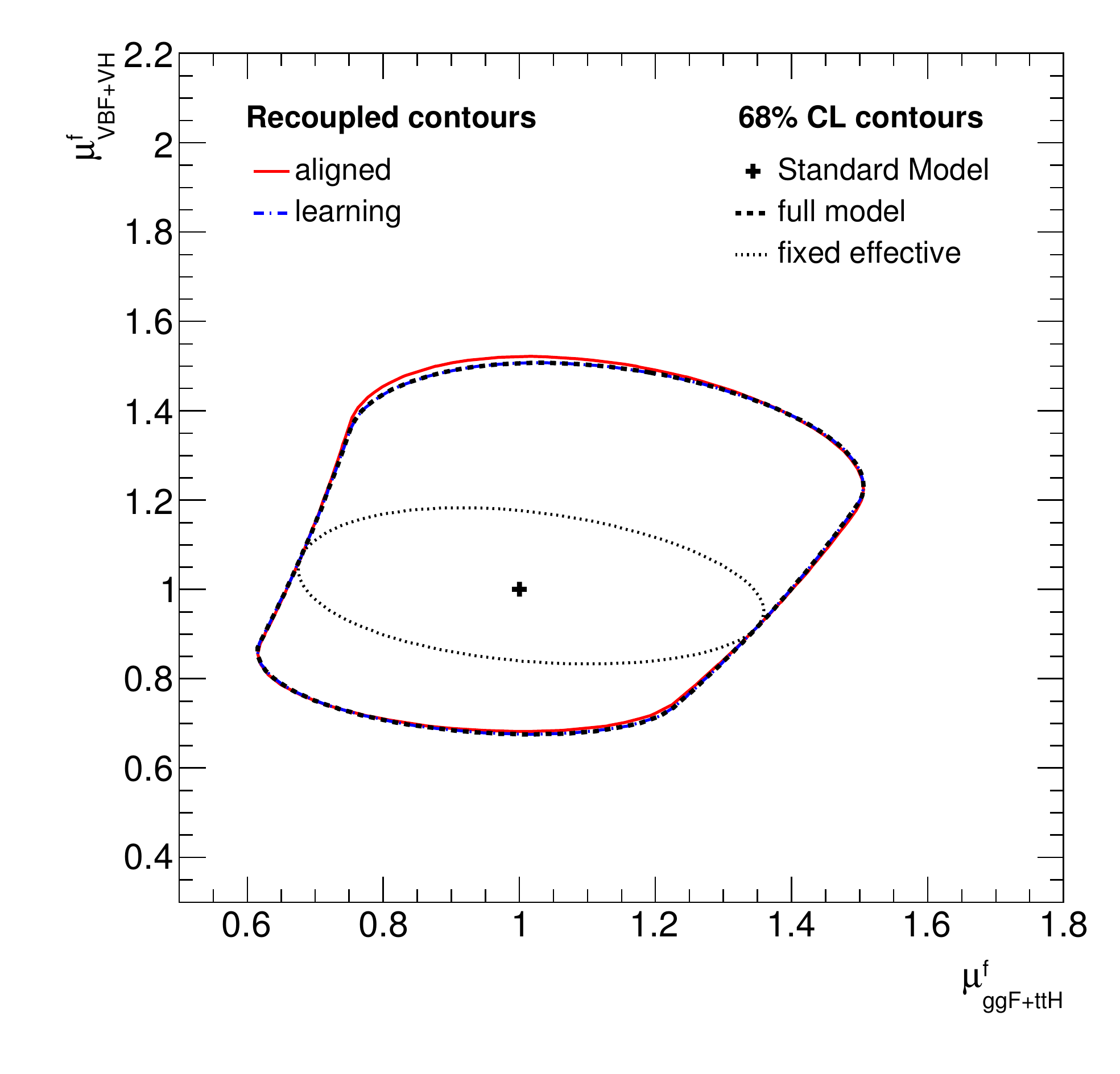} } 
	\subfigure{ \includegraphics[width=0.31\textwidth]{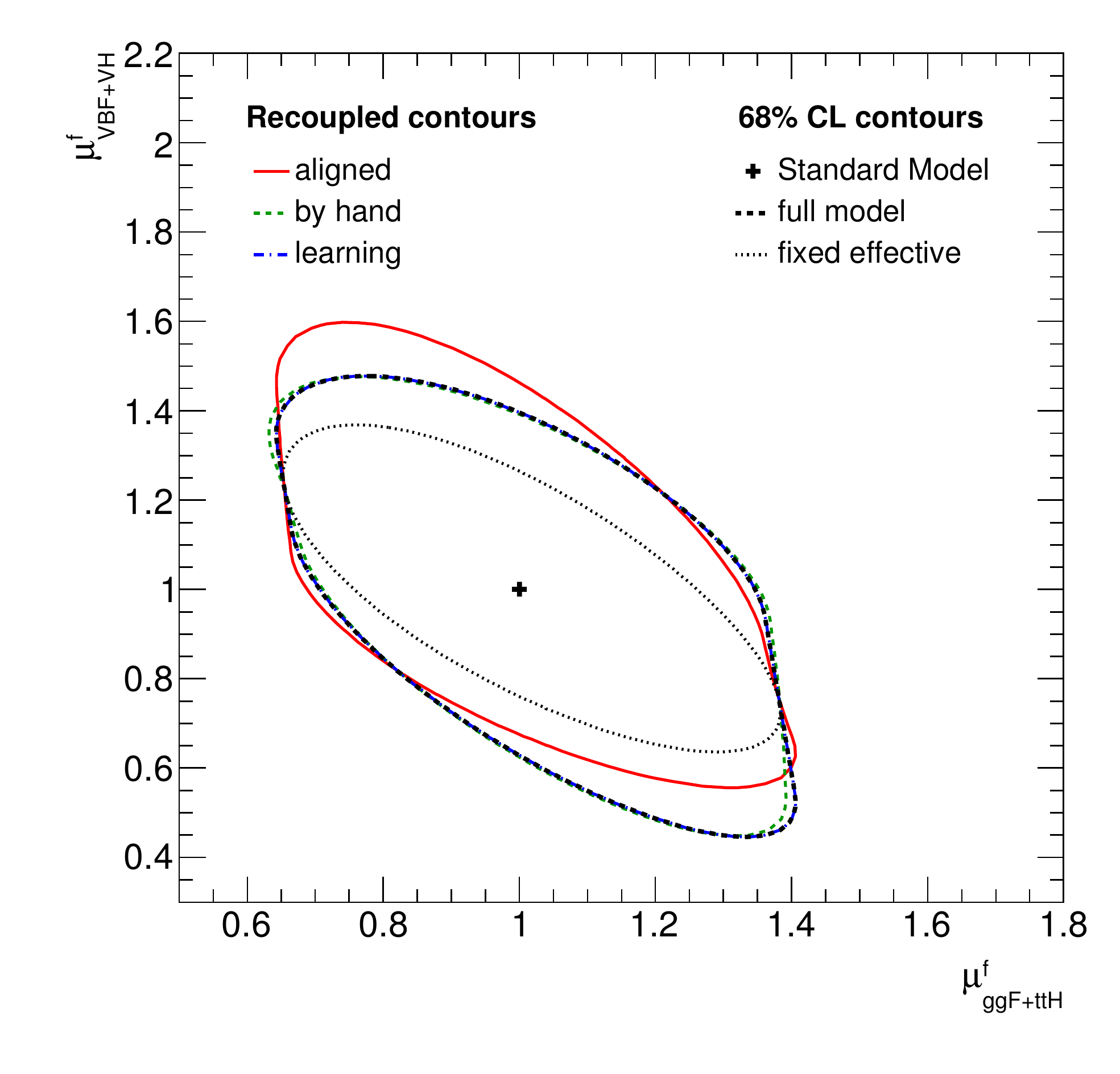} }
	\caption{Comparison of full likelihood (solid) and recouped (dashed) likelihood  for Scenarios A, B, and C. Scenario C illustrates the impact of using three templates `aligned' (red), `by hand' (green), and `learning' (blue) as described in the text. The top row is based on the nominal Gaussian constraint and the bottom row shows the result of replacing it with an alternative \textsc{Rfit} constraint term.
	The effective likelihood with $\alpha=0$ is shown as a dotted line.}
		\label{fig:2bin}
\end{figure}

Figure~\ref{fig:2bin} shows a comparison of the full and recoupled likelihood using the `aligned' template of Eq.\eqref{Eq:inclusiveProdTemplateMu} and the solutions to the coefficients in Eq.\eqref{eq:exEtaSolutions}.  Scenario A is reproduced exactly, the non-category-universal property of Scenario B leads to a slight discrepancy, and the extreme non-category-universal property of Scenario C leads to a substantial discrepancy.  In particular, while the effect of the uncertainty in Scenario C is to change the inferred value of $\mu_\text{VBF}$, the size of this effect should scale with $\mu_\text{ggF}$. The agreement can be improved dramatically by moving to the more general template of Eq.\eqref{Eq:generalTemplate} that includes $\eta_{pi}^{p'}$. One approach is to fix `by hand' the coefficients of the template $\eta_\text{ggF}^\text{ggF}=0$, $\eta_\text{VBF}^\text{VBF}=0$ and $\phi=0$ and determine the two remaining coefficients using the local covariance matrix (which proceeds as above with a different Jacobian transformation).  An alternate approach is to use the unrestricted template of  Eq.\eqref{Eq:generalTemplate} and utilizing the `learning' approach of Eq.\eqref{eq:loss}.  

Finally, the second row of Fig.~\ref{fig:2bin} shows a comparison of the full and decoupled likelihood with a modified constraint term.  In particular, the Gaussian constraint is replaced with an \textsc{Rfit}  constraint term: $\Gaus(0|\alpha,1) \to \text{Uniform}(-1,1)$.  The coefficients for the templates are the same as for both rows of Fig.~\ref{fig:2bin}.

\clearpage
\section{New Physics Models}
\label{app:new_physics}

In this appendix we will give a more detailed picture of the new
physics models and their features briefly discussed in
Section~\ref{sec:new_physics}. In particular, we will motivate and
discuss the description of new physics effects by a single parameter
$\xi$, defined as the modification of the Higgs couplings to massive
gauge bosons, \ie $\Delta_V \simeq -\xi^2/2$. Note that this unified
definition of $\xi$ differs from Ref.~\cite{sfitter_2hdm} for some of the new
physics models. All signal strength deviations we compute 
by rescaling the SM production cross section, branching ratio and total width~\cite{hxsec},
while for the MSSM case we use {\sc FeynHiggs}~\cite{feynhiggs}.

\paragraph{Dark singlet}

\begin{figure}[b!]
\begin{center}
\includegraphics[width=0.35\textwidth]{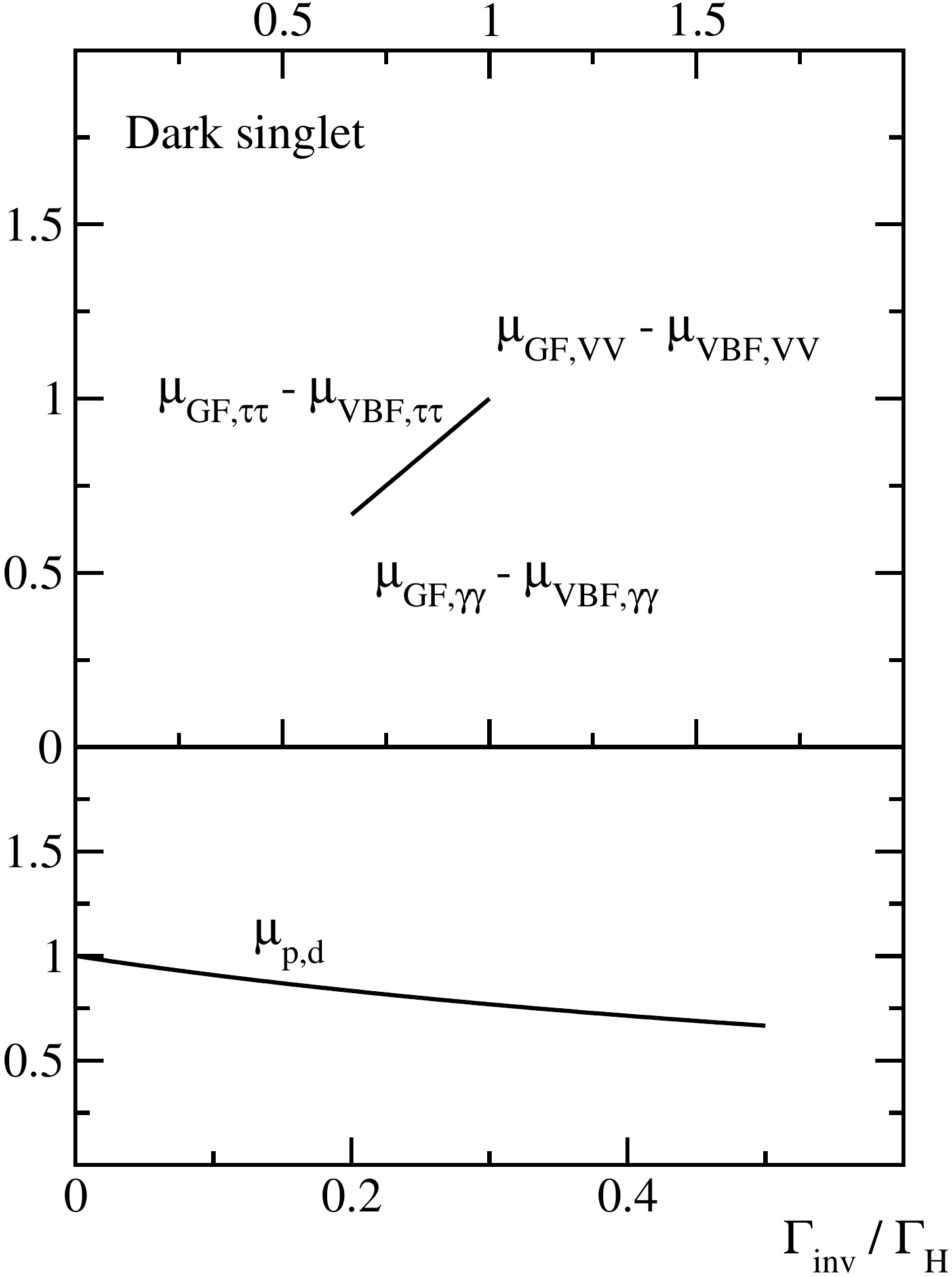} 
\hspace*{0.16\textwidth}
\includegraphics[width=0.35\textwidth]{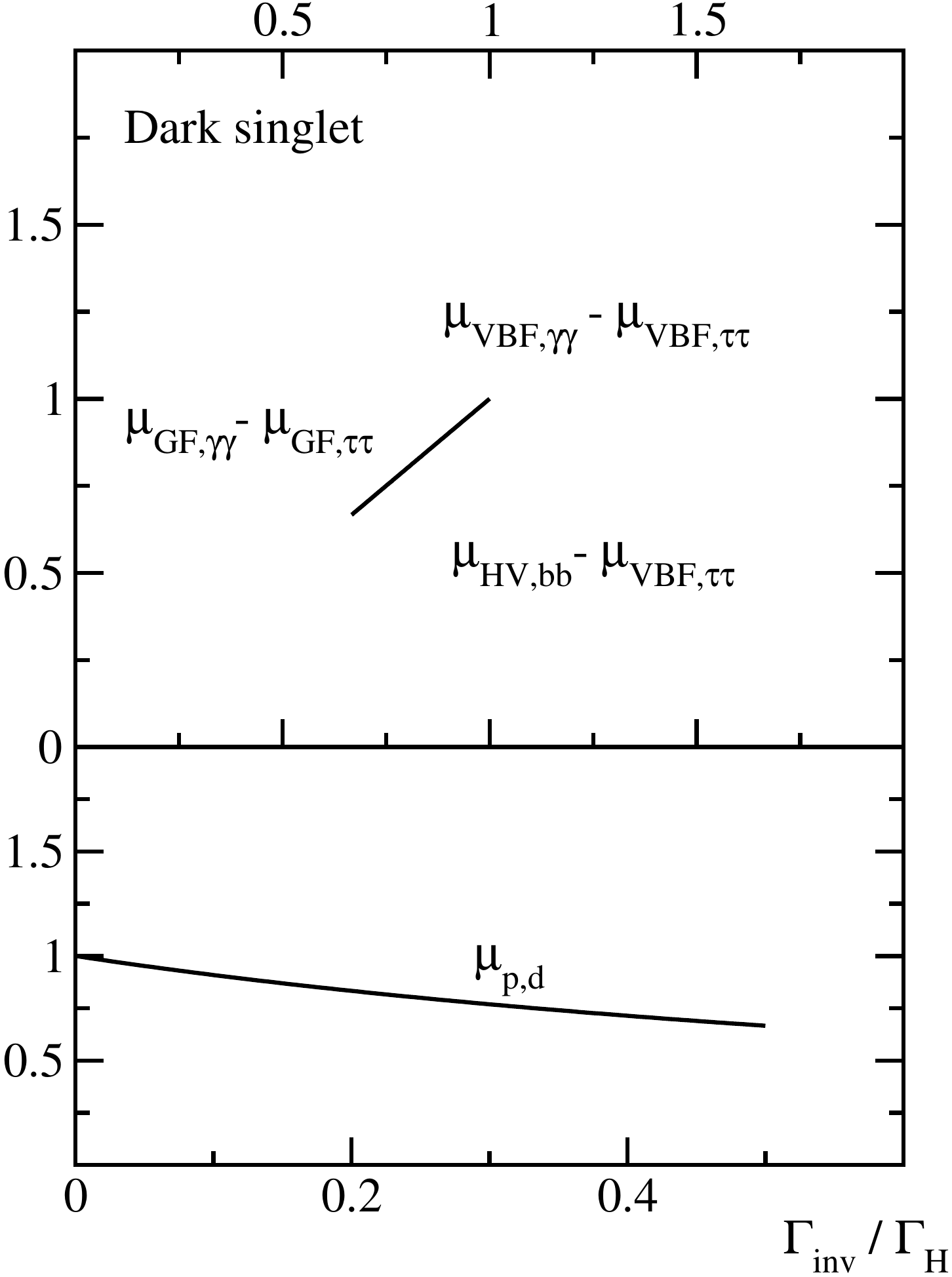} 
\end{center}
\vspace*{-7mm}
\caption{Dark singlet: correlated Higgs signal strengths for the
  decay--diagonal channels (left) and non--diagonal channels (right).
  The first of the two signal strengths in the notation
  $\mu_{p_1,d_1} - \mu_{p_2,d_2}$ is shown on the vertical axis,
  the second on the horizontal axis.  In the lower panels we give the
  different signal strengths as a function of
  $\Gamma_\text{inv}/\Gamma_\text{SM} = \xi^2$.}
\label{fig:correl-ds}
\end{figure}

A dark singlet is defined as a model with an additional scalar
particle $S$ which does not have a vacuum expectation value and hence
cannot mix with the Higgs boson. In addition, we assume that its only
interaction with the Standard Model will be the dimension-4
\emph{portal interactions} in the combined scalar
potential~\cite{dark_singlet_orig},
\begin{alignat}{5}
V(\Phi,S) = 
  \mu^2_1\,(\Phi^\dagger\,\Phi) 
+ \lambda_1\,|\Phi^{\dagger}\Phi|^2 
+ \lambda_3\,|\Phi^{\dagger}\,\Phi|S^2 \; .
\label{eq:pot_ds}
\end{alignat}
This interaction with strength $\lambda_3$ can, if kinematically
allowed, lead to an invisible Higgs decay
width~\cite{dark_singlet_coll} directly linked to a possible dark
matter agent~\cite{dark_singlet_all}
\begin{alignat}{5}
\Gamma_{\text{inv}} = \Gamma(h \to ss)
&= \frac{ \lambda^2_3 v^2}{32 \pi\,m_h}\; 
   \sqrt{1-\cfrac{4m_s^2}{m_h^2}}   
\equiv \xi^2 \; \Gamma_\text{SM} \; .
\label{eq:invwidth-text}
\end{alignat}
Such an invisible Higgs width contributes to the total Higgs width and
hence to the number of predicted Higgs events at the LHC. This leads
to an apparent reduction of all couplings, including the Higgs
coupling to massive gauge bosons, shown in Eq.\eqref{eq:scale_ds}.
Such a universal modification of all predicted event numbers
corresponds to a diagonal pattern, for example in the two--dimensional
$\mu_{\text{VBF},d}$ vs $\mu_{\text{GF},d}$ plane. In the upper panels of
Figure~\ref{fig:correl-ds} we show these identical diagonal
correlations for a set of decay--diagonal channels (left panels) and
non--diagonal channels (right panels). The correlations are all
identical. In the lower panels we show the expected signal strengths
relative to the Standard Model as a function of the invisible width.

\paragraph{Additional singlet}

\begin{figure}[t]
\begin{center}
\includegraphics[width=0.35\textwidth]{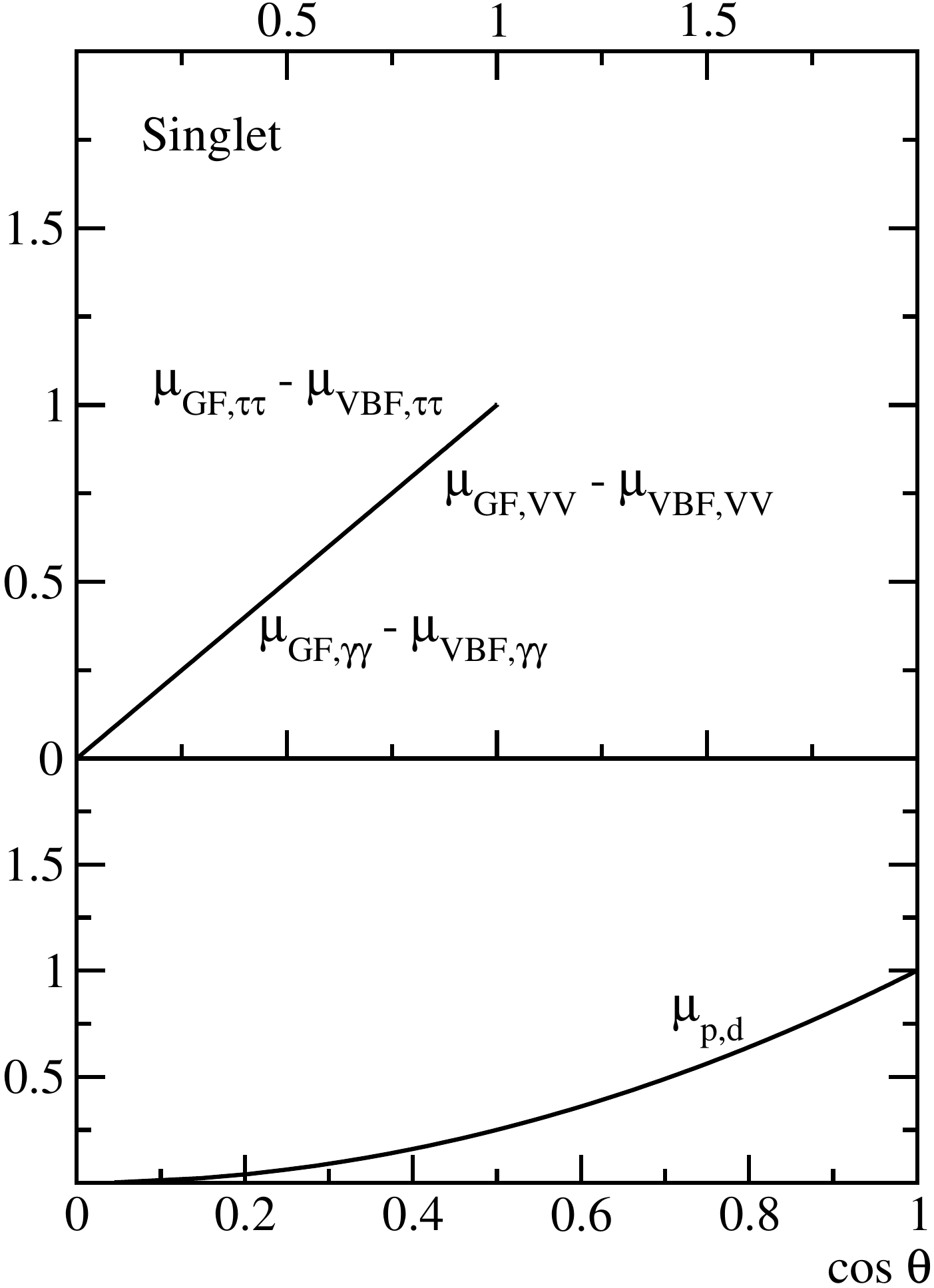} 
\hspace*{0.16\textwidth}
\includegraphics[width=0.35\textwidth]{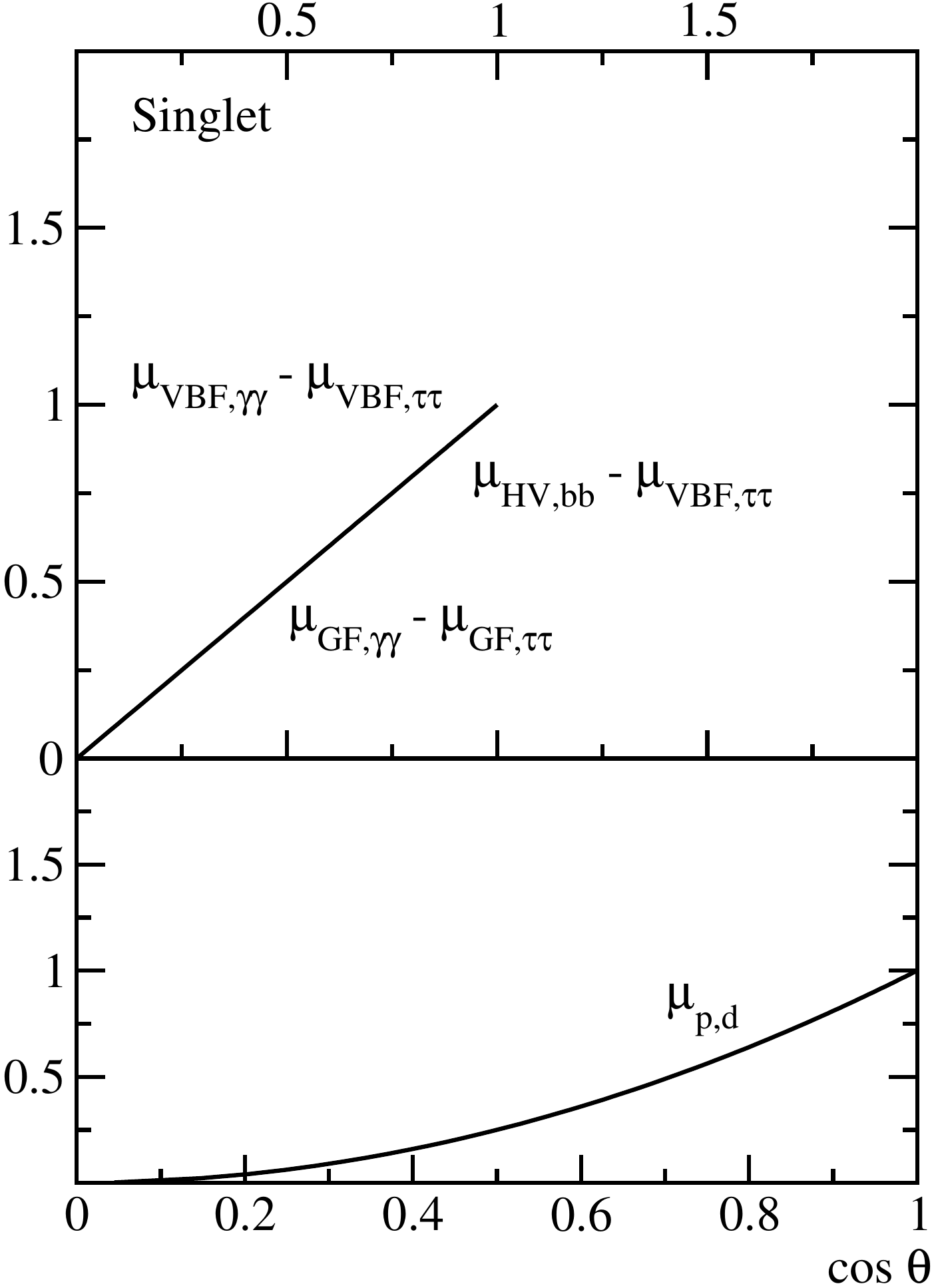} 
\end{center}
\vspace*{-7mm}
\caption{Singlet mixing: correlated Higgs signal strengths for the
  decay--diagonal channels (left) and non--diagonal channels (right).
  The first of the two signal strengths in the notation
  $\mu_{p_1,d_1} - \mu_{p_2,d_2}$ is shown on the vertical axis,
  the second on the horizontal axis.  In the lower panels we give the
  different signal strengths as a function of $\cos \theta =
  \sqrt{1-\xi^2}$.}
\label{fig:correl-singlet}
\end{figure}

If the additional $SU(2)_L$ singlet acquires a finite VEV $v_S$ the
combined Higgs potential~\cite{portal_orig,singlet_early}
\begin{alignat}{5}
V(\Phi,S) = 
  \mu^2_1\,(\Phi^\dagger\,\Phi) 
+ \lambda_1\,|\Phi^{\dagger}\Phi|^2 
+ \mu^2_2\,S^2 
+ \kappa S^3 
+ \lambda_2\,S^4 
+ \lambda_3\,|\Phi^{\dagger}\,\Phi|S^2 
\label{eq:singlet-potential}
\end{alignat}
with the portal interaction $\lambda_3$ leads to singlet--doublet
mixing. The rotation to mass eigenstates $h$ and $H$ defines the angle
\begin{equation}
\tan^2 (2\theta) = 
\dfrac{\lambda_3^2 v_H^2 v_S^2}
      {(\lambda_1 v_H^2-\lambda_2 v^2_S)^2} \; .
\label{eq:def_theta}
\end{equation}
All Higgs couplings to fermions and gauge bosons are rescaled by
$\cos\theta < 1$. In addition, the mostly Higgs state can decay into
two lighter, mostly singlet states~\cite{singlet_coll}
\begin{equation}
 \Gamma(H \to hh) = \frac{|\lambda_{Hhh}|^2}{32\pi m_H}\,\sqrt{1-\frac{4 m^2_h}{m^2_H}} \; .
\label{eq:singlet-width}
\end{equation}
The signature for such a decay depends on the lifetime and the decay
channels of the lighter state $h$. If sizeable, this additional decay
channel contributes to the total Higgs width, entering the predicted
number of LHC events as a second universal
modification.

This universal coupling modification again predicts diagonal lines for
all $\mu_{p_1,d_1} - \mu_{p_2,d_2}$ correlations, as shown in
Figure~\ref{fig:correl-singlet}. If we do not observe the new Higgs
decay modes for each of the additional singlet scenarios, the
degeneracy in the signal strength deviations makes it impossible to
distinguish a dark singlet, an additional singlet, and the simplest
strongly interacting form factor models (as discussed in the next
paragraph)~\cite{portal_us}.

\paragraph{Composite Higgs}

\begin{figure}[t]
\begin{center}
\includegraphics[width=0.35\textwidth]{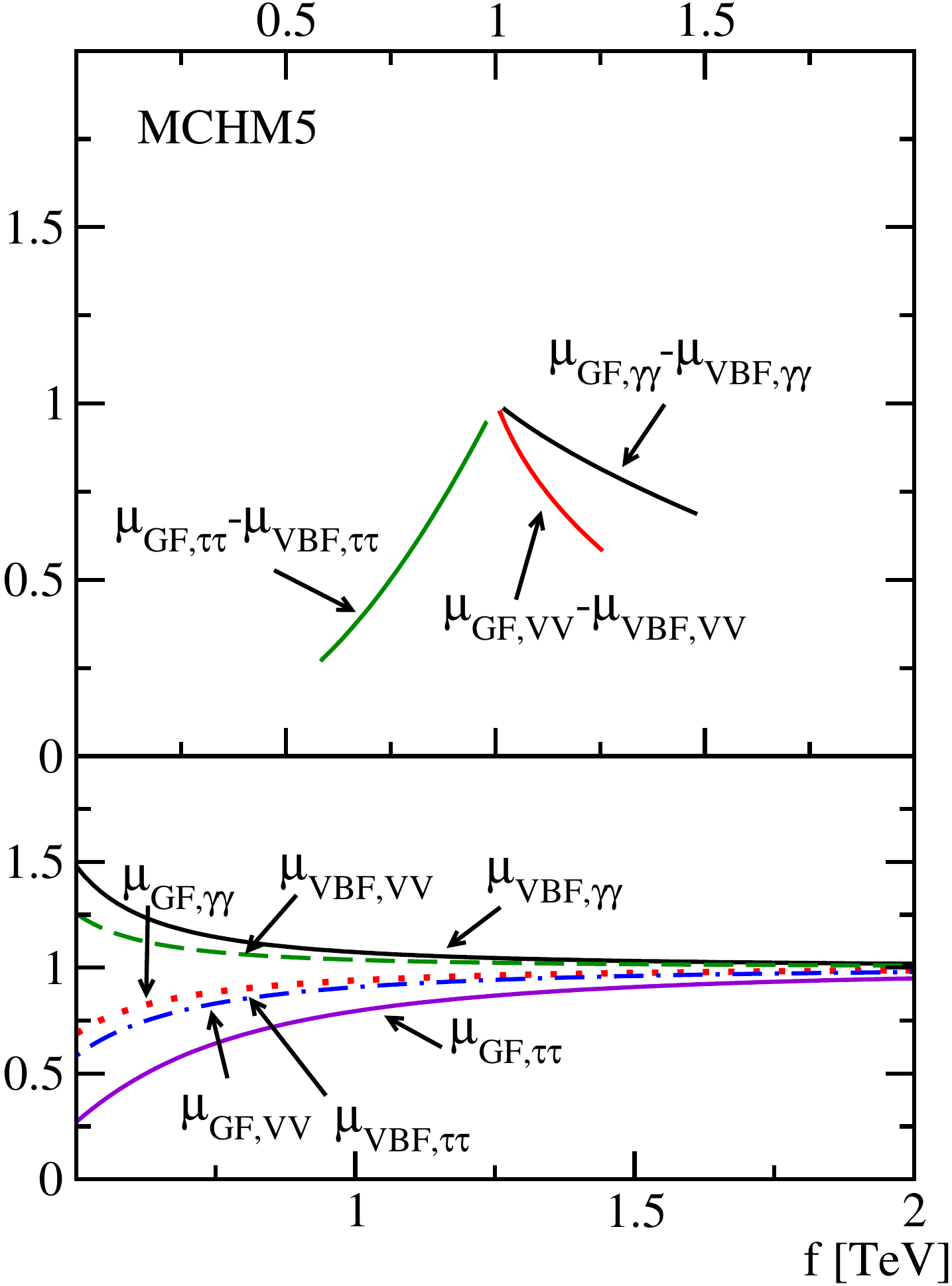} 
\hspace*{0.16\textwidth}
\includegraphics[width=0.35\textwidth]{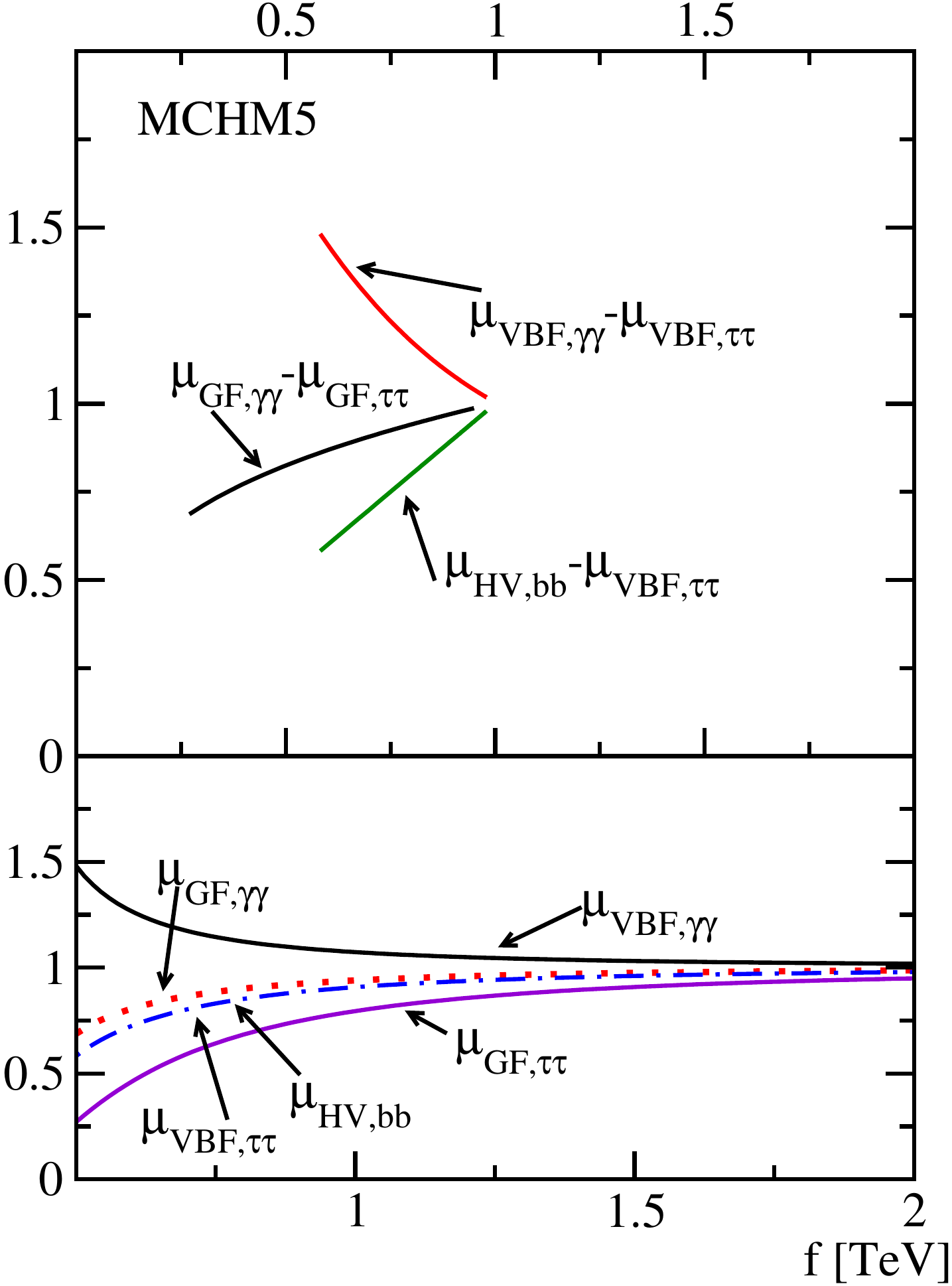} 
\end{center}
\vspace*{-7mm}
\caption{Composite Higgs: correlated Higgs signal strengths for the
  decay--diagonal channels (left) and non--diagonal channels (right).
  The first of the two signal strengths in the notation
  $\mu_{p_1,d_1} - \mu_{p_2,d_2}$ is shown on the vertical axis,
  the second on the horizontal axis.  In the lower panels we give the
  different signal strengths as a function of $f = v/\xi$.}
\label{fig:correl-mchm}
\end{figure}

Depending on the symmetry structure this strongly interacting but
Goldstone--protected Higgs sector predicts different coupling patterns
for fermions and gauge bosons~\cite{mchm}. The MCHM4 setup with
$\Delta_V = \Delta_f = \sqrt{1-\xi^2}-1$ is experimentally equivalent
to a mixing term in a singlet extension.  The reduced couplings of the
lightest Higgs state reflect the fact that at the energy scale $f$
there exist many strongly interacting Higgs fields which share the
unitarization of the usual $2 \to2$ scattering processes. The MCHM5
setup is more interesting. While $\Delta_V$ is identical to the MCHM4
case the fermions follow a different pattern, $1+ \Delta_f =
(1-2\xi^2)/\sqrt{1-\xi^2}$.  If we ignore the heavy states'
contributions to the the effective Higgs--photon and Higgs--gluon
couplings we find that the ratio of production rates scales like
\begin{alignat}{5}
\frac{\mu_{\text{VBF},d}}{\mu_{\text{GF},d}} 
&= \left( \frac{1 + \Delta_V}{1 + \Delta_f} \right)^2 
 = \frac{\phantom{xxxx} 1-\xi^2 \phantom{xxxx}}{\cfrac{(1-2\xi^2)^2}{1-\xi^2}} 
 = \left( \frac{1-\xi^2}{1-2\xi^2} \right)^2
 = 1 + 2 \xi^2 + \mathcal{O}(\xi^3) \notag \\
\frac{\mu_{\text{VBF},VV}}{\mu_{\text{GF},ff}} 
&= \left( \frac{1 + \Delta_V}{1 + \Delta_f} \right)^4 
 = 1 + 4 \xi^2 + \mathcal{O}(\xi^3) 
\qqqquad \qqquad
\frac{\mu_{\text{VBF},ff}}{\mu_{\text{GF},VV}} 
= 1 \; .
\label{eq:scale_mchm_app}
\end{alignat}
Unlike all previous models the MCHM5 setup accommodates 
$\mu_{p,d} > 1$, for example 
\begin{alignat}{5}
\mu_{\text{VBF},VV} 
&\simeq \dfrac{\left( 1 - \dfrac{\xi^2}{2} \right)^4}
              {0.3 \left( 1 - \dfrac{\xi^2}{2} \right)^2 
             + 0.7 \left( 1 - \dfrac{3\xi^2}{2} \right)^2} 
\simeq 1 + 0.4 \, \xi^2 \ + \mathcal{O}(\xi^3) > 1 
\label{eq:mchm-wbfvv}.
\end{alignat}
The sharper suppression in the fermion couplings still leads to 
depleted gluon fusion channels.
%
%
In Figure~\ref{fig:correl-mchm} we show a set of correlations between
different signal strengths. The different possibilities portrayed in
Eq.\eqref{eq:mchm-wbfvv} explain some of the patterns in the signal
strength planes. Corrections to the straight linear correlations arise
at $\mathcal{O}(\xi^3)$.

\paragraph{Additional doublet}

\begin{figure}[t]
\begin{center}
\includegraphics[width=0.24\textwidth]{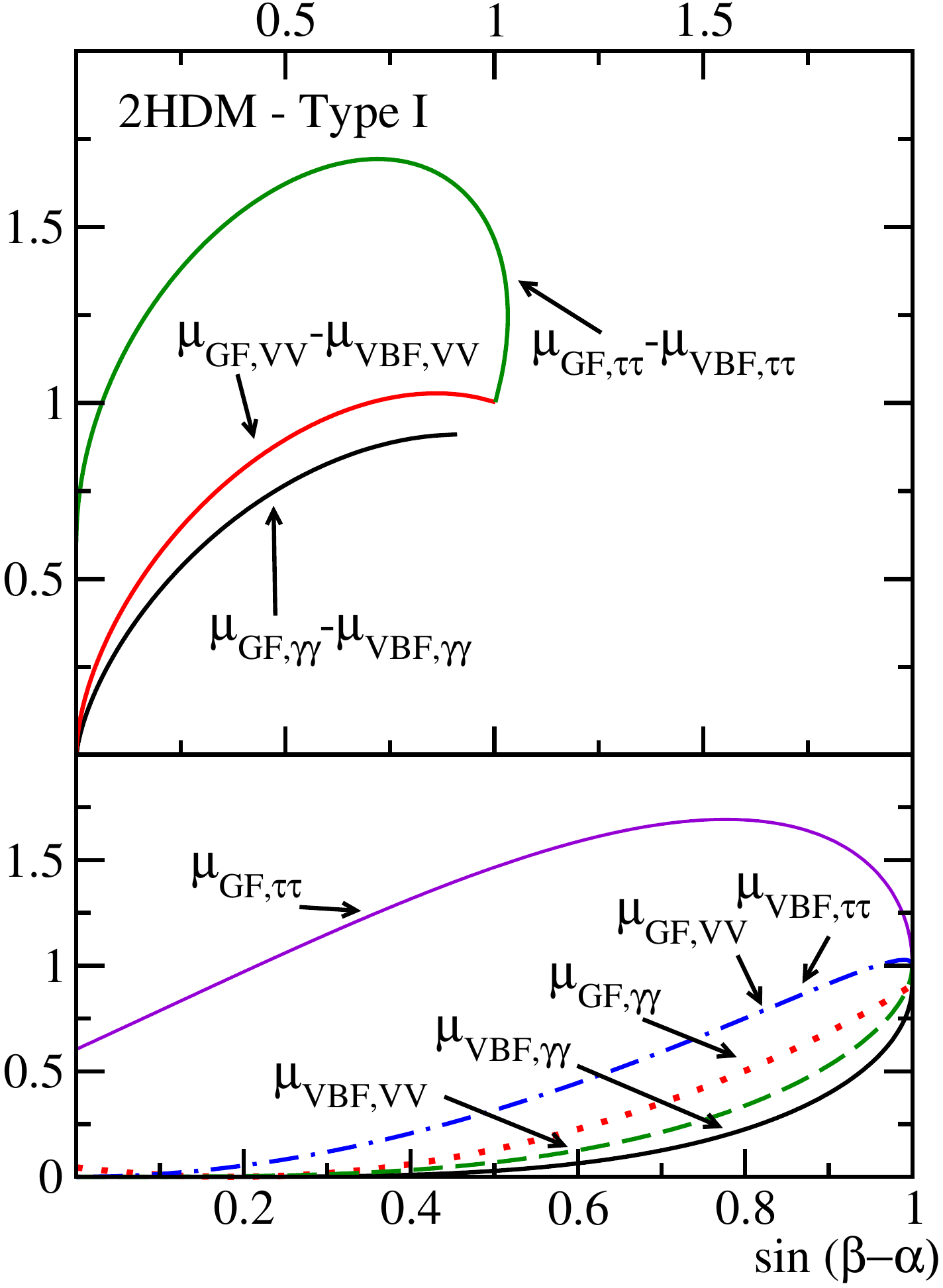} 
\includegraphics[width=0.24\textwidth]{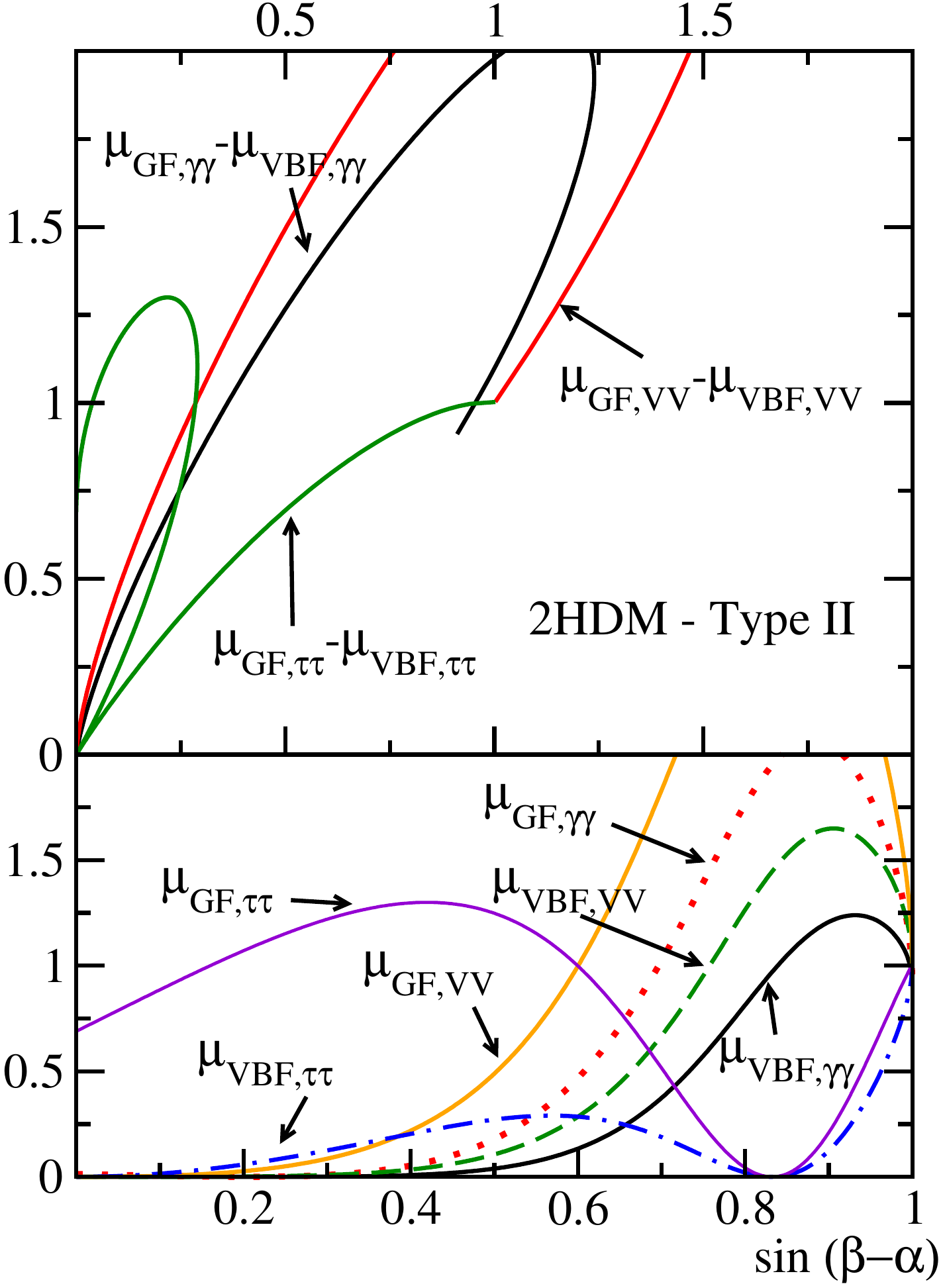} 
\includegraphics[width=0.24\textwidth]{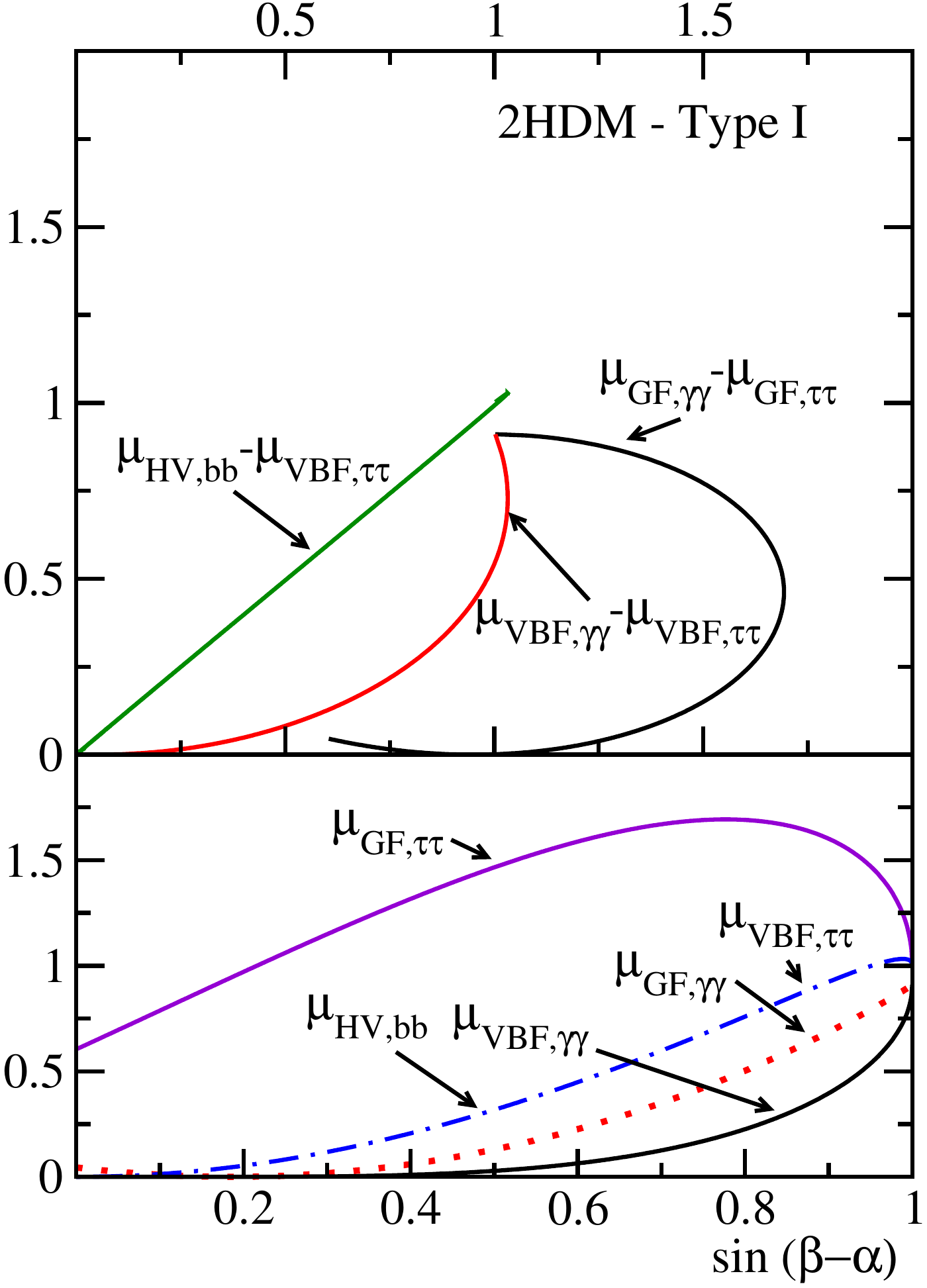} 
\includegraphics[width=0.24\textwidth]{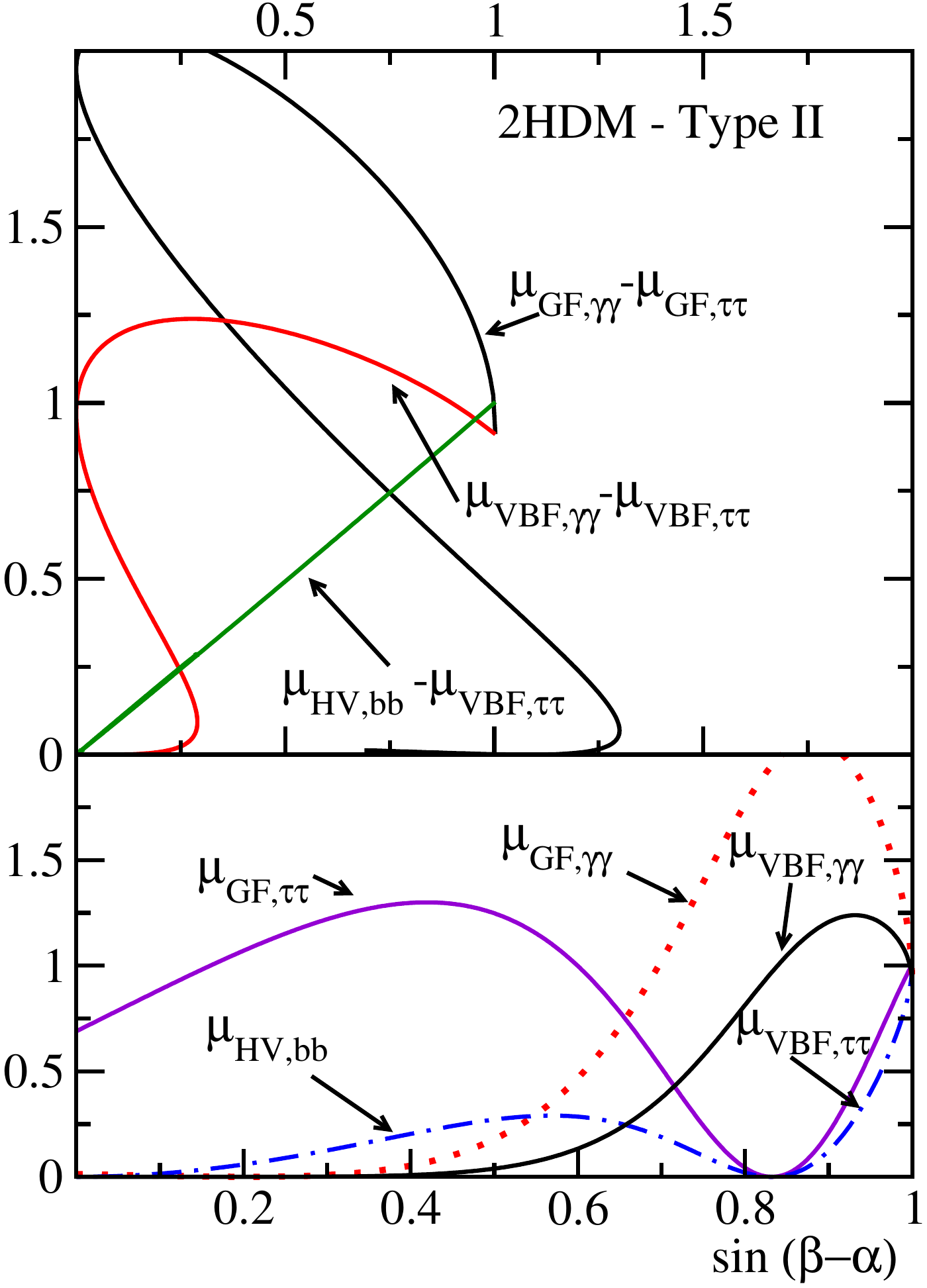} 
\end{center}
\vspace*{-7mm}
\caption{2HDM: correlated Higgs signal strengths for decay--diagonal
  channels (left two panels) and non--diagonal channels (right two
  panels), each for type-I and type-II setups.  The first of the two
  signal strengths in the notation $\mu_{p_1,d_1} - \mu_{p_2,d_2}$
  is shown on the vertical axis, the second on the horizontal axis.
  In the lower panels we give the different signal strengths as a
  function of $\cos (\beta - \alpha) = \xi$.}
\label{fig:correl-thdm}
\end{figure}

An even richer structure in particular in separating the gauge and
Yukawa couplings of the light Higgs appears when we add a second Higgs
doublet~\cite{2hdm,2hdm_flavor}. Up to terms of $\mathcal{O}(\xi^3)$
the modification to the LHC signal strengths
$(1+\Delta_p)^2\,(1+\Delta_d)^2 - 1$ reads
%
\begin{center}
\begin{tabular}{l|l|l|l} 
\hline
& $b\bar{b}$ & $\tau \tau$ & $VV$  \\ \hline 
\multirow{3}{1.cm}{GF} 
& $2\xi[\cot\beta-\tan(\beta-\gamma_b)]$  
& $2\xi[\cot\beta+\tan(\beta-\gamma_\tau)]$ 
& $2\xi\cot\beta$ \\    
& $ -\xi^2 \; [2-\tan^2(\beta-\gamma_b)-\cot^2\beta $ 
& $ -\xi^2 \; [2-\tan^2(\beta-\gamma_\tau)-\cot^2\beta $ 
& $-\xi^2 \; (2-\cot^2\beta)$ \\ 
& \qquad $+4\cot\beta\tan(\beta-\gamma_b)]$ 
& \qquad $+4\cos\beta\tan(\beta-\gamma_\tau)]$ 
&  \\ \hline
VBF  
& $-2\xi\tan(\beta-\gamma_b)$ 
& $-2\xi\tan(\beta-\gamma_\tau)$ 
& $-2\xi^2$ \\ 
& $-\xi^2 \; [2-\tan^2(\beta-\gamma_b)]$ &
  $-\xi^2 \; [2-\tan^2(\beta-\gamma_\tau)]$ &  \\ 
\hline
\end{tabular}
\end{center} 
%
These expressions hold for the general Yukawa--aligned
model~\cite{sfitter_2hdm,2hdm_align}. For example a type-I model
corresponds to $\gamma_b = \gamma_\tau = \pi/2$, while a type-II model
appears if we set $\gamma_b = \gamma_\tau = 0$. The main feature is
that the leading signal strength deviations involving fermions arise at order
$\xi$, while gauge couplings only vary with $\xi^2$~\cite{2hdm_coll}.

The detailed 2HDM signal strength patterns illustrated in Fig.~\ref{fig:correl-thdm} are to a large extend
model--dependent, as they are tied to the specific Yukawa structures.
In the type-I setup all Yukawas couplings and hence the effective
Higgs--gluon coupling are shifted by a common factor
$\cos\alpha/\sin\beta$. This way they are not suppressed for
$\sin(\beta-\alpha) \lesssim 1$. The Higgs couplings to gauge bosons
scale like $\sin(\beta-\alpha)$.  As a result, all gluon fusion
channels show an increased signal strength as compared to their weak
boson fusion counterparts. Type-II models link up--like and down--like
fermions to different Higgs doublets, implying separate Yukawa
modifications.  The bottom and tau final states are suppressed for a
wide range of $\sin(\beta-\alpha)$ -- eventually leading
to the unphysically large signal strength deviations visible
in the plots. For an even larger deviation from
$\sin(\beta-\alpha) \simeq 1$ models of type-II, lepton--specific models, and
flipped models feature sign--inverted Yukawas. These sign ambiguities are
visible in the ellipsoidal correlated variations in the signal
strength plane. The additional charged Higgs--mediated contribution
$\Delta_\gamma$ is responsible for 
the mild offset $\mu_{p,\gamma\gamma}\neq 1$ in the decoupling limit.

\paragraph{MSSM}

\begin{figure}[b!]
\begin{center}
\includegraphics[width=0.24\textwidth]{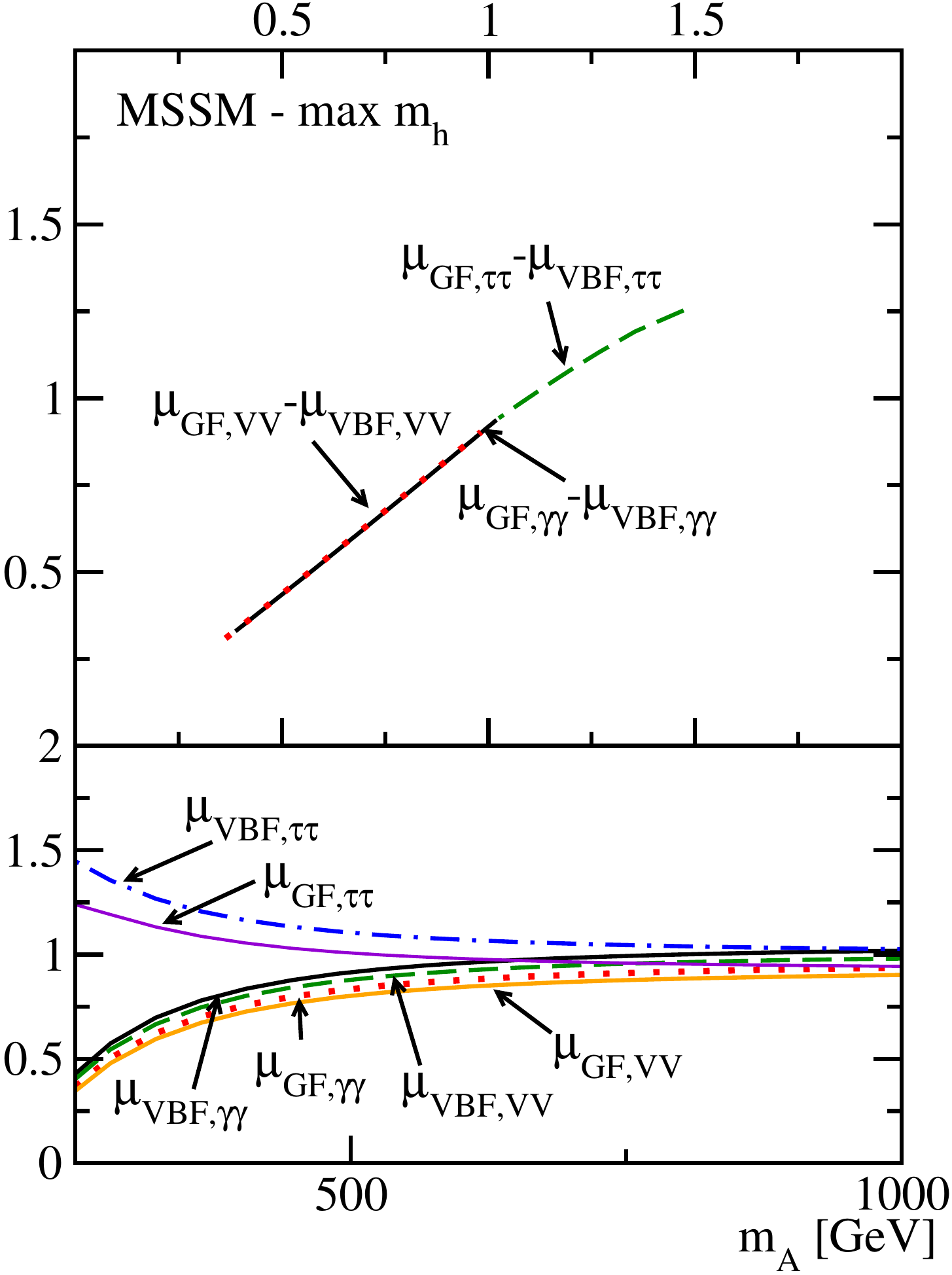} 
\includegraphics[width=0.24\textwidth]{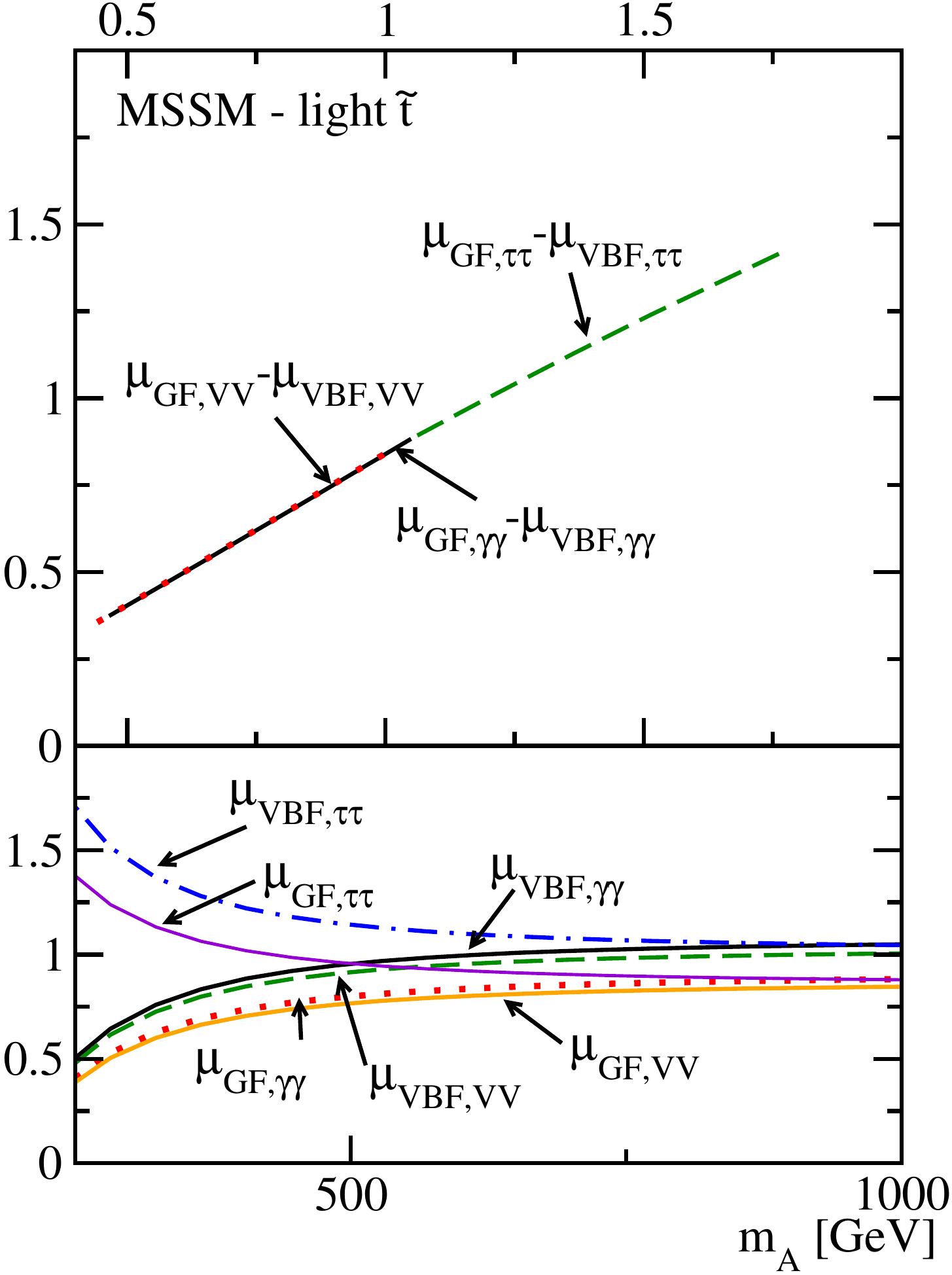} 
\includegraphics[width=0.24\textwidth]{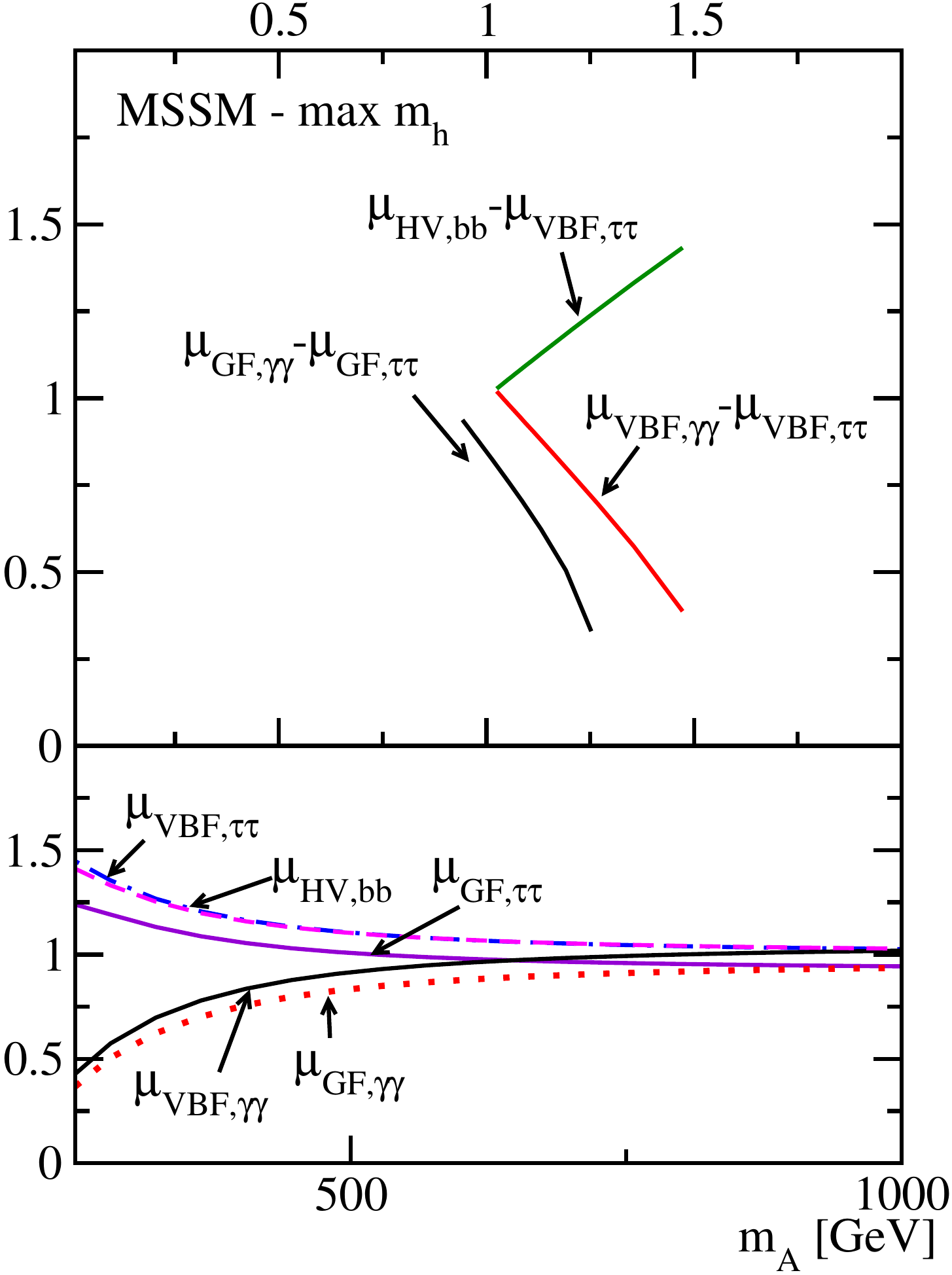} 
\includegraphics[width=0.24\textwidth]{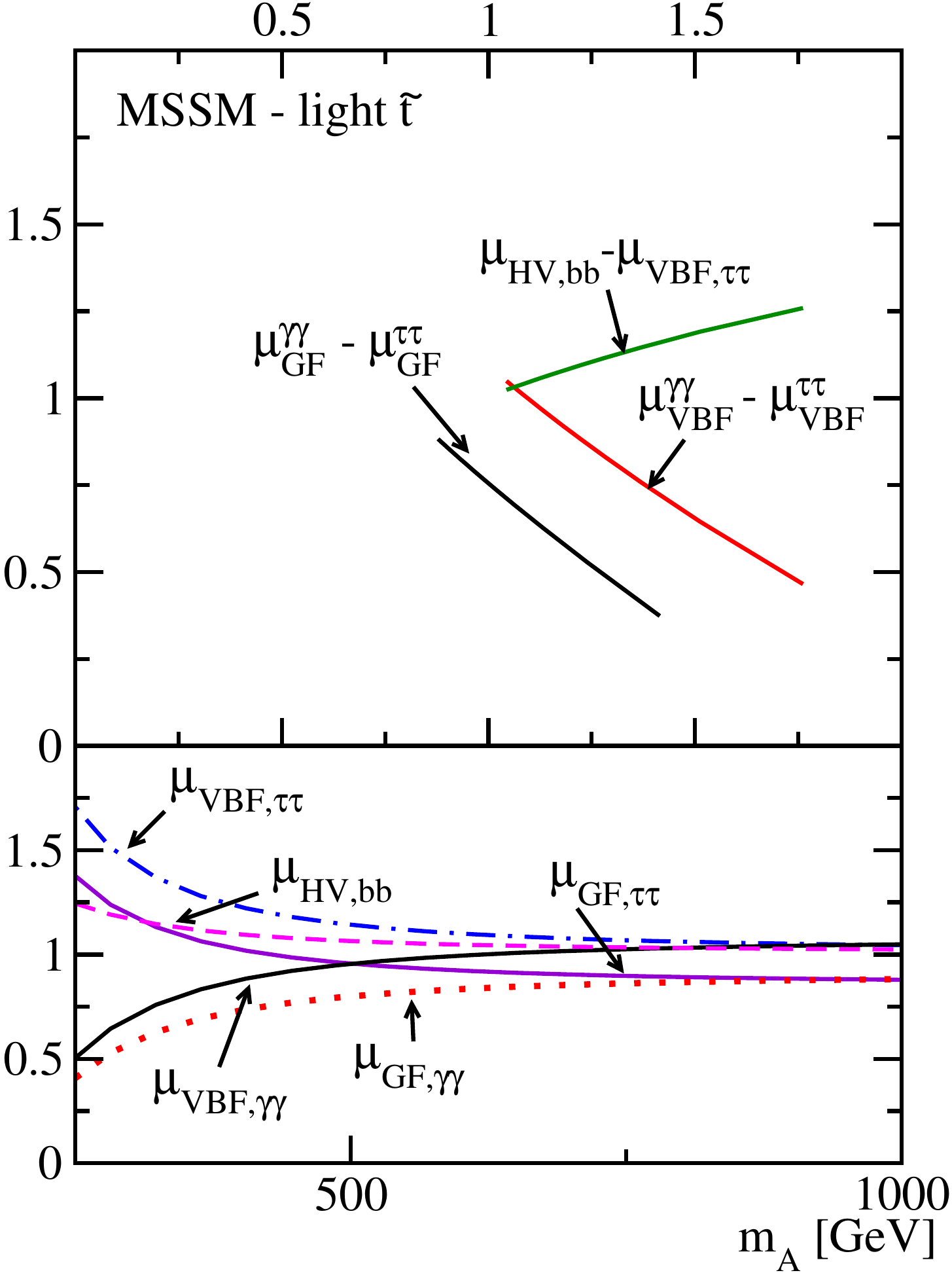} 
\end{center}
\vspace*{-7mm}
\caption{MSSM: correlated Higgs signal strengths for decay--diagonal
  channels (left two panels) and non--diagonal channels (right two
  panels), each for two MSSM benchmark spectra.  The first of the
  two signal strengths in the notation $\mu_{p_1,d_1} -
  \mu_{p_2,d_2}$ is shown on the vertical axis, the second on the
  horizontal axis.  In the lower panels we give the different signal
  strengths as a function of $m_A$, linked to $\xi$ via
  Eq.\protect\eqref{eq:mssm-decoup}.}
\label{fig:correl-mssm}
\end{figure}

A supersymmetric Lagrangian requires two Higgs doublets to give mass
to up-type and down-type fermions. At tree level the MSSM Higgs sector
is a type-II 2HDM, but with the different quartic Higgs couplings
fixed to gauge couplings. If we assume the lighter of the scalar Higgs
bosons to lie at 126~GeV the heavy Higgs masses are almost degenerate
at $m_{\Hzero} \simeq m_{\Azero} \simeq m_{\PHiggs^{\pm}}$ up to
$\mathcal{O}(m^2_Z/m^2_{\Azero})$ corrections.  Just like for the
general 2HDM the decoupling is described by $\xi = \cos (\beta -
\alpha)$.

At tree level the entire MSSM Higgs sector is fully described by
$m_{\Azero}$ and $\tan \beta$. Loop corrections in the top--stop
sector lead to significant corrections and yield an additional
parameter dependence for example on the stop trilinear coupling $A_t$.
The leading quantum corrections to the Higgs couplings can be accounted for
through an effective mixing angle $\alpha \to \alpha_\text{eff}$ which
we use in the definition of $\xi$.  Additional contributions may
further modify the effective Higgs coupling pattern. One--loop
triangle corrections to the Yukawa couplings are governed by
gluino--sbottom loops~\cite{deltab} and shift the bottom quark Yukawas
by $\Delta_b \sim \mathcal{O}(m_Z^2\,\tan^2\beta/m^2_{\Azero})$. They
give rise to a delayed decoupling~\cite{mssm}, meaning that the
decoupling limit now demands $m_Z^2\,\tan^2\beta \ll m^2_{\Azero}$ for
large $\tan\beta$.  These effects depend on the detailed
supersymmetric mass spectrum and have to be included in a full MSSM
parameter study~\cite{mssm_para}. In this discussion of the light
Higgs couplings we omit any constraint from the new particles outside
the Higgs sector.  This also means that we ignore the possibility of
light neutralinos $2 m_{\chi_1^0} < m_{\hzero}$ contributing to the
invisible Higgs width. In specific supersymmetric models the coupling
of the lightest neutralino to the light Higgs boson can be important,
leading to an accelerated dark matter annihilation in the early
universe. Nevertheless, even the couplings required for this light
Higgs funnel are unlikely to give a measurable invisible Higgs decay
rate at the LHC.\bigskip

Technically, we use the MSSM Higgs cross sections and branching ratios
given by {\sc FeynHiggs~2.9.5}~\cite{feynhiggs}.  The corresponding
Higgs signal strengths are obtained by normalizing the individual
production and decay rates to their SM counterparts~\cite{hxsec},
after identifying the (lightest) Higgs mass in both models
$m^\text{SM}_{\PHiggs} = m_{\hzero}^\text{MSSM}$.  We consider two
benchmarks relevant to LHC searches~\cite{mssm_benchmarks}, all
compatible with the observed $\sim 126$~GeV resonance: the
$m_h^\text{max}$ case with maximum stop mixing, heavy squarks, and
$\tan \beta = 10$ and the light stop case with maximum stop mixing,
but generally light squarks, and $\tan\beta = 35$. Both of them are
numerically similar. The difference is that only the light stop
scenario generates sizeable $\mathcal{O}(10\%)$ corrections to
$\Delta_g$ and at the same time gives relatively large negative
$\Delta_b$ corrections to the bottom Yukawa.  In
Fig.~\ref{fig:correl-mssm} we see that compared to the general 2HDM,
the possible departures from the linear correlations are milder.  The
two benchmarks predict very similar signal strengths.  All this can be
understood from the more constrained Higgs potential and the moderate
negative corrections to the total width. The largest deviations
obviously arise in the low-$m_{\Azero}$ regime.


\end{document}

%% file: declare.tex
\def\vec#1{\ifmmode
\mathchoice{\mbox{\boldmath$\displaystyle\bf#1$}}
{\mbox{\boldmath$\textstyle\bf#1$}}
{\mbox{\boldmath$\scriptstyle\bf#1$}}
{\mbox{\boldmath$\scriptscriptstyle\bf#1$}}\else
{\mbox{\boldmath$\bf#1$}}\fi}

\newcommand{\mpar}[1]{\rule{2pt}{10pt}
                      {\marginpar{\hbadness10000
                      \sloppy\hfuzz10pt\boldmath\bf\footnotesize#1}}
                       \typeout{marginpar: #1}\ignorespaces}
\def\mda{\mpar{\hfil$\downarrow$\hfil}\ignorespaces}
\def\mua{\mpar{\hfil$\uparrow$\hfil}\ignorespaces}
\def\mla{\marginpar[\boldmath\hfil$\rightarrow$\hfil]
                   {\boldmath\hfil$\leftarrow $\hfil}
                    \typeout{marginpar: $\leftrightarrow$}\ignorespaces}

\newcommand{\Pois}{{\ensuremath\text{Pois}}}
\newcommand{\Gaus}{{\ensuremath\text{Gaus}}}
\newcommand{\mueff}{{\ensuremath{\mu}^\text{eff}}}
\newcommand{\vecmueff}{{\ensuremath{\vec\mu}^\text{eff}}}
\newcommand{\alphaeff}{{\ensuremath{\alpha}^\text{eff}}}
\newcommand{\munoth}{{\ensuremath{\mu}_\text{exp}}}
\def\muefff#1{{\ensuremath{\mu}_{#1}^\text{eff}}}

\def\contentsname{{\normalsize Content}}
\def\tablename{Table}
\def\figurename{Figure}

\def\pveto{P_\text{veto}}
\def\nj{n_\text{jets}}
\def\meff{m_\text{eff}}
\def\ptmin{p_T^\text{min}}
\def\gtot{\Gamma_\text{tot}}
\def\as{\alpha_s}
\def\az{\alpha_0}
\def\gz{g_0}
\def\w{\vec{w}}
\def\sdag{\Sigma^{\dag}}
\def\s{\Sigma}
\newcommand{\psib}{\overline{\psi}}
\newcommand{\Psib}{\overline{\Psi}}
\newcommand\one{\leavevmode\hbox{\small1\normalsize\kern-.33em1}}
\newcommand{\Mpl}{M_\mathrm{Pl}}
\newcommand{\p}{\partial}
\newcommand{\lag}{\mathcal{L}}
\newcommand{\ord}{\mathcal{O}}
\newcommand{\qqquad}{\qquad \qquad}
\newcommand{\qqqquad}{\qquad \qquad \qquad}

\newcommand{\qb}{\bar{q}}
\newcommand{\matx}{|\mathcal{M}|^2}
\newcommand{\really}{\stackrel{!}{=}}
\newcommand{\msbar}{\overline{\text{MS}}}
\newcommand{\qns}{f_q^\text{NS}}
\newcommand{\lqcd}{\Lambda_\text{QCD}}
\newcommand{\met}{\slashchar{p}_T}
\newcommand{\pmiss}{\slashchar{\vec{p}}_T}

\newcommand{\sq}{\tilde{q}}
\newcommand{\go}{\tilde{g}}
\newcommand{\st}[1]{\tilde{t}_{#1}}
\newcommand{\stb}[1]{\tilde{t}_{#1}^*}
\newcommand{\nz}[1]{\tilde{\chi}_{#1}^0}
\newcommand{\cp}[1]{\tilde{\chi}_{#1}^+}
\newcommand{\cm}[1]{\tilde{\chi}_{#1}^-}
\newcommand{\CP}{CP}

\providecommand{\mg}{m_{\tilde{g}}}
\providecommand{\mst}{m_{\tilde{t}}}
\newcommand{\msn}[1]{m_{\tilde{\nu}_{#1}}}
\newcommand{\mch}[1]{m_{\tilde{\chi}^+_{#1}}}
\newcommand{\mne}[1]{m_{\tilde{\chi}^0_{#1}}}
\newcommand{\msb}[1]{m_{\tilde{b}_{#1}}}
\newcommand{\vsm}{\ensuremath{v_{\rm SM}}}

\newcommand{\mev}{{\ensuremath\rm MeV}}
\newcommand{\gev}{{\ensuremath\rm GeV}}
\newcommand{\tev}{{\ensuremath\rm TeV}}
\newcommand{\fb}{{\ensuremath\rm fb}}
\newcommand{\ab}{{\ensuremath\rm ab}}
\newcommand{\pb}{{\ensuremath\rm pb}}
\newcommand{\sign}{{\ensuremath\rm sign}}
\newcommand{\ifb}{{\ensuremath\rm fb^{-1}}}
\newcommand{\ipb}{{\ensuremath\rm pb^{-1}}}

\def\slashchar#1{\setbox0=\hbox{$#1$}           
   \dimen0=\wd0                                 
   \setbox1=\hbox{/} \dimen1=\wd1               
   \ifdim\dimen0>\dimen1                        
      \rlap{\hbox to \dimen0{\hfil/\hfil}}      
      #1                                        
   \else                                        
      \rlap{\hbox to \dimen1{\hfil$#1$\hfil}}   
      /                                         
   \fi}
\newcommand{\dslash}{\slashchar{\partial}}
\newcommand{\Dslash}{\slashchar{D}}

\def\eg{{\sl e.g.} \,}
\def\ie{{\sl i.e.} \,}
\def\etal{{\sl et al} \,}

\setlength{\floatsep}{0pt}
\setcounter{topnumber}{1}
\setcounter{bottomnumber}{1}
\setcounter{totalnumber}{1}
\renewcommand{\topfraction}{1.0}
\renewcommand{\bottomfraction}{1.0}
\renewcommand{\textfraction}{0.0}
\renewcommand{\thefootnote}{\fnsymbol{footnote}}

\newcommand{\rig}{\rightarrow}
\newcommand{\lrig}{\longrightarrow}
\renewcommand{\d}{{\mathrm{d}}}
\newcommand{\be}{\begin{eqnarray*}}
\newcommand{\ee}{\end{eqnarray*}}
\newcommand{\gl}[1]{(\ref{#1})}
\newcommand{\ta}[2]{ \frac{ {\mathrm{d}} #1 } {{\mathrm{d}} #2}}
\newcommand{\bee}{\begin{eqnarray}}
\newcommand{\eee}{\end{eqnarray}}
\newcommand{\beeq}{\begin{equation}}
\newcommand{\eeeq}{\end{equation}}
\newcommand{\mc}{\mathcal}
\newcommand{\mr}{\mathrm}
\newcommand{\ep}{\varepsilon}
\newcommand{\emt}{$\times 10^{-3}$}
\newcommand{\emfo}{$\times 10^{-4}$}
\newcommand{\emfi}{$\times 10^{-5}$}

\newcommand{\revision}[1]{{\bf{}#1}}

\newcommand{\hzero}{h^0}
\newcommand{\Hzero}{H^0}
\newcommand{\Azero}{A^0}
\newcommand{\PHiggs}{H}
\newcommand{\PW}{W}
\newcommand{\PZ}{Z}

\newcommand{\sw}{\ensuremath{s_w}}
\newcommand{\cw}{\ensuremath{c_w}}
\newcommand{\swd}{\ensuremath{s^2_w}}
\newcommand{\cwd}{\ensuremath{c^2_w}}

\newcommand{\mhhd}{\ensuremath{m^2_{\Hzero}}}
\newcommand{\mhh}{\ensuremath{m_{\Hzero}}}
\newcommand{\mlhd}{\ensuremath{m^2_{\hzero}}}
\newcommand{\Mlh}{\ensuremath{m_{\hzero}}}
\newcommand{\mad}{\ensuremath{m^2_{\Azero}}}
\newcommand{\mhpd}{\ensuremath{m^2_{\PHiggs^{\pm}}}}
\newcommand{\mhp}{\ensuremath{m_{\PHiggs^{\pm}}}}

 \newcommand{\sa}{\ensuremath{\sin\alpha}}
 \newcommand{\ca}{\ensuremath{\cos\alpha}}
 \newcommand{\cad}{\ensuremath{\cos^2\alpha}}
 \newcommand{\sad}{\ensuremath{\sin^2\alpha}}
 \newcommand{\sbd}{\ensuremath{\sin^2\beta}}
 \newcommand{\cbd}{\ensuremath{\cos^2\beta}}
 \newcommand{\cb}{\ensuremath{\cos\beta}}
 \renewcommand{\sb}{\ensuremath{\sin\beta}}
 \newcommand{\tanbd}{\ensuremath{\tan^2\beta}}
 \newcommand{\cotbd}{\ensuremath{\cot^2\beta}}
 \newcommand{\tanb}{\ensuremath{\tan\beta}}
 \newcommand{\tb}{\ensuremath{\tan\beta}}
 \newcommand{\cotb}{\ensuremath{\cot\beta}}

\newcommand{\GeV}{\ensuremath{\rm GeV}}
\newcommand{\MeV}{\ensuremath{\rm MeV}}
\newcommand{\TeV}{\ensuremath{\rm TeV}}